\documentclass[aps,prx,twocolumn,superscriptaddress]{revtex4-2}
\usepackage{amsmath,amssymb,graphicx,bm}
\usepackage[utf8]{inputenc}
\usepackage{multirow}
\usepackage{chemmacros}
\usepackage{mathtools}
\usepackage[lining,semibold]{libertine} % a bit lighter than Times--no osf in math
\usepackage[libertine, cmintegrals, bigdelims, vvarbb]{newtxmath}
\usepackage{amsmath}
\usepackage{amsfonts}
\usepackage{mathrsfs}
\usepackage{gensymb}
\usepackage{bbm}
\usepackage{dsfont}

\usepackage{chemformula}
\usepackage[caption=false]{subfig}
\usepackage{chemfig}
\usepackage[version=3]{mhchem}

\usepackage{tikz}
\usetikzlibrary{matrix,positioning,decorations.pathreplacing}

\usepackage{braket}

\usepackage{soul}
\usepackage{cancel}
\usepackage{xcolor}

\usepackage{scalerel} % To change size things in math environment \scaleto{...}{1pt}
\usepackage{comment}

\usepackage{blkarray}

\usetikzlibrary{arrows, automata}
\usepackage{algorithm}
\usepackage{algpseudocode}

\definecolor{webgreen}{rgb}{0,.5,0}
\definecolor{webbrown}{rgb}{.6,0,0}
\definecolor{WEBBROWN}{rgb}{.6,0,0}
\definecolor{grigio}{rgb}{.85,.85,.85} 
\definecolor{RoyalBlue}{rgb}{0.0, 0.14, 0.4}
\definecolor{skyblue1}{rgb}{0.45,0.62,0.81}
\definecolor{skyblue2}{rgb}{0.2,0.39,0.64}
\definecolor{skyblue3}{rgb}{0.13,0.29,0.53}
\definecolor{scarlet1}{rgb}{0.93,0.16,0.16}
\definecolor{scarlet2}{rgb}{0.8,0,0}
\definecolor{scarlet3}{rgb}{0.64,0,0}

\definecolor{g}{gray}{0.50}

\usepackage{hyperref}
\hypersetup{%
    colorlinks=true, linktocpage=true, pdfstartpage=1, pdfstartview=FitV,%
    breaklinks=true, pdfpagemode=UseNone, pageanchor=true, pdfpagemode=UseOutlines,%
    plainpages=false, bookmarksnumbered, bookmarksopen=true, bookmarksopenlevel=1,%
    hypertexnames=true, pdfhighlight=/O,%nesting=true,%frenchlinks,%
    urlcolor=webbrown, linkcolor=RoyalBlue, citecolor=webgreen, %pagecolor=RoyalBlue,%
    pdftitle={},%
    pdfauthor={Benedikt Remlein},%
    pdfsubject={},%
    pdfkeywords={},%
    pdfcreator={pdfLaTeX},%
    pdfproducer={LaTeX REVTeX}%
}

\usepackage[capitalise]{cleveref}

\usepackage{verbatim}

\usepackage{kbordermatrix}

\usetikzlibrary{calc}

\newcommand{\remark}{\textit{Remark. }}
\newcommand{\nremark}{\textit{Notation Remark. }}
\newcommand{\mremark}{\textit{Mathematical Remark. }}

\newcommand{\avg}[1]{\langle #1 \rangle}
\newcommand{\avgInf}[1]{\langle #1 \rangle_{\infty}^{(0)}}

\newcommand{\sys}{{A}}
\newcommand{\res}{{B}}

\newcommand{\kb}{k_{\mathrm{B}}}

\newcommand{\bn}{\boldsymbol n}

\newcommand{\bni}{\boldsymbol n}

\newcommand{\bmi}{\boldsymbol m}

\newcommand{\bnm}{\boldsymbol m}
\newcommand{\bL}{\boldsymbol L}

\newcommand{\Spe}{\mathcal S}
\newcommand{\Spei}{\mathcal S_{{i}}}
\newcommand{\Spec}{\mathcal S_{{c}}}
\newcommand{\Spex}{\mathcal S_{{x}}}
\newcommand{\Spey}{\mathcal S_{{y}}}
\newcommand{\SpeY}{\mathcal S_{{Y}}}
\newcommand{\SpeYY}{\mathcal{Y}}

\newcommand{\NYy}{N_{\SpeYY(y)}}

\newcommand{\rhoi}{\rho_{\mathrm i}}
\newcommand{\rhoe}{\rho_{\mathrm e}}

\newcommand{\Rct}{\mathcal R}
\newcommand{\Rint}{\mathcal R_{\mathrm{int}}}
\newcommand{\Rexc}{\mathcal R_{\mathrm{exc}}}

\newcommand{\colS}{\mathbf{S}}
\newcommand{\colSi}{\mathbf{S}_{\rhoi}}
\newcommand{\colSxi}{\mathbf{S}_{x,\rhoi}}
\newcommand{\colSyi}{\mathbf{S}_{y,\rhoi}}

\newcommand{\colSe}{\mathbf{S}_{\rhoe}}

\newcommand{\W}{W}

\newcommand{\Wexc}{\hat W_{\mathrm{exc}}}
\newcommand{\Wint}{\hat W_{\mathrm{int}}}

\newcommand{\piInf}{\pi_{\infty}^{(0)}}
\newcommand{\piInfy}{\pi_{\infty}^{(y)}}

\newcommand{\epri}{\braket{\dot \Sigma_{\mathrm{i}}}}
\newcommand{\epre}{\braket{\dot \Sigma_{\mathrm{e}}}}

\newcommand{\epriq}{\braket{\dot \Sigma^{(q)}_{\mathrm{i}}}}
\newcommand{\epreq}{\braket{\dot \Sigma^{(q)}_{\mathrm{e}}}}

\newcommand{\eprez}{\braket{\dot \Sigma^{(0)}_{\mathrm{e}}}}

\newcommand{\eprif}{\braket{\dot \Sigma^{(1)}_{\mathrm{i}}}}
\newcommand{\epref}{\braket{\dot \Sigma^{(1)}_{\mathrm{e}}}}

\renewcommand{\epsilon}{\varepsilon}

\newcommand{\cl}{\boldsymbol{\ell}}
\newcommand{\cle}{{\ell}}
\newcommand{\icle}{\overline{\ell}}

\makeatletter
\def\maketag@@@#1{\hbox{\m@th\normalfont\normalsize#1}}
\makeatother

\DeclareMathAlphabet{\mathpzc}{OT1}{pzc}{m}{it}

\begin{document}
	
	\preprint{AIP/123-QED}

    \title{Emergence of Open Chemical Reaction Network Thermodynamics within Closed Systems}

    \author{Benedikt Remlein}
	\email{benedikt.remlein@uni.lu}
    \affiliation{Complex Systems and Statistical Mechanics, Department of Physics and Materials Science, University of Luxembourg, 30 Avenue des Hauts-Fourneaux, L-4362 Esch-sur-Alzette, Luxembourg}%

	\author{Massimiliano Esposito}%
	\email{massimiliano.esposito@uni.lu}
    \affiliation{Complex Systems and Statistical Mechanics, Department of Physics and Materials Science, University of Luxembourg, 30 Avenue des Hauts-Fourneaux, L-4362 Esch-sur-Alzette, Luxembourg}%

	\author{Francesco Avanzini}%
	\email{francesco.avanzini@unipd.it}
	\affiliation{Department of Chemical Sciences, University of Padova, Via F. Marzolo, 1, I-35131 Padova, Italy}%

	\date{\today}
%%%%%%%%%%%%%%%%%%%%%%%%%%%%%%%%%%%%%%%%%%%%%%%%%%%%%%%
%%%%%%%%%%%%%%%%%%%%%%%%%%%%%%%%%%%%%%%%%%%%%%%%%%%%%%%
%%%%%%%%%%%%%%%%%%%%%%%%%%%%%%%%%%%%%%%%%%%%%%%%%%%%%%%
%%%%%%%%%%%%%%%%%%%%%%%%%%%%%%%%%%%%%%%%%%%%%%%%%%%%%%%

\newpage
\begin{abstract}
% -
% -
% -
Under which conditions do the dynamics and thermodynamics of open chemical reaction networks (CRNs),
grounded on the notion of idealized chemostats that exchange selected species,
emerge from underlying closed CRNs?
% -
While open CRNs provide the standard framework to describe out-of-equilibrium chemical systems,
real systems are finite and ultimately relax to equilibrium,
leaving the status of this description conceptually unresolved.
% -
% -
Here, we address this fundamental question.
% -
We show that open-CRN behavior arises as an asymptotic regime of closed CRNs
when two physically minimal and transparent conditions are met:
a time-scale separation,
whereby fast reactions effectively act as exchange mechanisms,
and an abundance separation,
whereby a subset of species behaves as chemostats with diverging chemical capacity.
% -
In this regime,
both the stochastic dynamics and the thermodynamic structure
\--- including local detailed balance, entropy production, and free-energy balance \---
emerge to leading order from the underlying closed CRN.
% -
We further predict that this regime persists up to a characteristic time
proportional to the scale of the abundance separation,
thereby providing an experimentally accessible criterion for estimating how long
a finite closed system behaves as an open CRN.
% -
Our results apply to arbitrary stoichiometries.
% -
They show that chemostats need not be introduced as external idealizations,
but instead arise as emergent thermodynamic structures within closed systems,
providing a unified and physically grounded foundation for the nonequilibrium thermodynamics of CRNs.
% -
% -
\end{abstract}
%%%%%%%%%%%%%%%%%%%%%%%%%%%%%%%%%%%%%%%%%%%%%%%%%%%%%%%
%%%%%%%%%%%%%%%%%%%%%%%%%%%%%%%%%%%%%%%%%%%%%%%%%%%%%%%

\maketitle

%%%%%%%%%%%%%%%%%%%%%%%%%%%%%%%%%%%%%%%%%%%%%%%%%%%%%%%
%%%%%%%%%%%%%%%%%%%%%%%%%%%%%%%%%%%%%%%%%%%%%%%%%%%%%%%
%%%%%%%%%%%%%%%%%%%%%%%%%%%%%%%%%%%%%%%%%%%%%%%%%%%%%%%
%%%%%%%%%%%%%%%%%%%%%%%%%%%%%%%%%%%%%%%%%%%%%%%%%%%%%%%
%%%%%%%%%%%%%%%%%%%%%%%%%%%%%%%%%%%%%%%%%%%%%%%%%%%%%%%
%%%%%%%%%%%%%%%%%%%%%%%%%%%%%%%%%%%%%%%%%%%%%%%%%%%%%%%
%%%%%%%%%%%%%%%%%%%%%%%%%%%%%%%%%%%%%%%%%%%%%%%%%%%%%%%
\section{Introduction}

When chemical systems are fueled
\--- namely,
when free energy is extracted from the interconversion
of high-energy species (commonly referred to as fuel) 
into low-energy species (commonly referred to as waste) \---
thermodynamically unfavorable reactions can be sustained,
thereby giving rise to complex out-of-equilibrium phenomena.
% -
% -
Such phenomena are a defining feature of living systems~\cite{yang2021}.
In particular, metabolic processes transfer energy from nutrients to energetic mediator molecules 
\--- primarily adenosine triphosphate \--- 
which are subsequently consumed to power essential nonequilibrium functions.
These functions include, for instance,
directional transport via molecular motors~\cite{kolomeisky2007}, 
intracellular communication via calcium waves~\cite{Berridge2000, Falcke2004},
local chemical organization through biomolecular condensates~\cite{Banani:2017condensates, Zwicker2025},
as well as large-scale mechanical organization of the cytoskeleton~\cite{Prost2015}.
% -
% -
However, out-of-equilibrium phenomena are no longer exclusive to living systems. 
Over the past two decades,
artificial systems have been systematically designed and synthesized to 
generate analogous behaviors~\cite{hermans2017}.
Like living systems, these artificial systems can exhibit
directional motion~\cite{Patino2024, Borsley2024},
information transfer~\cite{Roberts2025},
local organization~\cite{Munana2018, Das2021},
as well as large-scale mechanical organization~\cite{Wang2025}.
% -
% -

% -
% -
Our understanding of the mechanisms underlying these complex out-of-equilibrium phenomena relies
to a large extent on
the nonequilibrium thermodynamics of open chemical reaction networks (CRNs)~\cite{gasp04, qian2005, schm06, rao16, rao18, avan21, spec21}.
% -
% -
Chemical systems are described in terms of their constituent species
together with the chemical reactions by which they are interconverted,
thereby defining a CRN~\cite{Klamt2009}.
% -
% -
Fueling is modeled by treating CRNs as open systems,
which exchange specific species with the environment,
represented by chemostats~\footnote{
Here, the environment should be understood in the general statistical-mechanical sense:
everything that is, broadly speaking, coupled to a system of interest,
but is not explicitly described in detail.
% -
In this regard,
chemostats can be interpreted as physically distinct chemical reservoirs,
for instance external solutions containing the exchanged species
at effectively fixed chemical potentials.
% -
However,
physical separation from the system is not required.
% -
Chemostats can also be realized by chemical species
homogeneously mixed with the system of interest.
For instance,
a buffer solution can act as a chemostat for protons,
as discussed in Sec.~\ref{sec:GeneralChemostatting}.}.
% -
%.
% -
A chemostat is an \textit{infinitely} large molecular reservoir 
that acts analogously to a thermostat:
the latter controls the temperature of a system while remaining at equilibrium, 
just as the former controls the abundance of a species while remaining at equilibrium.
% -
% -
Over the past decades, 
this nonequilibrium thermodynamic theory of open CRNs has developed into 
a powerful and systematic framework, both conceptually and quantitatively.
% -
It has, for instance, 
shown that exchanging species is a necessary, but not sufficient, condition 
for maintaining a CRN out of equilibrium~\cite{rao18}:
fueling can emerge only when the open CRN operates 
as a catalyst for emergent effective reactions
that solely involve the chemostatted species,
while the internal species are continuously restored to their initial abundances.
% -
Furthermore, this theory has elucidated
phenomena such as oscillations and chaos~\cite{andr08b, gaspard2020},
the transition to homochirality~\cite{Laurent2021, Laurent2022},
and nonequilibrium phase transitions~\cite{vell09, ge10, nguy18, fala18, nguy20, reml24}.
It has also revealed 
thermodynamic constraints on out-of-equilibrium dynamics
\---
including the cost of maintaining coherent oscillations~\cite{mars19, ober22, reml22, Santolin2025}
and the emergence of thermodynamic speed limits~\cite{yoshimura2021a, yoshimura2021b}
\---
and clarified the response of open CRNs to perturbations,
both in species abundances~\cite{Chun2023}
and in reaction currents~\cite{Samoilov2002, altaner2015, falasco2019ndr, marehalli2023}.
% -
Beyond its conceptual formulation,
this theory has yielded quantitative predictions.
These include
partitioning the free-energy cost of maintaining chemical systems out of equilibrium
into energy and information flows~\cite{Shuntaro2022, Penocchio2022, Leighton2025},
and quantifying the thermodynamic efficiency of processes
from self-assembly~\cite{Penocchio2019} to metabolism~\cite{wachtel2022, voor2024, Bila25b}.
% -
% -

% -
% -
Yet this theory rests on a crucial idealization:
the environment is represented in terms of chemostats.
% -
In reality, chemical systems are fueled by a \textit{finite} amount of fuel,
which is eventually depleted,
and produce only a \textit{finite} amount of waste,
which eventually accumulates.
% -
In other words, real chemical systems are fundamentally closed
and ultimately relax to equilibrium.
% -
% -
This points to a fundamental conceptual issue:
whether the nonequilibrium thermodynamics of open CRNs
reflects an external modeling idealization
or instead corresponds to an emergent regime of closed out-of-equilibrium systems.
% -
% -
Recent studies have begun to investigate how to move beyond the use of chemostats.
% - 
Some works have considered alternative models for exchange mechanisms~\cite{blokhuis2018, avan22, mare24a, mare24b},
while others have examined how finite molecular reservoirs affect CRNs~\cite{frit20}.
% -
However, 
their fundamental setup remains rooted in 
the framework of open CRNs.
% -
% -
Hence, a fundamental question remains:
under which precise and general conditions 
can the dynamics and thermodynamics of open CRNs
genuinely emerge from an underlying closed CRN?
% -
In a previous contribution~\cite{reml25},
we showed that, in the limit in which the abundances of certain species become macroscopic,
\textit{some} CRNs behave effectively as open ones.
However, this required restrictive stoichiometric constraints.
% -
% -
In this paper,
we show that 
open-CRN dynamics and thermodynamics do emerge as a well-defined regime of closed CRNs, 
and we identify general conditions under which this emergence occurs.
% -
% -

% -
% -
We do so by uncovering
the fundamental physical properties that underpin chemostats,
which are directly analogous to those underpinning thermostats.
% -
% -
A thermostat controls the temperature of a system
through the combination of two properties:
a fast energy-exchange mechanism and a diverging heat capacity.
% -
The former ensures that
any energy variation in the system is rapidly balanced,
while the latter ensures that
the thermostat remains effectively unaffected by these exchanges
and thus at equilibrium.
% -
Analogously, for a chemostat to control the abundance of a species in a CRN,
any variation in that abundance resulting from chemical reactions
must be rapidly balanced by exchanges with the chemostat,
while these exchanges are negligible from the perspective of the chemostat,
which therefore remains at equilibrium.
% -
We characterize this later property by introducing the notion of \textit{chemical capacity},
which must diverge in chemostats in analogy with the diverging heat capacity of thermostats.
% -
% -
We show that closed CRNs can realize these properties,
thereby 
generating an emergent regime in which
the stochastic dynamics governed by a chemical master equation
and the associated thermodynamic structure
\--- including local detailed balance, entropy production, and free-energy balance \---
coincide to leading order with those of an open CRN.
% -
% -

% -
% -
The paper is organized as follows.
The main text presents both the physical picture and the essential derivations of our theory.
Each section begins with a broad, physics-based discussion,
followed by the key derivations, with emphasis on their physical meaning.
% -
% -
We focus on CRNs at the microscopic scale 
(represented using the notation introduced in Sec.~\ref{sec:notation}),
where reactions are stochastic events 
and the numbers of molecules are fluctuating variables,
but our approach also extends to CRNs at the macroscopic scale.
% -
% -
% -
For open-CRN dynamics and thermodynamics (Sec.~\ref{sec:openCRNs}) to emerge,
the two fundamental properties underpinning chemostats must be realized within closed CRNs 
(Sec.~\ref{sec:closedCRNs}).
% -
% -
These properties arise under specific dynamical conditions 
on the reactions and species (Sec.~\ref{sec:chemo_like_dynamics}).
% -
First, some reactions must act as the exchange mechanism with the chemostats:
they have to be much faster than the remaining reactions,
so that the latter are rapidly balanced
(Subs.~\ref{subs:ASSUMPTION_tss}).
Mathematically, 
this corresponds to a time-scale separation between two sets of reactions.
% -
Second, some species involved in the fast reactions must behave 
as if they had a diverging chemical capacity:
they have to be much more abundant than the remaining species,
so that their abundances are effectively unchanged by individual reaction events
(Subs.~\ref{subs:ASSUMPTION_as}).
Mathematically, 
this corresponds to an abundance separation between two sets of species.
% -
% -
As a result,
the leading-order dynamics of a closed CRN 
reduces to that of an open CRN 
(Sec.~\ref{sec:open_like_dynamics}).
% -
Specifically, this open-CRN dynamics 
emerges as a distinct dynamical regime
in an intermediate time window
that can span large time scales,
after the fast reactions have equilibrated 
(Subs.~\ref{subs:IMPLICATIONS_tss})
and before 
the slower reactions significantly affect the abundant
species 
(Subs.~\ref{subs:IMPLICATIONS_dcc}),
which occurs on a time scale proportional to the scale of the abundance separation
(Subs.~\ref{subs:T_deviation}).
% -
% -
% - 
% -
For instance,
the periodic oscillations predicted for the open Brusselator when driven far from equilibrium
emerge in a closed Brusselator over this  
time window (Sec.~\ref{Sec:Brusselator}).
% -
% -
Moreover,
the emergent open CRN 
inherits a fully consistent thermodynamic structure
(Sec.~\ref{sec:open_like_thermo}).
% -
The local detailed balance condition for open CRNs holds (Subs.~\ref{subs:emeLDB});
the dissipation associated with the emergent open-CRN dynamics is exactly equal to
the leading-order dissipation of the underlying closed CRN (Subs.~\ref{subs:emeEPR})
within the same intermediate time window,
after the fast reactions have equilibrated 
and before the slower reactions significantly affect the abundant species
(Sec.~\ref{Sec:LinearExample});
furthermore, the dissipation satisfies a balance equation 
identical to the second law for open CRNs (Sec.~\ref{subs:eme2LAW}).
% -
% -
Crucially, the emergence of open-CRN dynamics and thermodynamics
is independent of the stoichiometry of both the fast and the slower reactions
(Sec.~\ref{sec:GeneralChemostatting}).
% -
For particular fast reactions acting as a buffer solution,
our notion of chemical capacity is equivalent to the well-known notion of buffer capacity.
% -
% -
Conclusions are drawn in Sec.~\ref{Sec:Conclusion}.
% -
% -

% -
% -
Overall, we establish precise physical conditions under which
open-CRN behavior emerges from closed CRNs,
showing that the nonequilibrium thermodynamics of open CRNs
is not merely an idealization,
but an emergent regime of closed out-of-equilibrium systems.
% -
These conditions translate into measurable requirements
on reaction time scales and species abundances.
% -
Together with our prediction that the duration of the emergent regime
scales with the abundance separation,
they provide experimentally accessible criteria
for determining whether,
and for how long,
a finite closed chemical system can be quantitatively described as an open CRN.
% -
Together, these results provide
a unified and thermodynamically consistent foundation
for the effective description and experimental interpretation
of fueled chemical systems.
% -
% -
% -
% -

% -
% -
More technical details are presented in the Appendices.
% -
% -
Appendix~\ref{Appendix:CMETimeScale} analyzes the implications of the time-scale separation.
% -
Appendix~\ref{Appendix:IdealReservoirs} clarifies the origin of the diverging chemical capacity of chemostats
and its relation to the diverging heat capacity of thermostats.
% -
Appendix~\ref{App:LargeDeviation} investigates the consequences of abundance separation
using a Wentzel–Kramers–Brillouin (WKB) ansatz within a large-deviation framework.
% -
Appendix~\ref{App:EPR} proves the convergence of the leading-order dissipation of closed CRNs
to that of the corresponding emergent open CRN.
% -
Appendix~\ref{App:LinearReactions} provides the semi-analytical analysis of the CRN
examined in Sec.~\ref{Sec:LinearExample}.
% -
Finally, App.~\ref{app:GeneralChemostatting} discusses the generality of our results
with respect to reaction stoichiometry.
% -
% -

%%%%%%%%%%%%%%%%%%%%%%%%%%%%%%%%%%%%%%%%%%%%%%%%%%%%%%%
%%%%%%%%%%%%%%%%%%%%%%%%%%%%%%%%%%%%%%%%%%%%%%%%%%%%%%%
%%%%%%%%%%%%%%%%%%%%%%%%%%%%%%%%%%%%%%%%%%%%%%%%%%%%%%%
%%%%%%%%%%%%%%%%%%%%%%%%%%%%%%%%%%%%%%%%%%%%%%%%%%%%%%%

\section{Notation\label{sec:notation}}
We fix here the basic notation 
used throughout the manuscript.
% -
% -
Reacting chemical species are indexed by a label, e.g., $\alpha \in \Spe$.
% -
For each species~$\alpha$, 
its chemical symbol, number of molecules, concentration, and standard chemical potential
are denoted by 
$Z_\alpha$, $n_\alpha$, $[\alpha]$, and $\mu^\circ_\alpha$, respectively. 
% -
% -
We use distinct labels for different kinds of species
(i.e., $i \in \Spei$, $c \in \Spec$, $x \in \Spex$, $y \in \Spey$, and $Y \in \SpeY$).
% -
% -

% -
% -
Chemical reactions are also indexed by a label, e.g., $\rho \in \Rct$.
% -
For each reaction~$\rho$,
the stoichiometric coefficient of species~$\alpha$
is denoted $\nu_{\alpha,\rho}$.
% -
% -
All reactions are reversible: 
each forward reaction~$\rho$
has a corresponding backward counterpart~$-\rho \in \Rct$,
and together they are described by the chemical equation
\begin{equation}
    \sum_\alpha \, \nu_{\alpha,\rho} \, Z_\alpha 
    \ch{<=>[$\rho$][$-\rho$]} 
    \sum_\alpha \, \nu_{\alpha,-\rho} \, Z_\alpha 
    \,.
    \label{eq:CE}
\end{equation}
% -
% -
We use distinct labels for different kinds of reactions
(e.g., $\rhoi \in \Rint$ and $\rhoe \in \Rexc$).
% -
% -

% -
% -
The net variation of the number of molecules of species~$\alpha$ in reaction~$\rho$ 
is given by 
\begin{equation}
    S_{\alpha,\rho} = \nu_{\alpha,-\rho} -\nu_{\alpha,\rho} 
    \,.
\end{equation}
% -
The vector $\colS_\rho = (\dots\,, S_{\alpha,\rho}\,, \dots)$ collects
the net variation of the number of molecules of all species in reaction~$\rho$. 
% -
We use distinct vectors to collect the corresponding variations for each kind of species
(i.e., $\colS_{i, \rho} = (\dots\,, S_{i,\rho}\,, \dots)$,
$\colS_{c, \rho} = (\dots\,, S_{c,\rho}\,, \dots)$,
$\colS_{x, \rho} = (\dots\,, S_{x,\rho}\,, \dots)$,
$\colS_{y, \rho} = (\dots\,, S_{y,\rho}\,, \dots)$,
and
$\colS_{Y, \rho} = (\dots\,, S_{Y,\rho}\,, \dots)$).

%%%%%%%%%%%%%%%%%%%%%%%%%%%%%%%%%%%%%%%%%%%%%%%%%%%%%%%
%%%%%%%%%%%%%%%%%%%%%%%%%%%%%%%%%%%%%%%%%%%%%%%%%%%%%%%
%%%%%%%%%%%%%%%%%%%%%%%%%%%%%%%%%%%%%%%%%%%%%%%%%%%%%%%
%%%%%%%%%%%%%%%%%%%%%%%%%%%%%%%%%%%%%%%%%%%%%%%%%%%%%%%

\section{Open CRNs\label{sec:openCRNs}}

%%%%%%%%%%%%%%%%%%%%%%%%%%%%%%%%%%%%%%%%%%%%%%%%%%%%%%%
\subsection{Setup}
CRNs describe systems consisting of an ideal dilute solution
in which the solutes are reacting species 
interconverted via reversible chemical reactions.
% -
The solvent is instead a non-reacting species 
which acts as both a thermal and volume reservoir
maintaining the solution at constant temperature and constant volume~$V$.
% -
% -
A CRN is said to be \textit{open} when some of the solutes (labeled~$c \in \Spec$)
are exchanged with the environment 
(a CRN is said to be \textit{closed} otherwise).
The environment is modeled in terms of \textit{chemostats},
i.e., infinitely large reservoirs of chemical species,
which fix the abundances of the $\Spec$ species~\footnote{
In general, chemostats control the abundances of the $\Spec$ species
by imposing prescribed protocols.
Often, these protocols keep the abundances constant in time.
In this paper, we restrict ourselves to this situation.}.
Accordingly, 
the $\Spec$ species
and the corresponding exchange processes 
are usually referred to as 
\textit{chemostatted} species
and
\textit{chemostatting} procedure,
respectively.
% -
By contrast, the remaining solutes (labeled~$i \in \Spei$) 
are usually referred to as \textit{internal} species.
% -
% -
% -
% -

% -
% -
The chemical equation~\eqref{eq:CE} for open CRNs specializes to
\begin{equation}\small
    \sum_i \nu_{i,\rho} \, Z_i + \sum_c \nu_{c,\rho} \, Z_c 
    \ch{<=>[$\rho$][$-\rho$]} 
    \sum_i \, \nu_{i,-\rho} Z_i +\sum_c \, \nu_{c,-\rho} Z_c  \,.
    \label{eq:CRNopen}
\end{equation}
Throughout the manuscript, 
we refer to the reactions $\rho \in \Rct$ as \textit{internal} reactions.
This terminology emphasizes that these reactions occur within 
open CRNs,
whereas the chemostatting procedure involves exchanges 
with the environment.

%%%%%%%%%%%%%%%%%%%%%%%%%%%%%%%%%%%%%%%%%%%%%%%%%%%%%%%
\subsection{Dynamics}
Reactions~\eqref{eq:CRNopen} are assumed to be elementary stochastic events. 
% -
This means that the numbers of molecules~$\bni \equiv (\dots, n_i, \dots) \in \mathcal{N}$ of the internal species 
are fluctuating quantities
(with $\mathcal{N}$ being the state space of all possible numbers of molecules).
% -
The concentrations~$[\boldsymbol c] \equiv (\dots, [c],\dots)$ of the chemostatted species
are instead constant quantities fixed by the chemostatting procedure.
% -
% -
As a result, 
the dynamics of open CRNs is characterized in terms of
the probability~$p_t(\bni)$ of there being~$\bni$ molecules of internal species at time~$t$
which follows the chemical master equation
\begin{equation}\small
   {\mathrm{d}_t}
   p_{t}(\bni) =  
    \sum_{\rho} 
    \big\{
     \omega_{\rho}(\bni - \colS_{i,\rho}|[\boldsymbol c])p_t(\bni - \colS_{i,\rho})
    - \omega_{\rho}(\bni|[\boldsymbol c])p_t(\bni ) 
    \big\}
    \,,
    \label{eq:CME}
\end{equation}
where the rates of elementary reactions in ideal dilute solutions 
satisfy mass-action kinetics~\cite{gasp04},
namely,
\begin{equation}
    \omega_\rho(\bni|[\boldsymbol c]) =
    k_\rho V
    \prod_i\frac{n_i! \, \theta(n_i -\nu_{i,\rho})}
    {(n_i-\nu_{i,\rho})! \, V^{\nu_{i,\rho}}}
    \prod_c [c]^{\nu_{c,\rho}}
    \,,
    \label{eq:MassActionOpen}
\end{equation}
with $k_\rho > 0$ being the kinetic constant of reaction $\rho$
and $ \theta(n_i -\nu_{i,\rho})$  being the Heaviside step function 
(i.e., $ \theta(n_i -\nu_{i,\rho})= 1 $ if $ n_i \geq\nu_{i,\rho} $
and $ \theta(n_i -\nu_{i,\rho})= 0 $ otherwise).
% -
% -

% -
% -
\remark 
The chemical master equation~\eqref{eq:CME} can also be written as
\begin{equation}
    \mathrm d_t p_{t}(\bni) =
    \sum_{\bmi} \W(\bn,\bmi | [\boldsymbol c]) p_{t}(\bmi)
    \label{eq:CME2}
\end{equation}
using the stochastic generator
\begin{equation}
    \W(\bn,\bmi | [\boldsymbol c]) = 
    \sum_\rho
    \omega_\rho(\bmi|[\boldsymbol c])
    \big(
    \delta_{\bni - \colS_{i,\rho}, \bmi} - \delta_{\bni, \bmi}
    \big)
    \,.
    \label{eq:st_generator}
\end{equation}
% -
% -
For closed CRNs, the $[\boldsymbol c]$-dependence drops since there are no chemostats yielding
\begin{equation}
    \mathrm d_t p_{t}(\bni) =
    \sum_{\bmi} \W(\bn,\bmi) p_{t}(\bmi)
    \label{eq:CME3}
\end{equation}
with the stochastic generator
\begin{equation}
    \W(\bn,\bmi ) = 
    \sum_\rho
    \omega_\rho(\bmi)
    \big(
    \delta_{\bni - \colS_{i,\rho}, \bmi} - \delta_{\bni, \bmi}
    \big)
    \,.
    \label{eq:st_generator3}
\end{equation}
% -
We use the formulation of the chemical master equation in Eq.~\eqref{eq:CME3}
in Subs.~\ref{subs:ASSUMPTION_tss} and Subs.~\ref{subs:IMPLICATIONS_tss}.

%%%%%%%%%%%%%%%%%%%%%%%%%%%%%%%%%%%%%%%%%%%%%%%%%%%%%%%
\subsection{Thermodynamics\label{sub:openThermo}}

Thermodynamic consistency imposes that 
all elementary reactions satisfy the local detailed balance condition~\cite{rao18}
\begin{equation}
	\ln 
    \frac{\omega_\rho(\bni|[\boldsymbol c])}{
    \omega_{-\rho}(\bni +\colS_{i,\rho}|[\boldsymbol c])}
    =
    -\Big\{\Delta_\rho g(\bni) + 
    \sum_c \mu_c([c]) S_{c, \rho}\Big\}
    \,,
	\label{eq:CRNLDB}
\end{equation}
where $\Delta_\rho g(\bni) \equiv g(\bni + \colS_{i,\rho}) - g(\bni)$ 
is the change in Gibbs free energy
along reaction $\rho$
due to the internal species,
\begin{equation}
    g(\bni) = 
    \sum_i 
    \big\{ 
    ( \mu_i^\circ - \ln n_s ) n_i + \ln n_i!
    \big\}
    \label{eq:gibbs}
\end{equation}
is the Gibbs free energy
(in units of temperature and Boltzmann constant)
of an ideal solution with $\bni$ molecules of internal species, and 
$\mu_c([c]) S_{c, \rho}$
is the change in Gibbs free energy
along reaction $\rho$
due to the chemostatted species $c$
whose chemical potential
(in units of temperature and Boltzmann constant) reads
\begin{equation}
    \mu_c([ c]) \equiv \mu_c^\circ + \ln ( {[c]}/{[s]} ) 
    \,.
    \label{eq:chem_pot}
\end{equation}
% -
Finally, $n_s$ and $[s] = n_s / V$
specify the number of molecules and the concentration of the solvent, respectively.
% -
% -

% -
% -
The local detailed balance condition~\eqref{eq:CRNLDB}
ensures that
the average dissipated free energy associated with the dynamics~\eqref{eq:CME}
is quantified by the average (total) entropy production rate reading 
(in units of Boltzmann constant)
\begin{equation}\small
    \braket{\dot \Sigma}_t = 
    \sum_{\rho, \bni} 
    \omega_\rho(\bni|[\boldsymbol c]) p_t(\bni) 
    \ln \frac
    {\omega_\rho(\bni|[\boldsymbol c]) p_t(\bni)}
    {\omega_{-\rho}(\bni+\colS_{i,\rho}|[\boldsymbol c]) p_t(\bni+\colS_{i,\rho})} 
    \geq0
    \,.
    \label{eq:EPR}
\end{equation}
% -
% -

% -
% -
Furthermore, the second law of thermodynamics can be written 
in terms of a balance equation for the average Gibbs potential
(in units of temperature and Boltzmann constant)
\begin{equation}
    \braket{G}_t = 
    \sum_{\bni} p_t(\bni) 
    \big\{\underbrace{
    g(\bni) + \ln p_t(\bni)}_{\equiv G_t(\bn)}
    \big\}
    \label{eq:avg_Gpot}
\end{equation}
and reads 
\begin{equation}
    \braket{\dot \Sigma}_t  
    = 
    - \mathrm d_t
    \braket{G}_t
    + \braket{\dot W_{\mathrm{chm}}}_t \geq 0
    \,,
    \label{eq:2law}
\end{equation}
with $\braket{\dot W_{\mathrm{chm}}}_t = 
\sum_c \mu_c([c]) 
\sum_{\rho, \bni} (-S_{c,\rho}) \omega_{\rho}(\bni|[\boldsymbol c]) p_t(\bni)$ being
the chemical work quantifying the Gibbs free energy exchanged with the environment. 
% -
% -
% -
% -

%%%%%%%%%%%%%%%%%%%%%%%%%%%%%%%%%%%%%%%%%%%%%%%%%%%%%%%
%%%%%%%%%%%%%%%%%%%%%%%%%%%%%%%%%%%%%%%%%%%%%%%%%%%%%%%
%%%%%%%%%%%%%%%%%%%%%%%%%%%%%%%%%%%%%%%%%%%%%%%%%%%%%%%
%%%%%%%%%%%%%%%%%%%%%%%%%%%%%%%%%%%%%%%%%%%%%%%%%%%%%%%

\section{Closed CRNs: 
Partition of Species and Reactions for Emergent Open-CRN Behavior \label{sec:closedCRNs}}
We now turn to closed CRNs 
and 
introduce the fundamental framework
that enable the emergence of open-CRN behavior.
% -
% -
To this end, 
we partition the 
solutes 
into three subsets:
species 
$x \in \Spex$, 
$y \in \Spey$, 
and $Y \in \SpeY$, 
acting as 
internal species,
chemostatted species,
and chemostats, 
respectively, 
in the emergent open CRNs.
% -
% -
Furthermore,
we partition the
reactions 
into two subsets:
reactions 
$\rhoi \in \Rint$ 
and $\rhoe \in \Rexc$,
acting as 
internal reactions 
and exchange processes (i.e., the chemostatting procedure), 
respectively,
in the emergent open CRNs.
% -
% -

% -
% -
Each reaction $\rhoi$ interconverts only the $\Spex$ and $\Spey$ species 
with an arbitrary stoichiometry, 
according to
\begin{equation}\small
\sum_x \nu_{x,\rhoi} \ch{Z}_x 
+ 
\sum_y \nu_{y,\rhoi} \ch{Z}_y 
\ch{<=>[ $\rhoi$ ][ $-\rhoi$ ]}
\sum_x \nu_{x,-\rhoi} \ch{Z}_x 
+ 
\sum_y \nu_{y,-\rhoi} \ch{Z}_y
\,.
\label{eq:CRNclosed}
\end{equation}
Accordingly, 
the entries of $\colS_{\rhoi}$ 
corresponding to the $\SpeY$ species vanish, 
i.e., $S_{Y,\rhoi} = 0$ $\forall Y \in \SpeY$.
% -
% -
Every species $y$ is additionally interconverted
through a specific pair of forward and backward reactions $\pm \rhoe(y)$ 
into a corresponding $Y(y)$ species,
according to  
\begin{equation}
\ch{Z}_y 
\ch{<=>[ $\rho_{\mathrm{e}}(y)$ ][ $-\rho_{\mathrm{e}}(y)$ ]}
\ch{Z}_{Y(y)}
\,.
\label{eq:CRN_ExchangeLinear}
\end{equation}
Accordingly, 
the entries of $\colS_{\rhoe}$ 
corresponding to the $\Spex$ species vanish,
i.e., $S_{x,\rhoe} = 0$ $\forall x \in \Spex$,
while the other entries 
corresponding to the $\Spey$ and $\SpeY$ species
read
$S_{y, \rho_{\mathrm{e}}} =  
-\delta_{\rho_{\mathrm{e}}, \rho_{\mathrm{e}}(y)}
+ \delta_{\rho_{\mathrm{e}}, -\rho_{\mathrm{e}}(y)}$
and 
${S_{Y, \rho_{\mathrm{e}}} = - \sum_y S_{y, \rho_{\mathrm{e}}} \delta_{Y,Y(y)}}$,
respectively.
% -
% -

% -
% -
\remark
For simplicity,
we assume that the reactions $\pm\rhoe(y)$ are unimolecular in the species $y$ and $Y(y)$,
as commonly done in models of the chemostatting procedure 
(see, e.g., Refs.~\cite{Polettini2015, marehalli2023, DalCengio2023, Shimada2024}).
The case of general stoichiometry is treated in Sec.~\ref{sec:GeneralChemostatting}.
% -
% -
% -
% -

%%%%%%%%%%%%%%%%%%%%%%%%%%%%%%%%%%%%%%%%%%%%%%%%%%%%%%%
%%%%%%%%%%%%%%%%%%%%%%%%%%%%%%%%%%%%%%%%%%%%%%%%%%%%%%%
%%%%%%%%%%%%%%%%%%%%%%%%%%%%%%%%%%%%%%%%%%%%%%%%%%%%%%%
%%%%%%%%%%%%%%%%%%%%%%%%%%%%%%%%%%%%%%%%%%%%%%%%%%%%%%%

\section{Dynamical Conditions for the Emergence of Open CRNs\label{sec:chemo_like_dynamics}}

We now specify the  
dynamical conditions 
\--- namely, time-scale separation (Subs.~\ref{subs:ASSUMPTION_tss}) 
and abundance separation (Subs.~\ref{subs:ASSUMPTION_as}) \---
under which closed CRNs 
(with species and reactions partitioned as in Sec.~\ref{sec:closedCRNs}) 
behave as open CRNs.
% -
Under these conditions, the $\Rexc$ reactions act as the chemostatting procedure 
and the $\SpeY$ species act as chemostats.
% -
% -

%%%%%%%%%%%%%%%%%%%%%%%%%%%%%%%%%%%%%%%%%%%%%%%%%%%%%%%
\subsection{Time-Scale Separation \label{subs:ASSUMPTION_tss}}

We begin by assuming that the $\Rexc$ and $\Rint$ reactions occur on distinct time scales,
with the former being much faster than the latter. 
% -
This assumption reflects the physical requirement that the chemostatting procedure in open CRNs
occurs on a much faster time scale than the internal reactions.
Specifically,
any variation in the abundances of the chemostatted species induced by the internal reactions
is rapidly compensated by the chemostatting procedure, 
ensuring that the concentrations $[\boldsymbol c]$ remain effectively constant.
% -
% -
This assumption can be mathematically formalized
in terms of a small dimensionless parameter $\epsilon \ll 1$:
the kinetic constants $\{k_{\rhoe}\}$ and $\{k_{\rhoi}\}$
of the $\Rexc$ and $\Rint$ reactions, respectively,
satisfy
\begin{equation}
    k_{\rhoe} = \hat{\kappa}_{\rhoe}
    \quad
    \text{ and }
    \quad
    k_{\rhoi} = \epsilon \, \hat{\kappa}_{\rhoi}
    \label{eq:epsilon_tss}
\end{equation}
with 
$\hat{\kappa}_{\rhoe} = \mathcal O(1)$
and $ \hat{\kappa}_{\rhoi} = \mathcal O(1)$  
as $\epsilon \to 0$.
% -
% -

% -
% -
Because of the time-scale separation, 
the chemical master equation~\eqref{eq:CME3} 
describing the evolution of 
the probability~$p_t(\bn)$ of there being~$\bn = (\dots, n_x, \dots, n_y, \dots, n_Y, \dots) \in \mathcal{N}$
molecules at time~$t$
can be systematically re-formulated in terms of a hierarchy of equations 
of successive orders in~$\epsilon$.
% -
The $\mathcal O(1)$ equation of the hierarchy 
(in the limit $\epsilon \to 0$)
describes the fastest time scale
and
can be analytically solved as we show in the following.
% -
% -

% -
% -
To do so,
we first recognize that
the stochastic generator in Eq.~\eqref{eq:st_generator3}
can be expressed as the sum of two generators according to
\begin{equation}
    \W(\bn,\bmi) = 
    \Wexc(\bn,\bnm) + 
    \epsilon\, \Wint(\bn,\bnm)
    \,,
    \label{eq:Wepsilon}
\end{equation}
where
\begin{subequations}
\begin{align}
    \Wexc(\bn, \boldsymbol m) &\equiv 
    \sum_{\rhoe}
    \hat \omega_{\rhoe}(\bnm) 
    \big(\delta_{\bn - \colS_{\rhoe}, \bnm} - \delta_{\bn, \bnm}\big)
    \,,
    \label{eq:Wexc}
    \\
    \Wint(\bn, \bnm) 
    &\equiv 
    \sum_{\rhoi}
    \hat \omega_{\rhoi}(\bnm) 
    \big(\delta_{\bn - \colS_{\rhoi}, \bnm} - \delta_{\bn, \bnm}\big)
    \,,
    \label{eq:Wint}
\end{align}
\end{subequations}
account for
the $\Rexc$ reactions occurring on the fast time scale 
and 
the $\Rint$ reactions occurring on the slow time scale, 
respectively,
with 
$\hat \omega_{\rhoe}(\bnm) \equiv \omega_{\rhoe}(\bnm)$ 
and 
$\hat \omega_{\rhoi}(\bnm) \equiv \omega_{\rhoi}(\bnm) / \epsilon$.
Second, 
we characterize  
the evolution of the probability~$p_t(\bn)$
in terms of two time variables,
rather than just one.
% -
That is,
we introduce the time variable $\tau = t$, capturing the evolution on the fast time scale,
and the time variable $T = \epsilon t$, capturing the evolution on the slow time scale, 
such that
$p_t(\bn) = p_{\tau, T}(\bn)$ 
and $\mathrm d_t p_t(\bn)=
\mathrm d_\tau p_{\tau, T}(\bn) 
+ \epsilon \mathrm d_T p_{\tau, T}(\bn)$.
% -
% -
Third, 
we express the probability~$p_{\tau,T}(\bn)$ 
using a multiscale expansion in powers of $\epsilon$:
\begin{equation}
    p_{\tau, T}(\bn) = 
    p_{\tau, T}^{(0)}(\bn) 
    + \sum_{q=1}^\infty p_{\tau, T}^{(q)}(\bn) \epsilon^q
    \,.
    \label{eq:PerturbAnsatz}
\end{equation}
% -
% -
% -
% -
As shown in App.~\ref{Appendix:CMEFastTimeScale},
the result is that 
the $\mathcal O(1)$ equation of the hierarchy 
(in the limit $\epsilon \to 0$)
reads
\begin{equation}
    \mathrm d_\tau p_{\tau, T}^{(0)}(\bn) 
    = 
    \sum_{\bnm} \Wexc(\bn,\bnm) \, p_{\tau, T}^{(0)}(\bnm)
    \,.
    \label{eq:CME_fast}
\end{equation}
We now examine the main properties of Eq.~\eqref{eq:CME_fast}
and derive its analytical solution in the long-time limit 
of the fast time-scale,
i.e., $\tau \to \infty$.
% -
% -

% -
% -
The stochastic generator $\Wexc(\bn,\bnm)$ 
appearing in Eq.~\eqref{eq:CME_fast} 
and given in Eq.~\eqref{eq:Wexc}
encodes the fast dynamics arising from the $\Rexc$ reactions.
% -
Such dynamics,
according to the stoichiometry given in Eq.~\eqref{eq:CRN_ExchangeLinear},
interconverts 
each species $y$ (resp. $Y(y)$) into species $Y(y)$ (resp. $y$)
via reaction $\rhoe(y)$ (resp. $-\rhoe(y)$)
with a one-to-one stoichiometry: 
one molecule of $y$ is interconverted into one molecule of $Y(y)$ and vice versa.
% -
Hence, 
the total numbers of molecules $\bL^y (\bn) = (\dots, L^y(\bn), \dots)$,
with
\begin{equation}
    L^y(\bn) \equiv n_y + n_{Y(y)} 
    \,,
    \label{eq:Ly}
\end{equation}
are conserved quantities. 
Furthermore, such dynamics, 
according to the stoichiometry given in Eq.~\eqref{eq:CRN_ExchangeLinear},
does not interconvert the $\Spex$ species.
% -
% -
Hence, 
the numbers of molecules $\bL^x (\bn) = (\dots, L^x(\bn), \dots)$,
with
\begin{equation} 
    L^x(\bn) \equiv n_x 
    \,,
    \label{eq:Lx}
\end{equation}
are conserved quantities too. 
% -
% -
% -
% -
This means that,
on the fast time scale described by Eq.~\eqref{eq:CME_fast}
and captured by the variable~$\tau$,
the numbers of molecules~$\bn$ 
(characterized by the specific values $\bL^x$ and $\bL^y$ 
of the quantities $\bL^x(\bn)$ and $\bL^y(\bn)$, respectively)
evolve within the following stoichiometric compatibility class
\begin{equation}\small
    \mathcal N (\bL^x, \bL^y) =
    \big\{
    \bn \in \mathcal N : \bL^x(\bn) = \bL^x 
    \text{ and } 
    \bL^y(\bn) = \bL^y
    \big\} 
    \,.
    \label{eq:StoichiometricCompabtibilityClass}
\end{equation}
% -
It is only on the slow time scale captured by the variable~$T$ 
that the quantities $\bL^x$ and $\bL^y$ evolve. 
% -
% -
Consequently,
we can factorize the 
zero-order probability $p_{\tau, T}^{(0)}(\bn)$ (with $\bn \in \mathcal N (\bL^x, \bL^y)$)
in terms of
a probability distribution for $\bn$ conditioned on the quantities $\bL^x$ and $\bL^y$
whose evolution is captured by the variable $\tau$
and 
a probability distribution for the quantities $\bL^x$ and $\bL^y$
whose evolution is captured by the variable $T$:
\begin{equation}\small
    p_{\tau, T}^{(0)}(\bn) = 
    \pi_{\tau}^{(0)}(\bn | \bL^x, \bL^y) \, 
    p_{T}^{(0)} (\bL^x, \bL^y)
    \,.
    \label{eq:p0_factorized}
\end{equation}
% -
% -

% -
% -
Equation~\eqref{eq:CME_fast} describes the evolution of $\pi_{\tau}^{(0)}(\bn | \bL^x, \bL^y)$ only.
% - 
As shown in App.~\ref{Appendix:CMEFastTimeScale}, 
in the long-time limit of the fast time scale, i.e.,  $\tau \to \infty$,
the probability 
$\piInf(\bn|\bL^x,\bL^y) \equiv \lim_{\tau \to \infty} \pi_{\tau}^{(0)}(\bn | \bL^x, \bL^y)$,
representing the steady state distribution of the fast dynamics, 
reads
\begin{equation}\small
    \piInf(\bn|\bL^x,\bL^y) =
    \prod_{y}
    \piInfy(n_y|L^y)\,
    \delta_{n_y+n_{Y(y)}, L^y}  \, 
    \prod_{x}
    \delta_{n_x, L^x}
    \,,
    \label{eq:Binomial}
\end{equation}
where $\delta_{\bullet,\bullet}$ is the Kronecker delta and
\begin{equation}\small
    \piInfy(n_y|L^y) =
    \frac{L^y!}{n_y! (L^y - n_y)!}
    \big(\mathfrak p_y\big)^{n_y}
    \big(\mathfrak p_{Y(y)}\big)^{L^y - n_y}
    \,,
    \label{eq:Binomialy}
\end{equation}
with 
\begin{equation}
    \mathfrak p_{Y(y)} + \mathfrak p_y= 1
    \quad\text{ and }\quad
    \mathfrak p_{Y(y)} / \mathfrak p_y = \hat{\kappa}_{\rhoe(y)} / \hat{\kappa}_{-\rhoe(y)}
    \,.
    \label{eq:py_prop}
\end{equation}
Note that this implies that
the average numbers of molecules of species $y$ and $Y(y)$
in the long-time limit of the fast time scale,
i.e.,
$\avgInf{n_y | L^y} \equiv \sum_{n_y = 0}^{ L^y} n_y \, \piInfy(n_y|L^y)$
and 
$\avgInf{n_{Y(y)} | L^y} \equiv L^y - \avgInf{n_y | L^y}$,
read
\begin{subequations}
\begin{align}
    \avgInf{n_y | L^y} & = L^y \mathfrak p_y \,, \label{eq:avg_ny}\\
    \avgInf{n_{Y(y)} | L^y} & = L^y \mathfrak p_{Y(y)} \,,
\end{align}
\label{eq:avg_ny_infy}%
\end{subequations}
respectively.
% -
% -

% -
% -
\remark
In Subs.~\ref{subs:IMPLICATIONS_tss},
we consider the long-time limit of the fast time scale, i.e.,  $\tau \to \infty$,
to derive a closed master equation (Eq.~\eqref{eq:CME_post_tss}) for the probability 
$p_{T}^{(0)} (\bL^x, \bL^y)$,
describing the evolution of the quantities $\bL^x$ and $\bL^y$
on the slow time scale captured by the variable~$T$.
% -
% -
Such master equation does not coincide with  
the chemical master equation of open CRNs (Eq.~\eqref{eq:CME}):
the time-scale separation is a necessary, but not sufficient, condition 
for the emergence of open-CRN dynamics. 
% -
The additional necessary condition is discussed in Subs.~\ref{subs:ASSUMPTION_as}.
% -
% -

% -
% -
\nremark
With a slight abuse of notation,
we use $\bL^x(\bn)$ and $\bL^y(\bn)$ to denote the values of the conserved quantities
corresponding to given numbers of molecules $\bn$,
while 
$\bL^x $ and $\bL^y$ denote arbitrary values of these conserved quantities. 
% -
% -
According to Eq.~\eqref{eq:StoichiometricCompabtibilityClass},
for arbitrary $\bL^x $ and $\bL^y$, 
we have $\bL^x(\bn) = \bL^x$ and $\bL^y(\bn) = \bL^y $
for any $\bn \in \mathcal N (\bL^x, \bL^y)$.
% -
Consequently,
the probability distribution $\pi_{\tau}^{(0)}(\bn | \bL^x, \bL^y)$ (with $\bn \in \mathcal N (\bL^x, \bL^y)$)
can equivalently be written as $\pi_{\tau}^{(0)}(\bn | \bL^x(\bn), \bL^y(\bn))$.

%%%%%%%%%%%%%%%%%%%%%%%%%%%%%%%%%%%%%%%%%%%%%%%%%%%%%%%

\subsection{Abundance Separation \label{subs:ASSUMPTION_as}}

We next assume that the $\SpeY$ species are, on average,
much more abundant than the $\Spey$ species. 
% -
This assumption reflects the physical requirement that chemostats remain essentially unaffected 
by the chemostatting procedure.
% -
As explained in App.~\ref{Appendix:IdealReservoirs_C}, if a chemostat is sufficiently large,
any finite exchange of molecules with the CRN
becomes negligible from the chemostat's perspective.
We refer to this property as \textit{diverging chemical capacity}, 
in analogy with the diverging heat capacity of thermostats 
(discussed in App.~\ref{Appendix:IdealReservoirs_T}).
% -
% -
This assumption can be mathematically formalized
in terms of a large dimensionless parameter $\Omega \gg 1$ as follows.
% -
First, 
the total number of molecules $L^y$ are $\mathcal O(\Omega)$ 
as $\Omega \to \infty$
for each species~$y$.
% -
Second, in the long-time limit of the fast time scale,
the average numbers of molecules
$\avgInf{n_{y} | L^y}$ and $\avgInf{n_{Y(y)} | L^y}$
given in Eq.~\eqref{eq:avg_ny_infy}
satisfy 
\begin{equation}
    \avgInf{n_{y} | L^y} = \mathcal O(1) 
    \quad
    \text{ and }
    \quad
    \avgInf{n_{Y(y)} | L^y} = \mathcal O(\Omega) 
    \,,
    \label{eq:omega_as}
\end{equation}
as $\Omega \to \infty$,
ensuring that $\avgInf{n_{y} | L^y} \ll \avgInf{n_{Y(y)} | L^y}$ .
Note that this implies that 
\begin{equation}
    \mathfrak p_y = \mathcal O(\Omega^{-1})
    \quad
    \text{ and }
    \quad
    \mathfrak p_{Y(y)} = \mathcal O(1) 
    \label{eq:py_order}
\end{equation}
and, therefore, 
\begin{equation}
    {\mathfrak p_{Y(y)} }/{ \mathfrak p_y} =
    {\hat{\kappa}_{\rhoe(y)} }/{ \hat{\kappa}_{-\rhoe(y)} }
    = {\mathcal O}(\Omega)
    \,,
\end{equation}
see Eqs.~\eqref{eq:py_prop} and~\eqref{eq:avg_ny_infy}.
Physically,
the abundance separation corresponds to a time-scale separation between each pair of
forward and backward reactions $\pm\rhoe(y)$.
This correspondence, however, is not general:
it arises from the unimolecular stoichiometry of the $\Rexc$ considered here,
and breaks down when
the $\Rexc$ reactions have multimolecular stoichiometries, as discussed in Sec.~\ref{sec:GeneralChemostatting}.
% -
% -

% -
% -
\remark
In Subs.~\ref{subs:IMPLICATIONS_dcc},
we apply the abundance separation
to the closed master equation
for the probability $p_{T}^{(0)} (\bL^x, \bL^y)$ (Eq.~\eqref{eq:CME_post_tss}),
in the long-time limit of the fast time scale, i.e.,  $\tau \to \infty$.
% -
% -
As a result, the master equation~\eqref{eq:CME_post_tss} converges to 
the chemical master equation of open CRNs (Eq.~\eqref{eq:CME}),
with the $\Spex$ and $\Spey$ species
acting as internal and chemostatted species, respectively.
% -
% -

%%%%%%%%%%%%%%%%%%%%%%%%%%%%%%%%%%%%%%%%%%%%%%%%%%%%%%% 
\subsection{Summary}
\begin{figure}[t]
    \centering
    \includegraphics[width=0.49\textwidth]{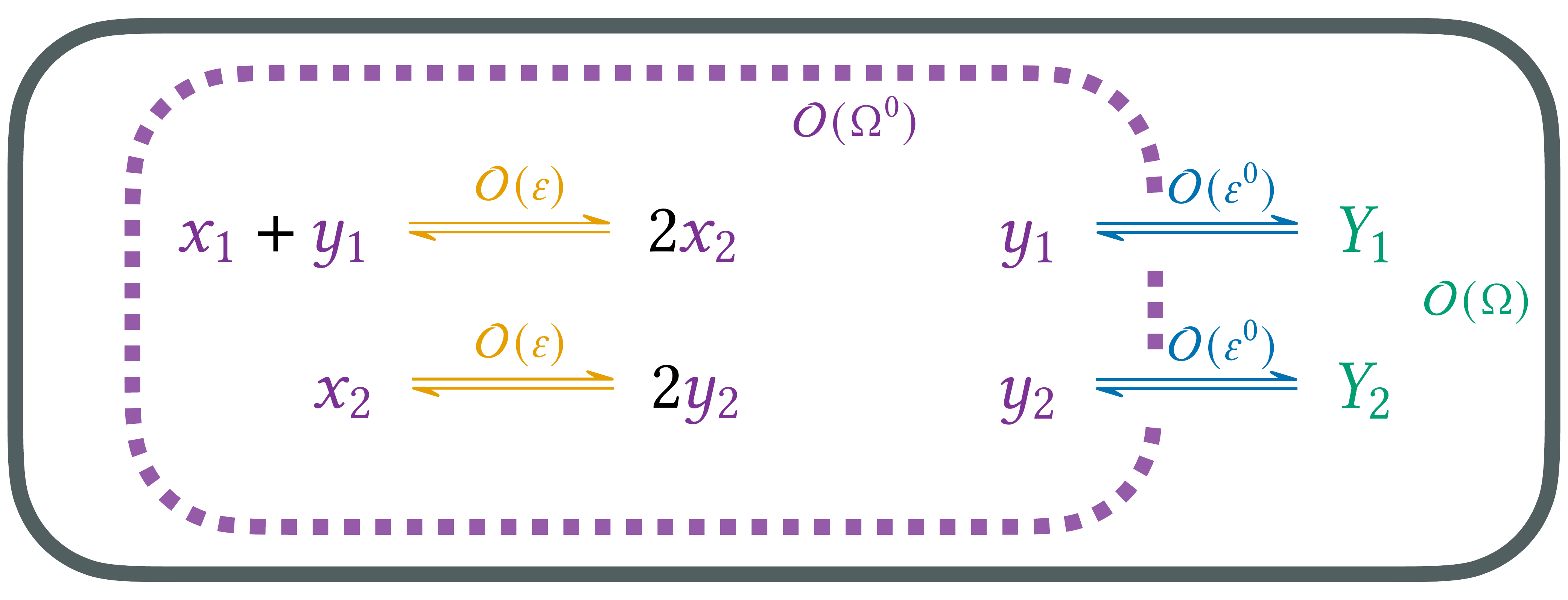}
    \caption{
    Illustration of the dynamical conditions yielding open-CRN behavior 
    for a particular closed CRN with species $\Spe = \{x_1, x_2, y_1, y_2, Y_1, Y_2\}$. 
    Time-scale separation:
    the kinetic constants of the 
    $\Rint$ (orange arrows) and $\Rexc$ (blue arrows) reactions 
    are
    $\mathcal O(\epsilon)$ and $\mathcal O(\epsilon^0) = \mathcal O(1)$, 
    respectively.
    % -
    Abundance separation:
    the numbers of molecules of the 
    $\Spex\cup\Spey$ (purple symbols) and $\SpeY$ (dark geen symbols) species are
    $\mathcal O(\Omega^0) = \mathcal O(1)$ and $\mathcal O(\Omega)$, 
    respectively.
    }
    \label{fig:ill_assumptions}
\end{figure}

The dynamical conditions introduced here specify
how the kinetic constants $\{k_{\rhoe}\}$ and $\{k_{\rhoi}\}$ scale with $\epsilon$
(Subs.~\ref{subs:ASSUMPTION_tss})
and how the abundances of the $\SpeY$ species scale with $\Omega$
(Subs.~\ref{subs:ASSUMPTION_as}).
% -
They also imply that the abundances of the $\Spex\cup\Spey$ species
do not scale with $\epsilon$ or $\Omega$,
and that the kinetic constants $\{k_{\rhoi}\}$ do not scale with $\Omega$.
% -
% -
In the following, we first take the time-scale-separation limit ($\epsilon \to 0$)
and then the abundance-separation limit ($\Omega \to \infty$).
% -
These limits do not commute.
% -
% -
These dynamical conditions are illustrated in Fig.~\ref{fig:ill_assumptions}.
% -
% -

% -
% -
\remark
The opposite ordering of the limits,
in which the abundance-separation limit ($\Omega \to \infty$)
is taken before the time-scale-separation limit ($\epsilon \to 0$),
falls within the framework developed in Ref.~\cite{reml25}.
% -
In that framework,
open-CRN behavior emerges only when restrictive stoichiometric conditions are satisfied.
% -
% -

%%%%%%%%%%%%%%%%%%%%%%%%%%%%%%%%%%%%%%%%%%%%%%%%%%%%%%%
%%%%%%%%%%%%%%%%%%%%%%%%%%%%%%%%%%%%%%%%%%%%%%%%%%%%%%%
%%%%%%%%%%%%%%%%%%%%%%%%%%%%%%%%%%%%%%%%%%%%%%%%%%%%%%%
%%%%%%%%%%%%%%%%%%%%%%%%%%%%%%%%%%%%%%%%%%%%%%%%%%%%%%%

\section{Emergent Open-CRN Dynamics\label{sec:open_like_dynamics}}

We now show  that open-CRN dynamics emerges naturally from closed CRNs
(with species and reactions partitioned as in Sec.~\ref{sec:closedCRNs})
under the
dynamical conditions introduced in Sec.~\ref{sec:chemo_like_dynamics}
and specified by Eqs.~\eqref{eq:epsilon_tss} and~\eqref{eq:omega_as}.
% -
To this end, we derive the master equation for the probability $p_{T}^{(0)} (\bL^x, \bL^y)$,
which describes the evolution of the quantities $\bL^x$ and $\bL^y$ 
on the slow time scale captured by the variable~$T$. 
% -
Crucially, 
we show that this master equation converges to the chemical master equation for open CRNs 
when both 
the time-scale separation 
(Eq.~\eqref{eq:epsilon_tss} in Sub.~\ref{subs:ASSUMPTION_tss})
and the abundance separation 
(Eq.~\eqref{eq:omega_as} in Sub.~\ref{subs:ASSUMPTION_as})
are satisfied.

%%%%%%%%%%%%%%%%%%%%%%%%%%%%%%%%%%%%%%%%%%%%%%%%%%%%%%% 

\subsection{Time-scale Separation \label{subs:IMPLICATIONS_tss}}

We begin by examining the implications of the time-scale separation
between the $\Rexc$ and $\Rint$ reactions,
introduced 
to reflect the fast chemostatting procedure in open CRNs.
% -
To do so, we focus on the $O(\epsilon)$ equation of the hierarchy
resulting from collecting terms of order $\epsilon$ 
of the chemical master equation~\eqref{eq:CME3} (or, equivalently, Eq.~\eqref{eq:CME_all_orders}).
% -
According to App.~\ref{Appendix:CMEFastTimeScale},
such equation reads
\begin{equation}
    \begin{split}
    \mathrm d_\tau p_{\tau, T}^{(1)}(\bn) 
    + \mathrm d_T p_{\tau, T}^{(0)}(\bn)  
    =&\sum_{\bnm}\Wexc(\bn,\bnm)p_{\tau, T}^{(1)}(\bnm)\\ 
    &+\sum_{\bnm}\Wint(\bn,\bnm)p_{\tau, T}^{(0)}(\bnm)\,,
    \end{split}
    \label{eq:CMEOrder1}
\end{equation}
and couples 
the zero-order probability $p_{\tau, T}^{(0)}(\bn)$ 
and 
the first-order probability $p_{\tau, T}^{(1)}(\bn)$
of the multiscale expansion in Eq.~\eqref{eq:PerturbAnsatz}.
% -
We now exploit the properties of the fast dynamics discussed in Subs.~\ref{subs:ASSUMPTION_tss}
to transform Eq.~\eqref{eq:CMEOrder1} into 
a closed equation for the zero-order probability 
$p_{T}^{(0)} (\bL^x, \bL^y)$.
% -
% -

% -
% -
To do so,
we first recall that the fast dynamics arising from the $\Rexc$ reactions 
occurs within the stoichiometric compatibility class $\mathcal N (\bL^x, \bL^y)$ 
given in Eq.~\eqref{eq:StoichiometricCompabtibilityClass}.
% -
This implies that
the $\mathcal O(1)$ equation, as $\epsilon \to 0$, 
of the hierarchy in Eq.~\eqref{eq:CME_fast} 
conserves the probability within each stoichiometric compatibility class $\mathcal N (\bL^x, \bL^y)$
or, equivalently, 
$ \sum_{\bn \in \mathcal N (\bL^x, \bL^y)} \Wexc(\bn,\bnm) = 0 $.
Consequently, by summing over ${\bn \in \mathcal N (\bL^x, \bL^y)}$,
Eq.~\eqref{eq:CMEOrder1} becomes
\begin{align}
    \begin{split}
    &\sum_{\bn \in \mathcal N (\bL^x, \bL^y)}
    \bigg(
    \mathrm d_\tau p_{\tau, T}^{(1)}(\bn) 
    + \mathrm d_T p_{\tau, T}^{(0)}(\bn)  
    \bigg)
\\ 
    &=\sum_{\bn \in \mathcal N (\bL^x, \bL^y)}
    \sum_{\bnm}\Wint(\bn,\bnm)p_{\tau, T}^{(0)}(\bnm)\,.
    \end{split}
    \label{eq:CMEOrder1_a}
\end{align}
% -
% -

% -
% -
Second, 
we consider the long-time limit of the fast time scale, i.e., $\tau \to \infty$.
% -
% -
This physically means that
we consider a slow time scale such that
the fast dynamics arising from the $\Rexc$ reactions has already relaxed to steady state.
% -
Furthermore, on such a slow time scale, 
any quantity evolving on the fast time scale captured by the variable~$\tau$ 
is averaged out.
% -
% -
Mathematically,
i) 
$\lim_{\tau\to\infty}\, \mathrm d_\tau p_{\tau, T}^{(1)}(\bn) = 0$,   
ii) 
$\lim_{\tau\to\infty}\, p_{\tau, T}^{(0)}(\bn) = 
\piInf(\bn | \bL^x, \bL^y) \, 
p_{T}^{(0)} (\bL^x, \bL^y)$
for any $\bn \in \mathcal N (\bL^x, \bL^y)$,
and iii) any function $f(\bn)$ is averaged over the fast dynamics
leading to
$\avgInf{f | \bL^x, \bL^y}
\equiv
\sum_{\bn \in \mathcal N (\bL^x, \bL^y)}
f (\bn)
\,
\piInf(\bn | \bL^x, \bL^y)$. 
% -
% -
As shown in App.~\ref{Appendix:CMESlowTimeScale},
Eq.~\eqref{eq:CMEOrder1_a}
thus becomes
\begin{widetext}
\begin{equation}
   \mathrm d_T p_{T}^{(0)} (\bL^x, \bL^y) =  
    \sum_{\rhoi} 
    \big\{
    \avgInf{\hat \omega_{\rhoi} | \bL^x - \colS_{x,\rhoi}, \bL^y - \colS_{y,\rhoi}} \, 
    p_{T}^{(0)} (\bL^x - \colS_{x,\rhoi}, \bL^y - \colS_{y,\rhoi})  
    -  
    \avgInf{\hat \omega_{\rhoi} | \bL^x, \bL^y} \,  
    p_{T}^{(0)} (\bL^x, \bL^y)
    \big\}
    \,,
\label{eq:CME_post_tss}
\end{equation}    
\end{widetext}
where the average reaction rates $\{\avgInf{\hat \omega_{\rhoi} | \bL^x, \bL^y}\}$ can be analytically determined 
and read
\begin{equation}\small
\begin{split}
    \avgInf{\hat \omega_{\rhoi} | \bL^x, \bL^y}
    =
    \hat{\kappa}_{\rhoi} V 
    &\prod_x
    \frac{L^x ! \, \theta(L^x -\nu_{x,\rhoi})}
    {(L^x - \nu_{x,\rhoi})! \, V^{\nu_{x,\rhoi}}}  \\
    \times&\prod_y
    \frac{L^y ! \, \theta(L^y -\nu_{y,\rhoi})}
    {(L^y - \nu_{y,\rhoi})! \, V^{\nu_{y,\rhoi}}}
    (\mathfrak p_y)^{\nu_{y,\rhoi}} \,.
\end{split}
\label{eq:avg_rr_tss} 
\end{equation}
% -
% -

% -
% -
Equation~\eqref{eq:CME_post_tss} is a closed master equation 
for the probability $p_{T}^{(0)}(\bL^x, \bL^y)$.
% -
It has the structure of a chemical master equation~\eqref{eq:CME}:
it describes the evolution of the probability of there being 
$\bL^x$ molecules of $\Spex$ species 
and 
$\bL^y$ total molecules of $\Spey$ and $\SpeY$ species,
arising from the $\Rint$ reactions whose rates have been averaged over the fast time scale. 
% -
% -
% -
However, these average reaction rates in Eq.~\eqref{eq:avg_rr_tss}
do not coincide with the reaction rates in Eq.~\eqref{eq:MassActionOpen} 
of the chemical master equation for open CRNs,
and
the quantities $\bL^y$ remain dynamical variables.
% -
As previously discussed,
the time-scale separation is a necessary, but not sufficient, condition for 
the emergence of open-CRN dynamics.
% - 
In Subs.~\ref{subs:IMPLICATIONS_dcc}, 
we show that imposing abundance separation as well 
leads to the full emergence of open-CRN dynamics.

%%%%%%%%%%%%%%%%%%%%%%%%%%%%%%%%%%%%%%%%%%%%%%%%%%%%%%%

\subsection{Abundance Separation \label{subs:IMPLICATIONS_dcc}}

We now examine the implications of the abundance separation
between the $\Spey$ and $\SpeY$ species,
introduced 
to reflect the diverging chemical capacity of chemostats.
% -
To do so, we focus on the master equation~\eqref{eq:CME_post_tss}.
% -
% -

% -
% -
We first recognize that 
the abundance separation, 
i.e., $L^y = \mathcal O(\Omega)$ and $\mathfrak p_y = \mathcal O(\Omega^{-1})$ 
(see Eq.~\eqref{eq:py_order}), 
implies that the average reaction rates 
given in Eq.~\eqref{eq:avg_rr_tss} and featuring Eq.~\eqref{eq:CME_post_tss}
can be split into a  
sum of terms of successive order in $\Omega^{-1}$.
By taking, as $\Omega \to \infty$,
the $\mathcal O(1)$ term only
and 
neglecting the successive orders,
the average reaction rates can be approximated according to 
\begin{subequations}\small
    \begin{align}
        \avgInf{\hat \omega_{\rhoi} | \bL^x, \bL^y}
        & \approx \hat R_{\rhoi}(\bL^x, \bL^y)
        \,,\\
        \avgInf{\hat \omega_{\rhoi} | \bL^x - \colSxi, \bL^y - \colSyi}
        & \approx \hat R_{\rhoi}(\bL^x- \colSxi, \bL^y)
        \,,
    \end{align}
    \label{eq:rr_dcc0}%
\end{subequations}
where
\begin{equation}\small
    \hat R_{\rhoi}(\bL^x, \bL^y)
    \equiv 
    \hat{\kappa}_{\rhoi} V 
    \prod_x
    \frac{L^x ! \, \theta(L^x -\nu_{x,\rhoi})}
    {(L^x - \nu_{x,\rhoi})! \, V^{\nu_{x,\rhoi}}} 
    \prod_y
    \bigg(\frac{L^y \, \mathfrak p_y}{V} \bigg)^{\nu_{y,\rhoi}} 
    \label{eq:rr_dcc1}
\end{equation}
with $L^y \, \mathfrak p_y = \mathcal O(1)$.
% -
% -

According to Eq.~\eqref{eq:avg_ny},
each factor $L^y \, \mathfrak p_y$ is
the average number of molecules of species~$y$ for a given~$L^y$
when the fast dynamics, captured by the variable~$\tau$,
has already relaxed to steady state.
% -
Each factor $L^y \, \mathfrak p_y / {V}$ is thus the corresponding average concentration
entering Eq.~\eqref{eq:rr_dcc1} 
in the same functional way the concentration of each chemostatted species enters 
the expression of the reaction rates of open CRNs in Eq.~\eqref{eq:MassActionOpen}.
% -
However, at this stage, 
the equivalence between the master equation~\eqref{eq:CME_post_tss} 
endowed with the rates in Eq.~\eqref{eq:rr_dcc1}
and the chemical master equation~\eqref{eq:CME} is not complete:
the missing ingredient is the time independence of the average concentrations of the $\Spey$ species, 
which reflects the time independence of the concentrations of the chemostatted species in open CRNs.
% -
Such time independence emerges, to leading order, as another implication of the abundance separation.
% -
% -

% -
% -
Indeed, we recognize that the large abundance $L^y = \mathcal O(\Omega)$ further ensures that 
the probability $p_{T}^{(0)} (\bL^y | \bL^x)$ 
of there being $\bL^y$ total molecules for the $\Spey$ and $\SpeY$ species
conditioned on there being $\bL^x$ molecules for the $\Spex$ species
satisfies a large deviation principle.
% -
% -
In this context, as proved in App.~\ref{App:LargeDeviation},
this means that $p_{T}^{(0)} (\bL^y | \bL^x)$ 
is sharply peaked around time-independent and $\bL^x$-independent
most probable values 
given by $\boldsymbol \eta^y_{\ast} \Omega$
with $\boldsymbol \eta^y_{\ast} = (\dots, \eta^y_{\ast}, \dots) = \mathcal{O} (1)$
as $\Omega \to \infty$~\footnote{
Note that, for a given initial probability distribution of the form
$p_{0}(\bn) = \delta(\bn - \bn^{0})$,
the most probable values of $p_{T}^{(0)}(\bL^y \mid \bL^x)$
are $\boldsymbol \eta^y_{\ast} \Omega = \overline{\bL}^y$,
with $\overline{L}^y = n^{0}_y + n^{0}_{Y(y)}$,
since they are time independent and are thus entirely determined
by the parameters of the initial probability
}.
% -
Equation~\eqref{eq:CME_post_tss} thus boils down to a master equation 
for the sole leading-order probability 
$P_T(\bL^x) \equiv \lim_{\Omega\to \infty} \sum_{\bL^y} p_{T}^{(0)} (\bL^x, \bL^y)$ 
of there being $\bL^x$ molecules of the $\Spex$ species.
% -
Such master equation reads (see App.~\ref{App:LargeDeviation} for a detailed derivation)
% -
% -
\begin{equation}
\begin{split}
   \mathrm d_T P_{T}(\bL^x) =  
    \sum_{\rhoi} 
    \big\{&
     \hat r_{\rhoi}(\bL^x - \colSxi|[\boldsymbol y])P_{T}(\bL^x - \colSxi)\\
    &- \hat r_{\rhoi}(\bL^x|[\boldsymbol y])P_{T}(\bL^x) 
    \big\}
    \,,
\end{split}
    \label{eq:emeCME}
\end{equation}
where
\begin{equation}
    \hat r_{\rhoi}(\bL^x|[\boldsymbol y]) =
    \hat{\kappa}_{\rhoi} V
    \prod_x\frac{L^x! \, \theta(L^x -\nu_{x,\rhoi})}
    {(L^x-\nu_{x,\rhoi})! \, V^{\nu_{x,\rhoi}}}
    \prod_y [y]^{\nu_{y,\rhoi}}
    \,,
    \label{eq:emeMassActionOpen}
\end{equation}
and $[\boldsymbol y] = (\dots, [y], \dots)$ with
\begin{equation}
    [y]\equiv 
    \frac{\eta^y_{\ast} \Omega \, \mathfrak p_y}{V}
    \,.
    \label{eq:y_conc}
\end{equation}
Note that $\Omega \, \mathfrak p_y = \mathcal{O}(1)$ 
as $\Omega \to \infty$.
% -
% -

% -
% -
The master equation~\eqref{eq:emeCME} is 
mathematically equivalent to 
the chemical master equation for open CRNs in Eq.~\eqref{eq:CME}.
% -
% -
Physically, the $\Spex$ species play the role of internal species 
whose numbers of molecules are given by $\bL^x$.
% -
The $\Spey$ species play the role of the chemostatted species 
whose time-independent concentrations are given by $[\boldsymbol y]$ in Eq.~\eqref{eq:y_conc}.
% -
The $\Rint$ reactions play the role of the internal reactions 
whose rates~\eqref{eq:emeMassActionOpen} follow mass-action kinetics.
% -
% -
This mapping establishes a precise correspondence between the closed CRN quantities and reactions 
and those of the emergent open CRN.
% -
% -

% -
% -
\remark
The master equation~\eqref{eq:emeCME} describes the evolution of an emergent open CRN
on the slow time scale captured by the variable~$T$.
% -
% -
Accordingly, the reaction rates in Eq.~\eqref{eq:emeMassActionOpen}
measure the propensity of reactions to occur on this slow time scale.
% -
They therefore differ from the rates of the underlying closed CRN by a factor $\epsilon$
since the latter measure the propensity of reactions to occur on the fast time scale,
captured by $t = T / \epsilon$.
% -
This difference in scaling is explicitly shown in Eq.~\eqref{eq:epsilon_tss},
which compares  
the kinetic constants $\{\hat \kappa_{\rhoi}\}$ 
appearing on the right-hand side of Eq.~\eqref{eq:emeMassActionOpen}
with 
the kinetic constants $\{ k_{\rhoi}\}$
of the underlying closed CRN.

%%%%%%%%%%%%%%%%%%%%%%%%%%%%%%%%%%%%%%%%%%%%%%%%%%%%%%%

\subsection{Time Window of the Open-CRN Dynamics
at Finite Abundance Separation\label{subs:T_deviation}}
% -
In Subs.~\ref{subs:IMPLICATIONS_tss} and~\ref{subs:IMPLICATIONS_dcc},
we proved that the leading-order dynamics
of a closed CRN is that of an open CRN
by first taking the limit $\epsilon\to0$ 
and then the limit $\Omega\to\infty$.
% -
% -
However, at finite abundance separation, namely, at large but finite $\Omega$,
the dynamics of a closed CRN deviates
from that of the corresponding emergent open CRN
when the cumulative effect of the slower reactions on the total numbers of molecules $\bL^y$ 
becomes non-negligible.
% -
We now use our theory to predict the onset of this deviation.
% -
% -
% -
% -

% -
% -
To do so, 
we consider the master equation~\eqref{eq:CME_post_tss},
which emerges after the fast reactions have equilibrated
and still describes the time evolution of the total numbers of molecules $\bL^y$.
% -
Accordingly, 
the average values 
$\braket{\bL^y}_T = 
\sum_{\bL^x,\bL^y}\,
\bL^y \,
p_T^{(0)}(\bL^x,\bL^y)$ satisfy
\begin{equation}
    d_T \braket{\bL^y}_T =
        \sum_{\rhoi}\colSyi
        \braket{\hat{\omega}_{\rhoi}}
        \,,
    \label{eq:mean_Ly}
\end{equation}
with
$\braket{\hat{\omega}_{\rhoi}}
\equiv
\sum_{\bL^x,\bL^y}
\avgInf{\hat{\omega}_{\rhoi}|\bL^x,\bL^y}
p_T^{(0)}(\bL^x,\bL^y)$.
% -
% -
Crucially, 
$\braket{\hat{\omega}_{\rhoi}}$ and $\braket{\bL^y}_T$ scale differently with $\Omega$,
as a consequence of the results discussed in Subs.~\ref{subs:IMPLICATIONS_dcc}.
% -
On the one hand, 
$\braket{\hat{\omega}_{\rhoi}} =\mathcal O(1)$,
since $\avgInf{\hat{\omega}_{\rhoi}|\bL^x,\bL^y}$ 
can be split into a sum of terms of successive order in $\Omega^{-1}$,
whose leading contribution is $\mathcal O(1)$.
% -
On the other hand, $\braket{\bL^y}_T = \mathcal O(\Omega)$,
since $p_{T}^{(0)} (\bL^y | \bL^x)$ 
is sharply peaked around $\boldsymbol \eta^y_{\ast} \Omega$
with $\boldsymbol \eta^y_{\ast} = (\dots, \eta^y_{\ast}, \dots) = \mathcal{O} (1)$.
% -
% -
This implies, according to Eq.~\eqref{eq:mean_Ly},
that the characteristic time required for 
$\braket{\bL^y}_T$ to undergo a non-negligible, i.e., $\mathcal{O}(\Omega)$, change is at least
\begin{equation}
    T = \mathcal{O} (\Omega)
    \,.
    \label{eq:characteristic_time_deviation}
\end{equation}
% -
% -
From this characteristic time onward, 
the closed-CRN dynamics starts to deviate from the open-CRN dynamics.
% -
% -
Note that the scaling in Eq.~\eqref{eq:characteristic_time_deviation}
is consistent with the characteristic time required for 
a finite chemostat with a chemical capacity of order $\mathcal O(\Omega)$
to be significantly affected by exchanges with the system
occurring at rates of order $\mathcal O(1)$,
as shown in App.~\ref{Appendix:IdealReservoirs_C}.
% -
% -

% -
% -
Let us further comment on this scaling.
% -
% -
First,
closed CRNs (with species and reactions partitioned as in Sec.~\ref{sec:closedCRNs})
are characterized by three distinct time scales:
fast,
slow,
and \textit{ultra-slow}.
% -
On the fast time scale, captured by the variable $\tau = t$,
the $\Rexc$ reactions equilibrate;
on the slow time scale, captured by the variable $T = \epsilon t$,
the dynamics is that of an open CRN;
on the ultra-slow time scale, captured by the variable $\mathcal T = T / \Omega$,
the dynamics starts to deviate from that of an open CRN.
% -
According to Eq.~\eqref{eq:characteristic_time_deviation},
the larger $\Omega$,
the longer the time window over which
the closed-CRN and open-CRN dynamics are equivalent to leading order.
% -
% -
This scaling thus provides a direct practical guideline:
increasing $\Omega$ 
\--- which, from a thermodynamic perspective,
corresponds to increasing the chemical capacity of the species acting as chemostats
(App.~\ref{Appendix:IdealReservoirs_C}) \--- 
extends the time window
over which open-CRN predictions remain quantitatively valid.
% -
% -

% -
% -
Second, 
Eq.~\eqref{eq:characteristic_time_deviation} defines the characteristic time scale
below which the closed-CRN and open-CRN dynamics coincide to leading order.
It does not imply that the two dynamics necessarily deviate for $T=\mathcal O(\Omega)$.
Whether this deviation occurs also depends on the CRN topology.
% -
% -
Consider first the case in which the emergent open CRN is detailed balanced~\cite{avan24}.
% -
The $\Rint$ reactions, which occur on the slow time scale, equilibrate on that same time scale.
Hence, the CRN relaxes to equilibrium before entering the ultra-slow time scale, 
implying that the closed-CRN and open-CRN dynamics coincide for all times
(after the fast reactions have equilibrated).
% -
Conversely, if the emergent open CRN is not detailed balanced,
the $\Rint$ reactions cannot equilibrate on the slow time scale.
Hence, the dynamics can enter the ultra-slow time scale
where the closed-CRN and open-CRN dynamics start to deviate
because the evolution of the total numbers of molecules $\bL^y$
can no longer be neglected.
% -
% -
However, the closed CRN, which is detailed balanced, eventually equilibrates
on this ultra-slow time scale.
% -
In this case, the two dynamics remain equivalent for all times 
only in the singular limit $\Omega \to \infty$.
% -
% -

% -
% -
In Sec.~\ref{Sec:LinearExample},
we verify numerically 
that the closed-CRN description starts to deviate from the open-CRN description 
at a characteristic time scaling as $T \propto \Omega$,
in agreement with Eq.~\eqref{eq:characteristic_time_deviation}.

%%%%%%%%%%%%%%%%%%%%%%%%%%%%%%%%%%%%%%%%%%%%%%%%%%%%%%%

\subsection{Summary}

Closed CRNs (with species and reactions partitioned as in Sec.~\ref{sec:closedCRNs})
give rise, to leading order, to open-CRN dynamics 
when two complementary dynamical conditions are satisfied: 
a {time-scale separation} (Eq.~\eqref{eq:epsilon_tss}), 
ensuring that the $\Rexc$ reactions are much faster than the $\Rint$ reactions, 
and 
an {abundance separation} (Eq.~\eqref{eq:omega_as}),
ensuring that the $\SpeY$ species remain effectively unaffected by the slow dynamics.
% -
Together, these conditions allow the $\Rexc$ reactions to act as a chemostatting procedure 
and the $\Spey$ species to play the role of chemostatted species, 
while the $\Spex$ species and $\Rint$ reactions effectively become 
the internal species and reactions of the emergent open CRN.
% -
Under this regime, 
the chemical master equation of a closed CRN reduces to 
that of an open CRN, 
demonstrating that open-CRN dynamics emerges naturally from closed CRNs 
purely as a consequence of dynamical conditions 
(Sec.~\ref{sec:chemo_like_dynamics}).
% -
% -
For a large but finite abundance separation,
this reduction generally remains valid up to a characteristic time $T=\mathcal O(\Omega)$,
after which the slower reactions can significantly affect the abundant $\SpeY$ species.

%%%%%%%%%%%%%%%%%%%%%%%%%%%%%%%%%%%%%%%%%%%%%%%%%%%%%%%
%%%%%%%%%%%%%%%%%%%%%%%%%%%%%%%%%%%%%%%%%%%%%%%%%%%%%%%
%%%%%%%%%%%%%%%%%%%%%%%%%%%%%%%%%%%%%%%%%%%%%%%%%%%%%%%
%%%%%%%%%%%%%%%%%%%%%%%%%%%%%%%%%%%%%%%%%%%%%%%%%%%%%%%

\section{Illustrative Example of the Emergent Open-CRN Dynamics}
\label{Sec:Brusselator}
% -
% -
\begin{figure}[t]
    \centering
    \includegraphics[width=0.49\textwidth]{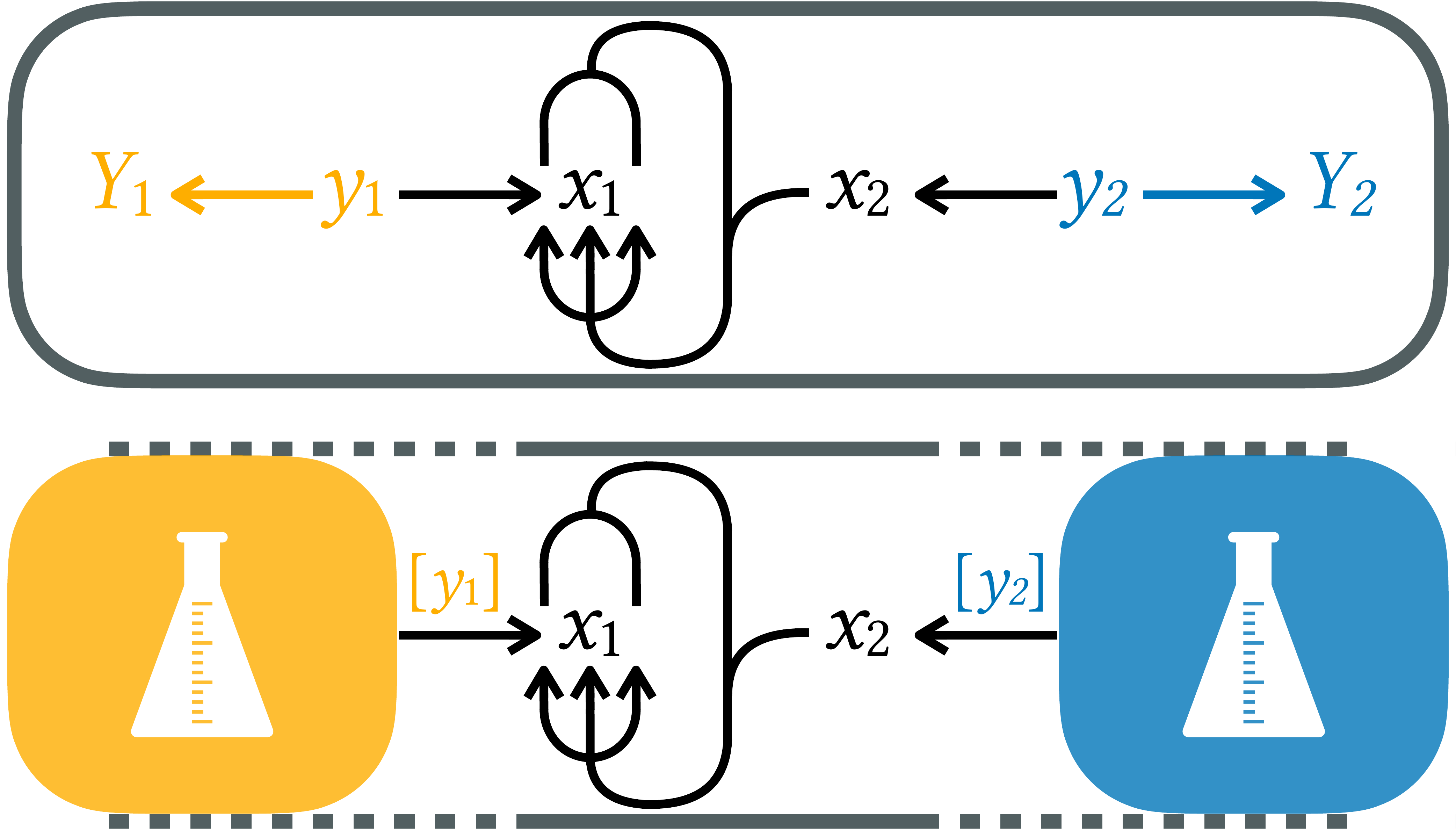}
    \caption{
    Hypergraph representation     
    of the closed Brusselator (upper closed panel) 
    and 
    its corresponding open counterpart (lower open panel).
    % -
    % -
    In the upper panel,
    colored chemical symbols and arrows
    represent the $\Spey\cup\SpeY$ species and $\Rexc$ reactions, respectively.
    % -
    In the lower panel,
    they are replaced by colored boxes with a flask, 
    representing chemostats that fix the corresponding concentrations.
    % -
    % -
    All reactions are reversible even though only the forward reactions are represented. }
    \label{fig:illustration_Brussel}
\end{figure}
% -
% -

We now illustrate our findings by comparing 
the dynamics of a closed CRN 
with 
the dynamics of the corresponding emergent open CRN
described by Eq.~\eqref{eq:emeCME}.
% -
% -
Specifically, we consider a modified version of the Brusselator 
\--- a minimal CRN capable of acting as a chemical clock~\cite{gasp02} \---
illustrated in Fig.~\ref{fig:illustration_Brussel}.
% -
We assume that the reacting species are $\{x_1, x_2, y_1, y_2, Y_1, Y_2\}$ 
and are classified as 
$\Spex = \{x_1, x_2\}$,
$\Spey = \{y_1, y_2\}$,
and
$\SpeY = \{Y_1, Y_2\}$.
% -
% -
Furthermore, 
species are interconverted by the 
$\Rint = \{\pm 1_{\mathrm{i}}, \pm 2_{\mathrm{i}}, \pm 3_{\mathrm{i}}\}$ reactions
\begin{equation}
\begin{split}
           y_1 &\ch{<=>[ $1_{\mathrm{i}}$ ][ $-1_{\mathrm{i}}$ ]} x_1 \,,\\
           2x_1 + x_2 &\ch{<=>[ $2_{\mathrm{i}}$ ][ $-2_{\mathrm{i}}$ ]} 3x_1\,, \\ 
           y_2 &\ch{<=>[ $3_{\mathrm{i}}$ ][ $-3_{\mathrm{i}}$ ]} x_2 \,,
\end{split}
\label{eq:BrusselatorIntern}
\end{equation}
constituting the minimal set of reactions of the Brusselator~\cite{schn79,qian02,nguy18,reml22},
together with the 
$\Rexc = \{\pm 1_{\mathrm{e}}, \pm 2_{\mathrm{e}}\}$ reactions 
\begin{equation}
\begin{split}
           y_1 &\ch{<=>[ $1_{\mathrm{e}}$ ][ $-1_{\mathrm{e}}$ ]} Y_1 \,,\\
           y_2 &\ch{<=>[ $2_{\mathrm{e}}$ ][ $-2_{\mathrm{e}}$ ]} Y_2 \,.
\end{split}
\label{eq:BrusselatorExchange}
\end{equation}
% -
% -

\begin{figure*}[t]
    \centering
        \includegraphics[width=2.\columnwidth]
        {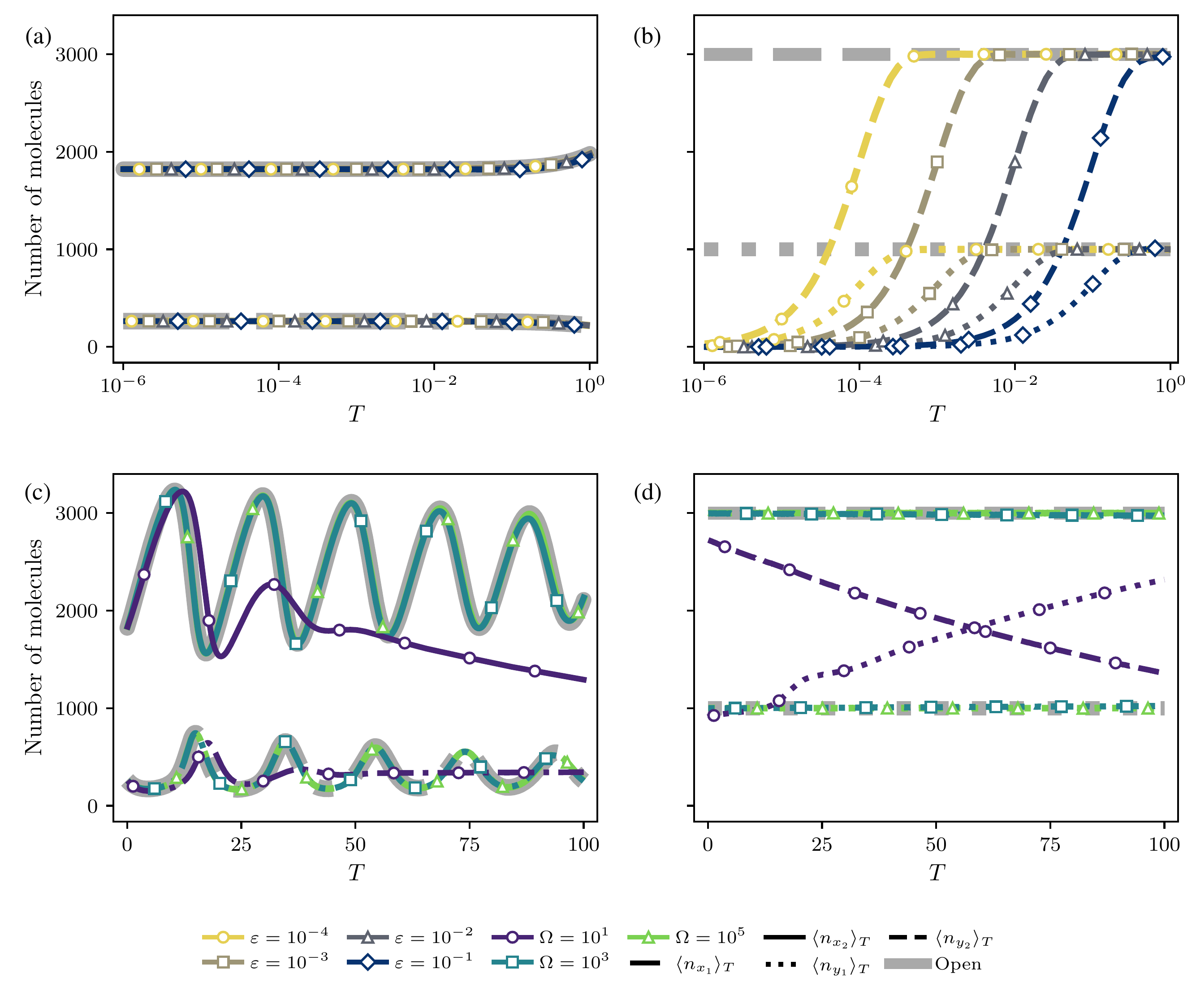}
    \caption{
    Time evolution of the average numbers of molecules
    $\avg{n_x}_T$ and $\avg{n_y}_T$ 
    with
    $x \in \Spex = \{x_1, x_2\}$ and $y \in \Spey = \{y_1, y_2\}$
    (with time expressed in terms of the variable $T = \epsilon t $).
    % -
    % -
    % -
    % -
    Colored lines with markers
    represent the dynamics for the closed Brusselator 
    for different values of $\epsilon$ and $\Omega$
    with 
    the initial probability written as 
    $p_{0}(\bn) = 
    \delta(n_{x_1}- V[x_1])
    \delta(n_{x_2}- V[x_2])
    \delta(n_{y_1})
    \delta(n_{y_2})
    \delta(n_{Y_1} - V[Y_1])
    \delta(n_{Y_2} - V[Y_2])$.
    % -
    Grey lines represent the dynamics for the corresponding emergent open CRN 
    described by Eq.~\eqref{eq:emeCME} 
    with 
    the initial probability written as 
    $p_{0}(\bn) = 
    \delta(n_{x_1}- V[x_1])
    \delta(n_{x_2}- V[x_2])$.
    % -
    % -
    Panels (a) and (b) show the time evolution of
    $\avg{n_{x_1}}_T$ and $\avg{n_{x_2}}_T$,
    and of
    $\avg{n_{y_1}}_T$ and $\avg{n_{y_2}}_T$,
    respectively,
    for different values of $\epsilon$ with $\Omega = 10^5$.
    % -
    Panels (c) and (d) show the corresponding time evolutions
    for different values of $\Omega$ with $\epsilon = 10^{-4}$.
    % -
    % -
    We use the following parameters (in arbitrary units).
    % -
    For the initial condition,  
    $[x_1] = 0.2643$, 
    $[x_2] = 1.8225$, 
    $[Y_1] = \Omega$,
    and  $[Y_2] = 3\Omega$.
    % -
    For the concentrations given in Eq.~\eqref{eq:y_conc} and entering Eq.~\eqref{eq:emeCME},
    $[y_1] = 1$ and $[y_2] = 3$.
    % -
    For the kinetic constants, 
    $\hat \kappa_{1_\mathrm{i}}  = \hat \kappa_{3_\mathrm{i}}= 0.1$, 
    $\hat \kappa_{-1_\mathrm{i}} = \hat \kappa_{2_\mathrm{i}} = \hat \kappa_{-2_\mathrm{i}} = 1$, 
    $\hat \kappa_{-3_\mathrm{i}}= 0.02469$,
    and $\hat \kappa_{1_\mathrm{e}} = 1/\hat \kappa_{-1_\mathrm{e}} = \hat \kappa_{2_\mathrm{e}} = 1/\hat \kappa_{-2_\mathrm{e}}= \sqrt{\Omega}$. 
    % -
    % -
    For the the volume,
    $V = 10^3$.
    % -
    % -
    Note that, given the above parameters, 
    the nonconservative thermodynamic force driving the emergent open CRN of the Brusselator 
    reads 
    $\Delta \mu \equiv  \ln([y_2]\kappa_{-1_\mathrm{i}}\kappa_{2_\mathrm{i}}\kappa_{3_\mathrm{i}}/[y_1]\kappa_{1_\mathrm{i}}\kappa_{-2_\mathrm{i}}\kappa_{-3_\mathrm{i}})= 4.8$
    (in units of temperature and Boltzmann constant).
    }
    \label{fig:Brussel}
\end{figure*}
% -
% -

% -
% -
Figure~\ref{fig:Brussel} shows
the time evolution of the average numbers of molecules of the $\Spex$ and $\Spey$ species,
computed using Gillespie's stochastic algorithm~\cite{gill77}.
% -
% -

% -
% -
Figures~\ref{fig:Brussel}a and~\ref{fig:Brussel}b show that the fast dynamics 
(even though it is measured by using the variable $T$) 
is characterized by an increasingly rapid convergence between 
the dynamics of the closed Brusselator and that of the corresponding emergent open CRN
as $\epsilon$ (the parameter quantifying the time-scale separation between the $\Rexc$ and $\Rint$ reactions) 
decreases.
% -
The relaxation of the averages $\avg{n_{y_1}}_T$ and $\avg{n_{y_2}}_T$ of the closed Brusselator
toward the corresponding averages of the emergent open CRN, shown in Fig.~\ref{fig:Brussel}b,
reflects the relaxation of the fast dynamics arising from the $\Rexc$ reactions 
toward steady state.
This relaxation occurs on increasingly shorter time scales as $\epsilon$ decreases.
% -
% -
The averages $\avg{n_{x_1}}_T$ and $\avg{n_{x_2}}_T$ of the closed Brusselator, 
shown in Fig.~\ref{fig:Brussel}a,
remain instead approximately constant and equal to the corresponding averages of the emergent open CRN.
% -
This reflects the fact that their evolution is determined by the $\Rint$ reactions,
which occur on the slow time scale.
% -
% -

% -
% -
Figures~\ref{fig:Brussel}c and~\ref{fig:Brussel}d show that the slow dynamics
is characterized by an increasingly long-lasting equivalence between 
the dynamics of the closed Brusselator and that of the corresponding open CRN
as $\Omega$ (the parameter quantifying the abundance separation between the $\Spey$ and $\SpeY$ species) 
increases.
% -
% -
For $\Omega = 10$, 
the two dynamics match for small $T$ (except for $\avg{n_{y_2}}_T$), 
but deviate for large $T$. 
% -
For $\Omega\geq 10^2$,
the two dynamics remain qualitatively equivalent throughout the entire time window:
the average numbers of molecules of the $\Spey$ species become effectively constant, 
while those of the $\Spex$ species oscillate 
as expected for the open Brusselator when driven far from equilibrium.
% -
This long-lasting equivalence results from the fact that the $\SpeY$ species become increasingly 
less affected by the dynamics arising from the $\Rint$ reactions 
on a longer time scale as $\Omega$ increases. 
% -
% -

% -
% -
\remark
We further verified that the evolution of the average numbers of molecules 
on the long time scale shown in Figs.~\ref{fig:Brussel}c and~\ref{fig:Brussel}d
remains practically the same for $\epsilon = 10^{-4}$, $\epsilon = 10^{-3}$, and $\epsilon = 10^{-2}$.
While $\epsilon$ strongly affects the relaxation of the fast dynamics, 
it weakly affects the dynamics on the long time scale 
(as long as a time-scale separation still exists between the $\Rexc$ and $\Rint$ reactions).

%%%%%%%%%%%%%%%%%%%%%%%%%%%%%%%%%%%%%%%%%%%%%%%%%%%%%%%
%%%%%%%%%%%%%%%%%%%%%%%%%%%%%%%%%%%%%%%%%%%%%%%%%%%%%%%
%%%%%%%%%%%%%%%%%%%%%%%%%%%%%%%%%%%%%%%%%%%%%%%%%%%%%%%
%%%%%%%%%%%%%%%%%%%%%%%%%%%%%%%%%%%%%%%%%%%%%%%%%%%%%%%

\section{Emergent Open-CRN Thermodynamics\label{sec:open_like_thermo}}

We now show that open-CRN thermodynamics emerges naturally from closed CRNs 
(with species and reactions partitioned as in Sec.~\ref{sec:closedCRNs})
under the 
dynamical conditions
introduced in Sec.~\ref{sec:chemo_like_dynamics}
and specified by Eqs.~\eqref{eq:epsilon_tss} and~\eqref{eq:omega_as}.
% -
To this end, we demonstrate that the emergent open-CRN dynamics 
is thermodynamically consistent.
% -
Specifically, we prove that:
i) the reaction rates in Eq.~\eqref{eq:emeMassActionOpen} satisfy a local detailed balance condition
identical to that of open CRNs (Subs.~\ref{subs:emeLDB});
ii) the dissipation of the emergent open CRN coincides, at leading order,
with that of the underlying closed CRN (Subs.~\ref{subs:emeEPR});
and
iii) the balance equation for the average Gibbs potential has the same structure and physical meaning
as the second law for open CRNs (Subs.~\ref{subs:eme2LAW}).

%%%%%%%%%%%%%%%%%%%%%%%%%%%%%%%%%%%%%%%%%%%%%%%%%%%%%%%
\subsection{Local Detailed Balance Condition \label{subs:emeLDB}}
We show here that 
the reaction rates in Eq.~\eqref{eq:emeMassActionOpen} satisfy a local detailed balance condition
which is formally equivalent to the one in Eq.~\eqref{eq:CRNLDB} for open CRNs.
% -
% -
To do so, 
we specialize Eq.~\eqref{eq:CRNLDB}
for the $\Rint$ reactions in Eq.~\eqref{eq:CRNclosed} of closed CRNs 
(with species and reactions partitioned as in Sec.~\ref{sec:closedCRNs}):
\begin{equation}\small
\begin{split}
    \ln 
    \frac{\omega_{\rhoi}(\bni)}{
    \omega_{-\rhoi}(\bni +\colS_{i,\rho})}
    =
    & -\sum_x\bigg\{ 
    ( \mu_x^\circ - \ln n_s ) S_{x,\rhoi} + \ln \frac{(n_x + S_{x,\rhoi})!}{n_x !} 
    \bigg\} \\
    &-\sum_y\bigg\{ 
     ( \mu_y^\circ - \ln n_s ) S_{y,\rhoi} + \ln \frac{(n_y + S_{y,\rhoi})!}{n_y !} 
     \bigg\}
\end{split}
\end{equation}
which, by using mass-action kinetics
(given in Eq.~\eqref{eq:MassActionOpen}) for closed CRNs
together with Eq.~\eqref{eq:epsilon_tss}, 
can be, equivalently, rewritten as
\begin{equation}
    \ln 
    \frac{\hat{\kappa}_{\rhoi}}{\hat{\kappa}_{-\rhoi}}
    = 
    -\sum_x ( \mu_x^\circ - \ln [s] ) S_{x,\rhoi}
    -\sum_y ( \mu_y^\circ - \ln [s] ) S_{y,\rhoi}
    \,.
    \label{eq:LDB_int_k}
\end{equation}
% -
Note that Eq.~\eqref{eq:LDB_int_k} establishes a correspondence between
the kinetic constants $\hat{\kappa}_{\pm\rhoi} = k_{\pm\rhoi} / \epsilon$, on the one hand,
and the standard chemical potentials $\{ \mu_x^\circ \}$ and $\{ \mu_y^\circ \}$, on the other hand, 
that must always be satisfied for thermodynamic consistency. 
% -
% -
Hence, by now using Eq.~\eqref{eq:LDB_int_k} in the ratio between 
$\hat r_{\rhoi}(\bL^x|[\boldsymbol y])$ 
and 
$\hat r_{-\rhoi}(\bL^x + \colSxi|[\boldsymbol y])$ 
given in Eq.~\eqref{eq:emeMassActionOpen},
we obtain that
\begin{equation}\small
    \ln
    \frac{\hat r_{\rhoi}(\bL^x|[\boldsymbol y])}
    {\hat r_{-\rhoi}(\bL^x + \colSxi|[\boldsymbol y])}
    = 
    -\Big\{\Delta_{\rhoi} \, {\hat{g}}(\bL^x) + 
    \sum_y \mu_y([y]) S_{y, \rhoi}\Big\}
    \,.
	\label{eq:emeLDB}
\end{equation}
Analogously to Subs.~\ref{sub:openThermo},
$\Delta_{\rhoi}\, {\hat{g}}(\bL^x) \equiv {\hat{g}}(\bL^x + \colSxi) - {\hat{g}}(\bL^x)$ 
is the change in Gibbs free energy
along reaction $\rhoi$
due to the $\Spex$ species
and 
\begin{equation}
    {\hat{g}}(\bL^x) =
    \sum_x 
    \big\{ 
    ( \mu_x^\circ - \ln n_s ) L^x + \ln L^x!
    \big\}
    \label{eq:eme_gibbs}
\end{equation}
is the Gibbs free energy
of an ideal solution with $\bL^x$ molecules of $\Spex$ species.
% -
Furthermore,
\begin{equation}
    \mu_y([ y]) \equiv \mu_y^\circ + \ln ( {[y]}/{[s]} ) 
    \label{eq:eme_chem_pot}
\end{equation}
corresponds to the chemical potential of species $y$
when its concentration is $[y]$ (see Eq.~\eqref{eq:chem_pot}).
% -
% -
Hence, 
Eq.~\eqref{eq:emeLDB} is equivalent to the local detailed balance of open CRNs given in Eq.~\eqref{eq:CRNLDB}.

%%%%%%%%%%%%%%%%%%%%%%%%%%%%%%%%%%%%%%%%%%%%%%%%%%%%%%%
\subsection{Dissipation \label{subs:emeEPR}}
We show here that the dissipation of the emergent open CRN
coincides with 
(the leading-order contribution to) the dissipation of the underlying closed CRN.
% -
% -
To do so, 
we apply the time-scale-separation and the abundance-separation assumption to 
the average entropy production rate in Eq.~\eqref{eq:EPR}.
% -
% -

% -
% -
We start by applying the time-scale separation 
between the $\Rexc$ and $\Rint$ reactions (Subs.~\ref{subs:ASSUMPTION_tss}).
We can thus characterize again the time evolution in terms of the two time variables $\tau$ and $T$,
capturing the fast time scale and slow time scale, respectively.
% -
The average entropy production rate in Eq.~\eqref{eq:EPR} of closed CRNs
can thus be equivalently rewritten as
\begin{equation}
    \braket{\dot \Sigma}_{\tau, T} = 
    \epre_{\tau, T} + 
    \epri_{\tau, T} \,,
    \label{eq:EPR_0}
\end{equation}
where $\epre_{\tau, T}$ and $\epri_{\tau, T}$ 
quantify the dissipation due to the $\Rexc$ and $\Rint$ reactions, respectively,
and read
\begin{subequations}\small
\begin{align}
    \epre_{\tau, T} & \equiv 
    \sum_{\rhoe, \bn} 
    \hat\omega_{\rhoe}(\bn) p_{\tau, T}(\bn) 
    \ln \frac
    {\hat\omega_{\rhoe}(\bn) p_{\tau, T}(\bn) }
    {\hat\omega_{-\rhoe}(\bn + \colSe) p_{\tau, T}(\bn + \colSe) } 
    \,, 
    \label{eq:EPRe_def}
    \\
    \epri_{\tau, T} & \equiv
    \epsilon 
    \sum_{\rhoi, \bn}
    \hat\omega_{\rhoi}(\bn) p_{\tau, T}(\bn) 
    \ln \frac
    {\hat\omega_{\rhoi}(\bn) p_{\tau, T}(\bn) }
    {\hat\omega_{-\rhoi}(\bn + \colSi) p_{\tau, T}(\bn + \colSi) } 
    \,,
    \label{eq:EPRi_def}
\end{align}
\label{eq:closedEPR}%
\end{subequations}
with
$\hat \omega_{\rhoe}(\bnm) \equiv \omega_{\rhoe}(\bnm)$ 
and 
$\hat \omega_{\rhoi}(\bnm) \equiv \omega_{\rhoi}(\bnm) / \epsilon$
as discussed in Subs.~\ref{subs:ASSUMPTION_tss}.
Note that, as $\epsilon \to 0$,  $\epre_{\tau, T} = \mathcal{O}(1)$, 
whereas $\epri_{\tau, T} = \mathcal{O}(\epsilon)$.
We now express $\epre_{\tau, T}$ and $\epri_{\tau, T}$  
using a multiscale expansion in powers of $\epsilon$ 
(similarly to the expansion used in Eq.~\eqref{eq:PerturbAnsatz} for $p_{\tau,T}(\bn)$):
\begin{subequations}
\begin{align}
    \epre_{\tau, T} & = \sum_{q = 0}^{\infty} \epreq_{\tau,T} \, \epsilon^q \,, \label{eq:epre_taylor}\\
    \epri_{\tau, T} & = \sum_{q = 1}^{\infty} \epriq_{\tau,T} \, \epsilon^q \,. \label{eq:epri_taylor}
\end{align}
\end{subequations}
% -
% -
Note that, consistently with Eq.~\eqref{eq:closedEPR},
as $\epsilon \to 0$,
the leading-order term of $\epre_{\tau,T}$ is $\mathcal{O}(1)$,
whereas the leading-order term of $\epri_{\tau,T}$ is $\mathcal{O}(\epsilon)$.
% -
% -

% -
% -
We now focus on the long-time limit of the fast time scale, i.e.,  $\tau \to \infty$, 
corresponding to the time scale where the open-CRN dynamics emerges. 
% -
% -
In such limit, as proven in App.~\ref{App:EPRe}, 
the zero-order and first-order contributions to the average entropy production rate 
due to the $\Rexc$ reactions vanish:
$\lim_{\tau \to \infty} \eprez_{\tau,T} = 0$
and 
$\lim_{\tau \to \infty} \epref_{\tau,T} = 0$,
which physical means that the $\Rexc$ reactions are 
(to leading order)
equilibrated.
% -
Hence, in such limit, the average entropy production rate in Eq.~\eqref{eq:EPR_0}
boils down to
\begin{equation}
    \lim_{\tau \to \infty} \braket{\dot \Sigma}_{\tau, T} = 
    \eprif_{\infty,T} \epsilon 
    + \mathcal{O}(\epsilon^2)
    \label{eq:EPRtot}
\end{equation}
where $\eprif_{\infty,T} \equiv \lim_{\tau \to \infty} \eprif_{\tau,T}$ and,
as proven in App.~\ref{App:EPRi},
reads
\begin{widetext}
\begin{equation}
    \eprif_{\infty,T} = 
    \sum_{\rhoi}
    \sum_{\bL^x, \bL^y}
    \avgInf{\hat \omega_{\rhoi} | \bL^x, \bL^y} \,  
    p_{T}^{(0)} (\bL^x, \bL^y)
    \ln
    \frac{\avgInf{\hat \omega_{\rhoi} | \bL^x, \bL^y} \,  
    p_{T}^{(0)} (\bL^x, \bL^y)}
    {\avgInf{\hat \omega_{-\rhoi} | \bL^x + \colS_{x,\rhoi}, \bL^y + \colS_{y,\rhoi}} \, 
    p_{T}^{(0)} (\bL^x + \colS_{x,\rhoi}, \bL^y + \colS_{y,\rhoi})} 
    \,,
    \label{eq:EPRi_epsilon}
\end{equation}
\end{widetext}
with 
$\{\avgInf{\hat \omega_{\rhoi} | \bL^x, \bL^y}\}$ given in Eq.~\eqref{eq:avg_rr_tss}
and 
$p_{T}^{(0)} (\bL^x, \bL^y)$ introduced in Eq.~\eqref{eq:p0_factorized}.
% -
% -

% -
% -
We apply now the abundance separation
between the $\Spey$ and the $\SpeY$ species (Subs.~\ref{subs:ASSUMPTION_as}).
We can thus approximate the average reactions rates featuring Eq.~\eqref{eq:EPRi_epsilon}
as in Eq.~\eqref{eq:rr_dcc0}.
% -
Furthermore, the probability $p_{T}^{(0)} (\bL^y| \bL^x)$ 
satisfies a large deviation principle. 
Namely, it is sharply peaked around time-independent and $\bL^x$-independent values 
given by $\boldsymbol \eta^y_{\ast} \Omega$,
implying that
% -
% -
the sum $\sum_{\bL^y}$ in Eq.~\eqref{eq:EPRi_epsilon} performs an average 
which, according to the Laplace's saddle-point approximation~\eqref{eq:LaplaceApprox},
yields, to leading order,
\begin{widetext}
    \begin{equation}
    \eprif_{\infty,T} = 
    {
    \sum_{\rhoi, \bL^x}
    \hat r_{\rhoi}(\bL^x|[\boldsymbol y])
    P_{T}(\bL^x)
    \ln \frac
    {\hat r_{\rhoi}(\bL^x|[\boldsymbol y]) P_{T}(\bL^x)}
    {\hat r_{-\rhoi}(\bL^x + \colSxi | [\boldsymbol y])P_{T}(\bL^x + \colSxi)}
    }\, \equiv\avg{\hat{\dot{\Sigma}}}_T \geq 0
    \,.
    \label{eq:emeEPR}
\end{equation}
\end{widetext}
with $[\boldsymbol y]$ given in Eq.~\eqref{eq:y_conc}.
% -
% -
Equation~\eqref{eq:emeEPR}, together with Eq.~\eqref{eq:EPRtot},
establishes the equivalence between
the leading-order contribution to the average entropy production rate 
of closed CRNs (left-most-hand side)
and
the average entropy production rate of the emergent open CRN
resulting from applying Eq.~\eqref{eq:EPR} to Eq.~\eqref{eq:emeCME},
denoted $\avg{\hat{\dot{\Sigma}}}_T$
(right-hand side).
% -
% -

% -
% -
\remark
The emergent open-CRN dynamics arises solely from the $\Rint$ reactions.
Nevertheless, it captures (to leading order) the entire dissipation of the underlying closed CRN 
on the slow time-scale $T$
because the $\Rexc$ reactions rapidly equilibrate.
% -
% -

% -
% -
\remark
The evolution of the emergent open CRN occurs on the slow time scale captured by the variable~$T$.
% -
% -
Accordingly, the average entropy production rate $\avg{\hat{\dot{\Sigma}}}_T$ 
measures the rate of dissipation on this slow time scale.
% -
It thus differs from the average entropy production rate of the underlying closed CRN 
by a factor $\epsilon$ 
since the latter measures the rate of dissipation on the fast time scale,
captured by $t = T / \epsilon$.
% -
This different scaling becomes explicit when combining 
Eq.~\eqref{eq:emeEPR} with Eq.~\eqref{eq:EPRtot}.

%%%%%%%%%%%%%%%%%%%%%%%%%%%%%%%%%%%%%%%%%%%%%%%%%%%%%%%
\subsection{Second Law of Thermodynamics \label{subs:eme2LAW}}
We derive
here the balance equation for the average Gibbs potential 
associated with the emergent open CRN  
by decomposing the corresponding average entropy production rate 
given in Eq.~\eqref{eq:emeEPR}.
Crucially, 
we show that such a balance equation has the same structure and physical meaning as 
the second law of thermodynamics of open CRNs given in Eq.~\eqref{eq:2law}.
% -
% -

% -
% -
We start from the average entropy production rate given in Eq.~\eqref{eq:emeEPR}
and 
use the local detailed balance condition~\eqref{eq:emeLDB}.
% -
We thus obtain 
\begin{equation}
\begin{split}
    \avg{\hat{\dot{\Sigma}}}_T = 
    - \sum_{\rhoi, \bL^x}
    &\hat r_{\rhoi}(\bL^x|[\boldsymbol y])
    P_{T}(\bL^x) \, 
    \Big\{ 
    \Delta_{\rhoi} \, {\hat{G}_{T}}(\bL^x) \\
    &+ \sum_y \mu_y([y]) S_{y, \rhoi}
    \Big\}\,.
    \label{eq:decomposition_EPR}
\end{split}
\end{equation}
Here, $\Delta_{\rhoi} {\hat{G}_{T}}(\bL^x) \equiv {\hat{G}_{T}}(\bL^x + \colSxi) - {\hat{G}_{T}}(\bL^x)$ 
is the change in Gibbs potential
along reaction $\rhoi$
due to the $\Spex$ species
and 
\begin{equation}
    {\hat{G}_{T}}(\bL^x) = 
    {\hat{g}}(\bL^x) + \ln P_T(\bL^x)
    \label{eq:avg_emeGpot}
\end{equation}
is the Gibbs free potential
(in units of temperature and Boltzmann constant)
of an ideal solution with $\bL^x$ molecules of $\Spex$ species
(with ${\hat{g}}(\bL^x)$ the corresponding Gibbs free energy given in Eq.~\eqref{eq:eme_gibbs}).
% -
Note that the Gibbs potential in Eq.~\eqref{eq:avg_emeGpot}
has the same structure of the Gibbs potential introduced in Eq.~\eqref{eq:avg_Gpot} 
for open CRNs.
% -
% -
We now recognize that Eq.~\eqref{eq:decomposition_EPR} can be written 
in terms of a balance equation for the average Gibbs potential
\begin{equation}
    \braket{\hat G}_{T} = 
    \sum_{\bL^x} P_T(\bL^x) 
    \big\{{\hat{g}}(\bL^x) + \ln P_T(\bL^x) 
    \big\}
    \label{eq:avg_emeGpot}
\end{equation}
and reads
\begin{equation}
    \avg{\hat{\dot{\Sigma}}}_T = 
    -\mathrm d_{T} \braket{\hat G}_{T}
    +\braket{\hat{\dot{W}}_{\mathrm{chm}}}_T \geq 0 
    \,,
    \label{eq:eme2law}
\end{equation}
where
\begin{equation}\small
    \braket{\hat{\dot{W}}_{\mathrm{chm}}}_T 
    \equiv
    \sum_y \mu_y([y]) 
    \sum_{\rhoi, \bL^x} (-S_{y, \rhoi})
    \hat r_{\rhoi}(\bL^x|[\boldsymbol y])
    P_{T}(\bL^x)
    \,.
\end{equation}
has the same functional form as the average chemical work introduced in Eq.~\eqref{eq:2law}.
This physically means that $\braket{\hat{\dot{W}}_{\mathrm{chm}}}_T$ can be interpreted as 
the free energy exchanges with the $\SpeY$ species,
which act as chemostats.
% -
% -

% -
% -
%%%%%%%%%%%%%%%%%%%%%%%%%%%%%%%%%%%%%%%%%%%%%%%%%%%%%%%

\subsection{Summary}
Closed CRNs
(with species and reactions partitioned as in Sec.~\ref{sec:closedCRNs})
give rise (to leading order) to open-CRN thermodynamics 
when the two complementary dynamical conditions giving rise to open-CRN dynamics
(namely, {time-scale separation} and {abundance separation}) are satisfied.
% -
% -
Indeed, under these conditions, the emergent open CRN is thermodynamically consistent:
its reaction rates satisfy a local detailed balance condition identical to that of open CRNs,
its dissipation coincides, at leading order, with that of the underlying closed CRN,
and 
the balance equation for its average Gibbs potential has the same structure and physical meaning as the second law for open CRNs.
% -
Accordingly, the thermodynamic structure of the closed CRN reduces to that of an open CRN,
demonstrating that open-CRN thermodynamics emerges naturally from closed CRNs
as a direct consequence of the imposed 
dynamical conditions (Sec.~\ref{sec:chemo_like_dynamics}).

%%%%%%%%%%%%%%%%%%%%%%%%%%%%%%%%%%%%%%%%%%%%%%%%%%%%%%%
%%%%%%%%%%%%%%%%%%%%%%%%%%%%%%%%%%%%%%%%%%%%%%%%%%%%%%%
%%%%%%%%%%%%%%%%%%%%%%%%%%%%%%%%%%%%%%%%%%%%%%%%%%%%%%%
%%%%%%%%%%%%%%%%%%%%%%%%%%%%%%%%%%%%%%%%%%%%%%%%%%%%%%%

\section{Illustrative Example of the Emergent Open-CRN Thermodynamics}
\label{Sec:LinearExample}
% -
% -
\begin{figure}[t]
    \centering
    \includegraphics[width=0.49\textwidth]{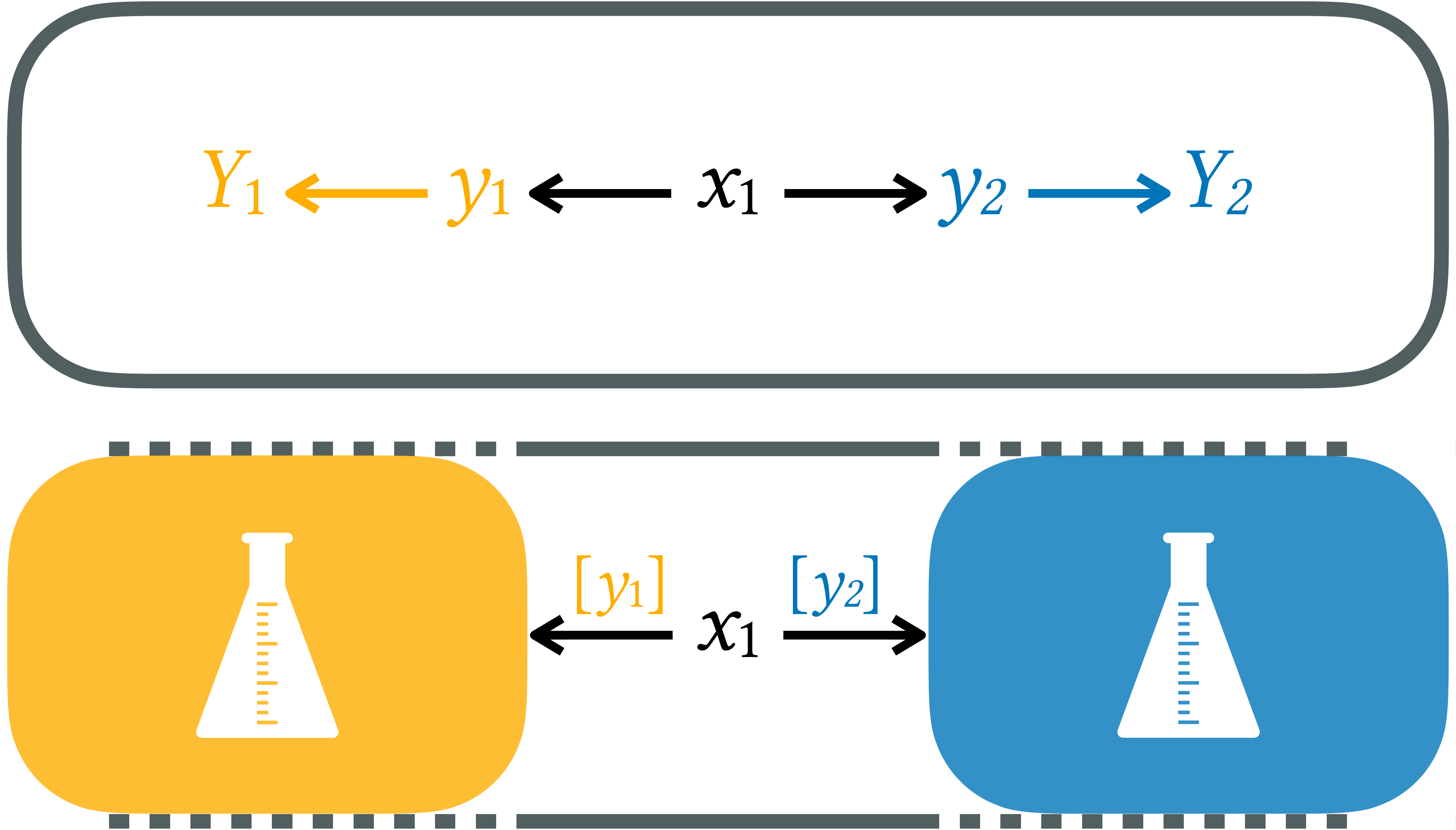}
    \caption{
    Hypergraph representation     
    of the closed CRN in Eqs.~\eqref{eq:LinearInternal} and~\eqref{eq:LinearExchange} (upper closed panel) 
    and 
    its corresponding open counterpart (lower open panel).
    % -
    % -
    In the upper panel,
    colored chemical symbols and arrows
    represent the $\Spey\cup\SpeY$ species and $\Rexc$ reactions, respectively.
    % -
    In the lower panel,
    they are replaced by colored boxes with a flask, 
    representing chemostats that fix the corresponding concentrations.
    % -
    % -
    All reactions are reversible even though only the forward reactions are represented. }
    \label{fig:illustration_linear}
\end{figure}
% -
% -

We now illustrate our findings by comparing 
the average entropy production rate of a closed CRN 
with 
the average entropy production rate of the corresponding emergent open CRN 
given in Eq.~\eqref{eq:emeEPR}.
% -
% -
To do so, 
we consider the 
CRN illustrated in Fig.~\ref{fig:illustration_linear}.
% -
We assume that the reacting species are $\{x, y_1, y_2, Y_1, Y_2\}$ 
and are classified as 
$\mathcal S_x = \{x\}$,
$\Spey = \{y_1, y_2\}$,
and
$\SpeY = \{Y_1, Y_2\}$.
% -
Furthermore, 
species are inteconverted by the
$ \Rint = \{\pm 1_\mathrm i, \pm 2_\mathrm i\}$ reactions
\begin{equation}
\begin{split}
           x &\ch{<=>[ $1_{\mathrm{i}}$ ][ $-1_{\mathrm{i}}$ ]} y_1 \,,\\
           x &\ch{<=>[ $2_{\mathrm{i}}$ ][ $-2_{\mathrm{i}}$ ]} y_2 \,,\\
\end{split}
\label{eq:LinearInternal}
\end{equation}
together with the 
$\Rexc = \{\pm 1_{\mathrm{e}}, \pm 2_{\mathrm{e}}\}$ reactions 
\begin{equation}
\begin{split}
           y_1 &\ch{<=>[ $1_{\mathrm{e}}$ ][ $-1_{\mathrm{e}}$ ]} Y_1 \,,\\
           y_2 &\ch{<=>[ $2_{\mathrm{e}}$ ][ $-2_{\mathrm{e}}$ ]} Y_2 \,.
\end{split}
\label{eq:LinearExchange}
\end{equation}
% -
% -

% - 
% -
% -
% -

\begin{figure}[h!]
    \centering
		\includegraphics[width=0.98\columnwidth]
        {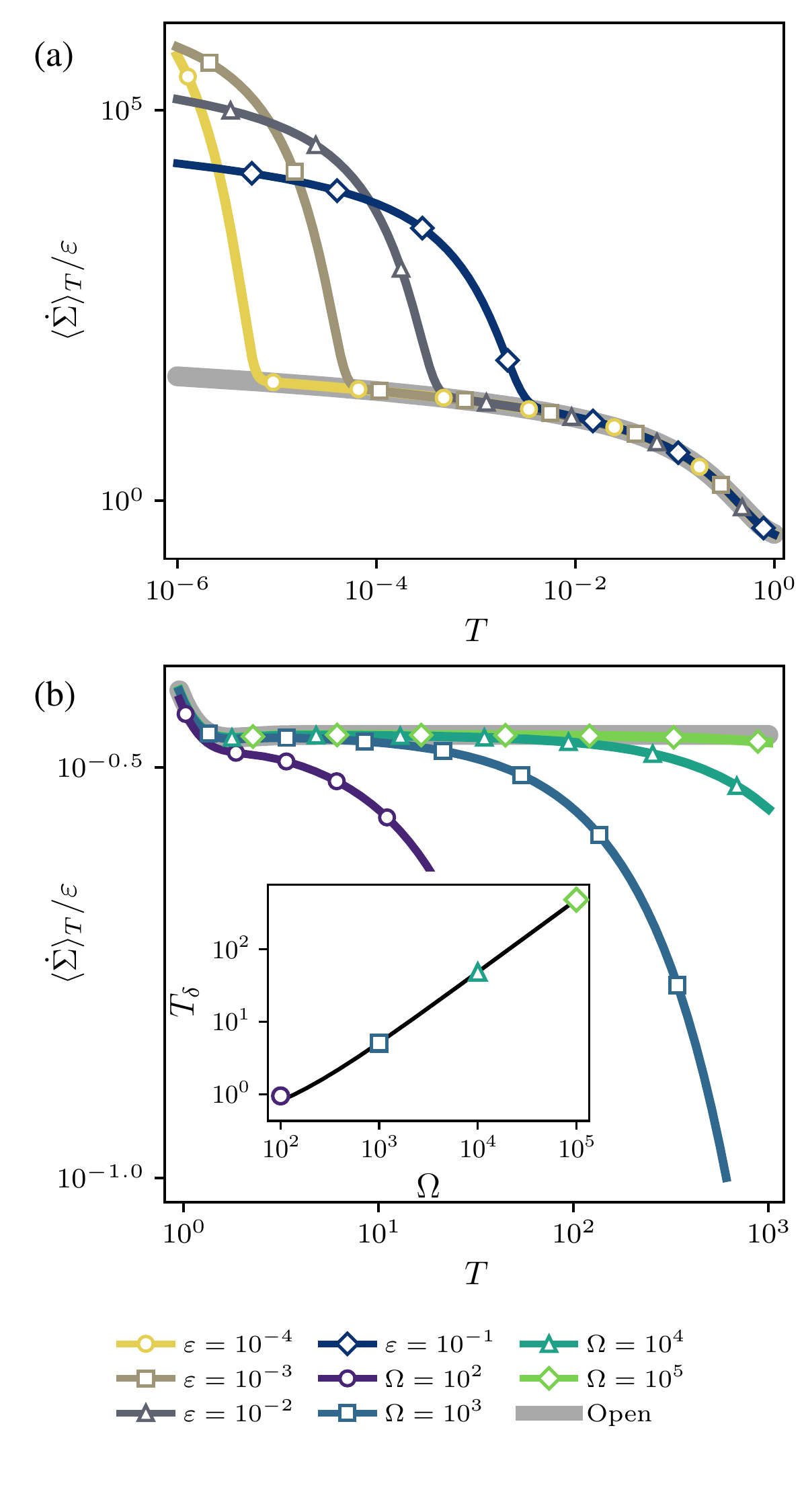}
    \caption{Time evolution of the average entropy production rate
    (with time expressed in terms of the variable $T = \epsilon t $).
    % -
    % -
    % -
    % -
    Colored lines with markers represent
    the average entropy production rate of the closed CRN
    for different values of $\epsilon$ and $\Omega$ 
    with 
    the initial probability written as 
    $p_{0}(\bn) = 
    \delta(n_{x})
    \delta(n_{y_1})
    \delta(n_{y_2})
    \delta(n_{Y_1} - V[Y_1])
    \delta(n_{Y_2} - V[Y_2])$. 
    % -
    The gray line represents
    the average entropy production rate of the emergent open CRN
    with 
    the initial probability written as 
    $p_{0}(\bn) = 
    \delta(n_{x})$.
    % -
    % -
    Panels (a) and (b) show the time evolution of $\avg{{\dot{\Sigma}}}_T$
    for different values of $\epsilon$ with $\Omega = 10^2$
    and 
    for different values of $\Omega$ with $\epsilon = 10^{-4}$,
    respectively.
    % -
    % -
    The inset in panel (b) shows the time $T_\delta$ (markers) at which 
    $|\avg{{\dot{\Sigma}}}_T / \epsilon - \avg{\hat{\dot{\Sigma}}}_T| / \avg{\hat{\dot{\Sigma}}}_T \geq \delta = 0.01$ for different values of $\Omega$ with $\epsilon = 10^{-4}$,
    together with the linear fit $T_\delta = 0.0048\, \Omega + 0.33$.
    % -
    % -
    We use the following parameters (in arbitrary units).
    % -
    For the initial condition,
    $[Y_1] = 2\Omega$
    and  $[Y_2] = \Omega$.
    % -
    For the concentrations given in Eq.~\eqref{eq:y_conc} and entering 
    Eq.~\eqref{eq:emeCME} and Eq.~\eqref{eq:emeEPR},
    $[y_1] = 2$ and $[y_2] = 1$.
    % -
    For the kinetic constants, 
    $\hat \kappa_{1_\mathrm{i}}  = \hat \kappa_{-1_\mathrm{i}}  = \hat \kappa_{2_\mathrm{i}}  = \hat \kappa_{-2_\mathrm{i}}  = 1$,
    $\hat \kappa_{1_\mathrm{e}} = \hat \kappa_{2_\mathrm{e}}= \Omega$, and 
    $ \hat \kappa_{-1_\mathrm{e}}=\hat \kappa_{-2_\mathrm{e}}=1$.
     % -
    % -
    For the the volume,
    $V =  1$
}
    \label{fig:LinearEPR}
\end{figure}

% -
% -

% -
% -
Figure~\ref{fig:LinearEPR} shows the evolution of the average entropy production rate,
obtained using the semi-analytical approach discussed in App.~\ref{App:LinearReactions}.
% -
% -

% -
% -
Figure~\ref{fig:LinearEPR}a focuses on the fast dynamics
(even though it is measured by using the variable $T$),
which is characterized by
an increasingly rapid convergence between 
the average entropy production rate of the closed CRN
and that of the corresponding emergent open CRN
as $\epsilon$ 
(the parameter quantifying the time-scale separation between the $\Rexc$ and $\Rint$ reactions) 
decreases.
% -
% -

% -
% -
Figure~\ref{fig:LinearEPR}b focuses on the slow dynamics.
It shows that
the average entropy production rate of the closed CRN given in Eq.~\eqref{eq:EPR_0}
and that of the corresponding emergent open CRN given in Eq.~\eqref{eq:emeEPR} 
are characterized by an increasingly long-lasting equivalence
as $\Omega$ (the parameter quantifying the abundance separation between the $\Spey$ and $\SpeY$ species) 
increases.
% -
% -
For $\Omega = 10^2$, 
the two average entropy production rates significantly deviate as $T $ increases
since the closed CRN equilibrates and, therefore, its entropy production rate vanishes. 
% -
For $\Omega =  10^5$,
the two average entropy production rates remain qualitatively equivalent throughout the entire time window.
% -
This long-lasting equivalence results from the fact that the $\SpeY$ species become increasingly 
less affected by the dynamics arising from the $\Rint$ reactions 
on a longer time scale as $\Omega$ increases
and, therefore, 
the closed CRN relaxes, in practice, to an emergent nonequilibrium steady state.
% -
% -

% -
% -
The inset in Fig.~\ref{fig:LinearEPR}b
focuses on the time $T_\delta$ at which the
average entropy production rate of the closed CRN deviates
from that of the corresponding emergent open CRN by the threshold value $\delta$,
according to
$|\avg{{\dot{\Sigma}}}_T / \epsilon - \avg{\hat{\dot{\Sigma}}}_T| / \avg{\hat{\dot{\Sigma}}}_T \geq \delta$.
It shows on a logarithmic scale that $T_\delta \propto \Omega$,
as expected from Subs.~\ref{subs:T_deviation}.

% -
% -
% -

% -
% -
\remark
We further verified two additional properties of the closed CRN.
% -
First, the correspondence between its dynamics and the dynamics of the emergent open CRN 
follows the same qualitative trends
already illustrated in Sec.~\ref{Sec:Brusselator}.
% -
Second, the average entropy production rate due to the internal reactions of the closed CRN
is the dominant contribution to its total entropy production rate since,
as analytically predicted in Subs.~\ref{subs:emeEPR} and numerically verified,
$\epre_t = \mathcal{O}(\epsilon^2)$ after a rapid relaxation of the fast dynamics.

%%%%%%%%%%%%%%%%%%%%%%%%%%%%%%%%%%%%%%%%%%%%%%%%%%%%%%%%%%%%
%%%%%%%%%%%%%%%%%%%%%%%%%%%%%%%%%%%%%%%%%%%%%%%%%%%%%%%%%%%%
%%%%%%%%%%%%%%%%%%%%%%%%%%%%%%%%%%%%%%%%%%%%%%%%%%%%%%%%%%%%

\section{Multimolecular $\Rexc$ Reactions\label{sec:GeneralChemostatting}}

In the context of open CRNs,
chemostatting procedures are often interpreted as resulting from reactions
of the form given in Eq.~\eqref{eq:CRN_ExchangeLinear}
between the CRN and the chemostats.
% -
% -
However, in practice, chemostatting procedures may be the result of more complex reactions.
% -
For instance,
protons $\ch{H+}$,
whose concentration is measured by the $\mathrm{pH} \approx -\log_{10}[\ch{H+}]$,
can be chemostatted by a \textit{buffer solution}~\cite{urba00, nels17},
implemented through the reaction
\begin{equation}
    \ch{H+ + A- <=>[ ][ ] HA} 
    \,,
    \label{eq:Hbuffer}
\end{equation}
involving a weak acid $\ch{HA}$ and its conjugate base $\ch{A-}$.
% -
Upon the addition of small amounts of acid or base,
reaction~\eqref{eq:Hbuffer} consumes or releases protons so that
the $\mathrm{pH}$ remains approximately constant.
% -
For example, the cytosolic $\mathrm{pH}$ of many cells is regulated by such
buffer solutions~\cite{nels17}.
% -
% -
The ability of buffer solutions to maintain 
an approximately constant $\mathrm{pH}$
is quantified by their buffer capacity $\mathcal B$,
which is equivalent to the chemical capacity
introduced in Subs.~\ref{subs:ASSUMPTION_tss} 
and App.~\ref{Appendix:IdealReservoirs_C}.
The buffer capacity $\mathcal B$ of reaction~\eqref{eq:Hbuffer}
is defined by
$1 / {\mathcal B}
= \mathrm d(\mathrm{pH})/ \mathrm d [\ch{A-}]
= - \mathrm d(\mathrm{pH})/ \mathrm d [\ch{HA}]$,
which has the same structure as Eq.~\eqref{eq:diverging_C_N},
defining the chemical capacity.
% -
Furthermore, the conditions for a diverging buffer capacity 
are analogous to those for a diverging chemical capacity:
diverging concentrations of the weak acid  and its conjugate base, i.e., 
$[\ch{HA}] \to \infty$ and $[\ch{A-}] \to \infty$.
% -
Using the Henderson-Hasselbalch equation~\cite{nels17},
i.e., 
$\mathrm{pH} \approx \mathrm{pK}_a + \log_{10} ([\ch{A-}] / [\ch{HA}])$
with $\mathrm{pK}_a$ being the negative logarithm of the acid dissociation constant,
together with the conservation 
$[L^{\ch{A}}] = [\ch{A-}] + [\ch{HA}]$,
one obtains $\mathcal B \approx ([\ch{A-}] [\ch{HA}]) / ([\ch{A-}] + [\ch{HA}])\log_{10}e$,
which is independent of $\mathrm{pK}_a$ and diverges when both concentrations diverge with the same scaling.
% -
% -
Under these conditions, 
a buffer solution acts as a chemostat fixing 
$\mathrm{pH} \approx \mathrm{pK}_a$.
% -
% -
This example provides a direct experimental realization
of the abundance-separation condition.
% -
% -

% -
% -
Motivated by reactions such as Eq.~\eqref{eq:Hbuffer}, which physically act as chemostatting procedures,
we now generalize the results derived in
Secs.~\ref{sec:chemo_like_dynamics}, \ref{sec:open_like_dynamics}, and~\ref{sec:open_like_thermo}
to the case of $\Rexc$ reactions with arbitrary molecularity. 
% -
We highlight
only on the main differences 
and omit the full derivations,
as most steps are entirely analogous to those presented in detail in 
Secs.~\ref{sec:chemo_like_dynamics}, \ref{sec:open_like_dynamics}, and~\ref{sec:open_like_thermo}.
% -
% -

% -
% -
\subsection{Stoichiometry and Conserved Quantities}
% -
Let us start by setting the stoichiometry of the $\Rexc$ reactions.
% -
We assume that every species $y$ is interconverted
through a specific pair of forward and backward reactions $\pm \rhoe(y) \in \Rexc$ 
involving the unique set of species $\SpeYY(y) \subset \SpeY$
(i.e., $\SpeYY(y) \bigcap \SpeYY(y') = \emptyset$ if $y \neq y'$)
according to the chemical equation
\begin{widetext}
\begin{equation}
\nu_{y,\rhoe(y)}\ch{Z}_y
+
\sum_{Y \in \SpeYY(y)} \nu_{Y,\rhoe(y)} \ch{Z}_{Y}
\ch{<=>[ $\rho_{\mathrm{e}}(y)$ ][ $-\rho_{\mathrm{e}}(y)$ ]}
\nu_{y,-\rhoe(y)}\ch{Z}_y
+
\sum_{Y \in \SpeYY(y)} \nu_{Y,-\rhoe(y)} \ch{Z}_{Y}\,.
\label{eq:ExchangeReaction}
\end{equation}    
\end{widetext}
Accordingly, 
the entries of $\colS_{\rhoe(y)}$ 
corresponding to any species $x \in \Spex$ and $Y' \in \SpeYY(y')$ (with $y'\neq y$) vanish,
i.e., $S_{x,\rhoe(y)} = 0$ and  $S_{Y', \rhoe(y)} = 0$.
On the other hand,
the entries 
corresponding to species $y$ and $Y \in \SpeYY(y)$
read
$S_{y, \rhoe(y)} =  
\nu_{y,-\rhoe(y)} - \nu_{y,\rhoe(y)}$
and 
$S_{Y, \rhoe(y)} = \nu_{Y,-\rhoe(y)} - \nu_{Y,\rhoe(y)}$,
respectively.
% -
% -

% -
% -
Such stoichiometry implies that 
the numbers of molecules $\bL^x (\bn) = (\dots, L^x(\bn), \dots)$
of the $\Spex$ species given in Eq.~\eqref{eq:Lx}
are conserved by the $\Rexc$ reactions.
% -
% -
Furthermore, 
by using the stoichiometric analysis illustrated in Ref.~\cite{avan24},
we recognize that each pair of reactions $\pm\rhoe(y)$ also conserves 
$\NYy$ quantities (with $\NYy$ being the number of $\SpeYY(y)$ species) 
identified by the label $\lambda \in \Lambda(y)$
of the form
\begin{equation}
L^\lambda(\bn)
= \ell^\lambda_y n_y
+ 
\sum_{Y \in \SpeYY(y)} \ell^\lambda_Y n_Y
\,,
\label{eq:Ll}
\end{equation}
where the so-called conservation law $\cl^\lambda = (\ell^\lambda_y, \dots, \ell^\lambda_Y,\dots)$ 
satisfies 
\begin{equation}
\ell^\lambda_y S_{y, \rhoe(y)}
+ 
\sum_{Y \in \SpeYY(y)} \ell^\lambda_Y S_{Y, \rhoe(y)} = 0 
\,.
\end{equation}
In the following,
all quantities defined in Eq.~\eqref{eq:Ll} 
which are conserved by all reactions $\rhoe\in\Rexc$
(and correspond to linearly independent conservation laws)
are collected in the vector $\bL^\lambda (\bn) = (\dots, L^\lambda(\bn), \dots)$.
% -
Note that the number of entries of $\bL^\lambda$ is thus $\sum_y \NYy$. 
% -
% -

% -
% -
Finally, 
because of the rank nullity theorem for the column matrix $\colS_{\rhoe(y)}$,
there is always a representation of the conservation laws such that 
the square matrix whose entries are $\{\ell^\lambda_Y\}$ 
with ${\lambda \in \Lambda(y)}$ and ${Y \in \SpeYY(y)}$ 
is invertible~\cite{avan24}.
Thus, by labeling $\{\icle_\lambda^Y\}$ 
with ${\lambda \in \Lambda(y)}$ and ${Y \in \SpeYY(y)}$ 
the entries of the corresponding inverse matrix,
the numbers of molecules $\{n_Y\}$ with ${Y \in \SpeYY(y)}$
can be expressed as a function 
of the number of molecules $n_y$
and 
of the quantities $\{L^\lambda\}$ 
with ${\lambda \in \Lambda(y)}$ according to
\begin{equation}
n_Y = \sum_{\lambda \in \Lambda(y)} \icle_\lambda^Y (L^\lambda - \cle^\lambda_y n_y)
\,.
\label{eq:nY_mm}
\end{equation}
% -
% -
For the sake of dealing with a compact notation, 
we rewrite Eq.~\eqref{eq:nY_mm} for every $Y \in \SpeYY(y)$ as
\begin{equation}
n_Y = a_Y(\bL^\lambda) - b_{Y,y} n_y
\label{eq:nYm}
\end{equation}
by introducing 
\begin{subequations}
\begin{align}
    a_Y(\bL^\lambda) &\equiv  \sum_{\lambda \in \Lambda(y)} \icle_\lambda^Y L^\lambda 
    \,, 
    \label{eq:aY}\\
    b_{Y,y} & \equiv  \sum_{\lambda \in \Lambda(y)} \icle_\lambda^Y \cle^\lambda_y 
    \,.
    \label{eq:bYy}
\end{align}
\end{subequations}
% -
% -

% -
% -
\remark
We anticipate that the main implication of the multimolecular stoichiometry of the $\Rexc$ reactions
is that 
the dependence on the quantities $\bL^y$ defined in Eq.~\eqref{eq:Ly},
which appears in the equations of
Secs.~\ref{sec:chemo_like_dynamics}, \ref{sec:open_like_dynamics}, and~\ref{sec:open_like_thermo},
must be replaced by 
a dependence on the quantities $\bL^\lambda$ defined in Eq.~\eqref{eq:Ll}.

% -
% -
\subsection{Dynamical Conditions for the Emergence of Open CRNs}
We now examine how the multimolecular stoichiometry of the $\Rexc$ reactions
modifies the results of Sec.~\ref{sec:chemo_like_dynamics}.
% -
% -

% -
% -
\paragraph*{Time-Scale Separation.}
By repeating the derivations of Subs.~\ref{subs:ASSUMPTION_tss},
which rely on the time-scale separation between the $\Rexc$ and $\Rint$ reactions,
we first find 
that the fast dynamics is confined within the stoichiometric compatibility class 
\begin{equation}\small
\mathcal N (\bL^x, \bL^\lambda) =
    \big\{
    \bn \in \mathcal N : \bL^x(\bn) = \bL^x 
    \text{ and } 
    \bL^\lambda(\bn) = \bL^\lambda
    \big\}
    \,,
\end{equation}
defined by the quantities $\bL^x$ and $\bL^\lambda$,
rather than by $\bL^x$ and $\bL^y$. 
% -
% -
Second,
the zero-order probability $p_{\tau, T}^{(0)}(\bn)$ (with $\bn \in \mathcal N (\bL^x, \bL^\lambda)$)
can still be written as in Eq.~\eqref{eq:p0_factorized},
upon replacing $\bL^y$ with $\bL^\lambda$.
% -
% -
Third, 
the main difference resulting from the multimolecular stoichiometry of the $\Rexc$ reactions concerns 
the steady state probability in the long time limit of the fast time scale
$\piInf(\bn|\bL^x,\bL^\lambda)$ which now reads
\begin{equation}
\begin{split}
    \piInf(\bn|\bL^x,\bL^\lambda) = 
    &\prod_{y}
    \bigg\{
    \piInfy(n_y|\bL^\lambda)
    \prod_{\lambda \in \Lambda(y)}
    \delta_{L^\lambda(\bn), L^\lambda}
    \bigg\} \\
    \times&\prod_{x}
    \delta_{n_x, L^x}
    \,,
    \label{eq:pss_mm}
\end{split}
\end{equation}
where, for the same reasons discussed in App.~\ref{Appendix:CMEFastTimeScale}, 
$\piInfy(n_y|\bL^\lambda)$ is a Poisson-like distribution of the form
\begin{equation}
    \piInfy(n_y|\bL^\lambda)
    \propto
    \frac{(\mathfrak c_y)^{n_y}}{n_y !}
    \prod_{Y \in \SpeYY(y)}
    \frac{(\mathfrak c_Y)^{{(a_Y(\bL^\lambda) - b_{Y,y} n_y)}}}{(a_Y(\bL^\lambda) - b_{Y,y} n_y)!}
    \,,
    \label{eq:piInfy_mm}
\end{equation}
with
\begin{equation}
    \bigg(\frac{\mathfrak c_{y}}{V}\bigg)^{S_{y,\rhoe(y)}}
    \prod_{Y \in \SpeYY(y)}
    \bigg(\frac{\mathfrak c_{Y}}{V}\bigg)^{S_{Y,\rhoe(y)}}
    = 
    \frac{k_{\rhoe(y)}}{k_{-\rhoe(y)}}
    \,,
    \label{eq:kconst_multimol}
\end{equation}
$L^\lambda = \ell^\lambda_y \mathfrak c_{y} + \sum_{Y \in \SpeYY(y)} \ell^\lambda_Y \mathfrak c_{Y}$, and $n_y \in \mathcal N_y (\bL^\lambda) =
\big\{ n_y \in \mathbb N : 
(a_Y(\bL^\lambda) - b_{Y,y} n_y) \geq 0\,\, \forall Y \in \SpeYY(y)
\big\}$ 
(with $\mathbb N$ denoting the set of non-negative integers).
% -
% -
Note that $N_y (\bL^\lambda)$ is the set of admissible values of $n_y$ such that,
for given  
$\bL^\lambda$,
the corresponding numbers of molecules of the $\SpeYY(y)$ are non-negative 
and therefore physically meaningful. 
Indeed, according to Eq.~\eqref{eq:nYm},
$(a_Y(\bL^\lambda) - b_{Y,y} n_y)$ represents the number of molecules of species $Y \in \SpeYY (y)$
for a given number of molecules $n_y$
and 
given values of the quantities $\bL^\lambda$.
% -
% -

% -
% -
\paragraph*{Abundance Separation.}
By repeating the reasoning of Subs.~\ref{subs:ASSUMPTION_as},
which relies on the abundance separation between the $\SpeY$ and $\Spey$ species,
we recognize that
the support of $\piInfy(n_y|\bL^\lambda)$ in Eq.~\eqref{eq:piInfy_mm} 
must be restricted to values of $n_y$ that are $\mathcal O(1)$
as $\Omega \to \infty$,
whereas the quantities 
$\{L^\lambda\}$
and
$\{ a_Y(\bL^\lambda) \}$
must be $\mathcal O(\Omega)$
(see Eqs.~\eqref{eq:Ll} and~\eqref{eq:nYm}, respectively).
% -
% -
This implies that each factor 
${(\mathfrak c_Y)^{(a_Y(\bL^\lambda) - b_{Y,y} n_y)}}/{(a_Y(\bL^\lambda) - b_{Y,y} n_y)!}$
in Eq.~\eqref{eq:piInfy_mm}
can be simplified by using
Stirling's approximation
together with 
a Taylor expansion in $ b_{Y,y} n_y / a_Y(\bL^\lambda) = \mathcal O(\Omega^{-1})$ 
(see App.~\ref{app:pss_mm} for details).
To leading order, this yields
$\exp\{-(\ln (\mathfrak c_Y) - \ln (a_Y(\bL^\lambda)))  b_{Y,y} n_y
+ a_Y(\bL^\lambda)(\ln (\mathfrak c_Y) - \ln (a_Y(\bL^\lambda)) + 1) \}$.
% -
% -
As a consequence, 
the distribution $\piInfy(n_y|\bL^\lambda)$ in Eq.~\eqref{eq:piInfy_mm} can be rewritten,
to leading order,
as a Poisson distribution:
\begin{equation}\small
    \piInfy(n_y|\bL^\lambda)
    = 
    \frac{(\tilde{\mathfrak c}_y (\bL^\lambda))^{n_y}}{n_y !} e^{-\tilde{\mathfrak c}_y (\bL^\lambda)}
    \,,
    \label{eq:piInfy_poisson_mm}
\end{equation}
where we define 
\begin{equation}\small
    \tilde{\mathfrak c}_y (\bL^\lambda) \equiv 
    \exp\Big\{\ln (\mathfrak c_y) - 
    \sum_{Y \in \SpeYY(y)}(\ln (\mathfrak c_Y) - \ln (a_Y(\bL^\lambda)))  b_{Y,y}\Big\}
    \,,
    \label{eq:cy_tilda}
\end{equation} 
and the normalization constant $\exp\{-\tilde{\mathfrak c}_y (\bL^\lambda)\}$ 
arises because the support of $\piInfy(n_y|\bL^\lambda)$ 
is effectively confined within its domain $N_y (\bL^\lambda)$,
so that, for normalization purposes, 
$N_y (\bL^\lambda)$ can be identified with the set of non-negative integers 
$\mathbb N$.
% -
% -
Note that this implies that
the average numbers of molecules of species $y$ and $Y \in \SpeYY(y)$
in the long-time limit of the fast time scale
read
\begin{subequations}
\begin{align}
    \avgInf{n_y | \bL^\lambda} & =
    \tilde{\mathfrak c}_y (\bL^\lambda) = \mathcal{O}(1) \,, \label{eq:avg_ny_m}\\
    \avgInf{n_Y | \bL^\lambda} & 
    = a_Y(\bL^\lambda) - b_{Y,y} \tilde{\mathfrak c}_y (\bL^\lambda)  = \mathcal{O}(\Omega)\,, \label{eq:avg_nYY_m}
\end{align}
\label{eq:avg_ny_infy_m}%
\end{subequations}
respectively, as $\Omega \to \infty$.
% -
% -
The $\mathcal{O}(1)$ scaling in Eq.~\eqref{eq:avg_ny_m}
follows from the abundance separation imposing that 
the support of $\piInfy(n_y|\bL^\lambda)$
is restricted to values of $n_y$ that are $\mathcal O(1)$.
% -
% -
The $\mathcal{O}(\Omega)$ scaling in Eq.~\eqref{eq:avg_nYY_m}
follows from the abundance separation imposing that 
$a_Y(\bL^\lambda) = \mathcal{O}(\Omega)$
together with 
Eq.~\eqref{eq:avg_ny_m} and $\{b_{Y,y}\}$ being $\mathcal{O}(1)$ by definition (see Eq.~\eqref{eq:bYy}).
% -
In contrast to the case where the $\Rexc$ reactions have unimolecular stoichiometries
(Subs.~\ref{subs:ASSUMPTION_as}),
the scaling in Eq.~\eqref{eq:avg_ny_infy_m} is not directly related to the kinetic constants $k_{\pm\rhoe(y)}$
in Eq.~\eqref{eq:kconst_multimol}.
% -
This means that the abundance separation does not, in general, correspond to
a time-scale separation between each pair of
forward and backward reactions $\pm\rhoe(y)$.

% -
% -
\subsection{Emergent Open-CRN Dynamics}
We now examine how the multimolecular stoichiometry of the $\Rexc$ reactions
modifies the results of Sec.~\ref{sec:open_like_dynamics}.
% -
% -

% -
% -
\paragraph*{Time-Scale Separation.}
We notice that the derivations of Subs.~\ref{subs:IMPLICATIONS_tss},
which rely on the time-scale separation between the $\Rexc$ and $\Rint$ reactions,
are independent of the stoichiometry of the $\Rexc$ reactions.
% -
Hence, in the presence of multimolecular $\Rexc$ reactions,
Eq.~\eqref{eq:CME_post_tss} can be rewritten as
\begin{widetext}
\begin{equation}
   \mathrm d_T p_{T}^{(0)} (\bL^x, \bL^\lambda) =  
    \sum_{\rhoi} 
    \big\{
    \avgInf{\hat \omega_{\rhoi} | \bL^x - \colS_{x,\rhoi}, \bL^\lambda - \colS_{\lambda,\rhoi}} \, 
    p_{T}^{(0)} (\bL^x - \colS_{x,\rhoi}, \bL^\lambda - \colS_{\lambda,\rhoi})  
    -  
    \avgInf{\hat \omega_{\rhoi} | \bL^x, \bL^\lambda} \,  
    p_{T}^{(0)} (\bL^x, \bL^\lambda)
    \big\}
    \,,
\label{eq:CME_post_tss_mm}
\end{equation}    
\end{widetext}
where we introduce the vector $\colS_{\lambda,\rhoi} = (\dots, S_{\lambda,\rhoi}, \dots)$ 
whose entries are defined as $S_{\lambda,\rhoi} \equiv \cle^\lambda_y S_{y, \rhoi}$.
% -
This follows from the fact that 
$L^\lambda(\bn - \colSyi) = L^\lambda(\bn) - \ell^\lambda_y S_{y, \rhoi}$
as a result of Eq.~\eqref{eq:Ll} and 
of the stoichiometry of the $\Rint$ reactions in Eq.~\eqref{eq:CRNclosed}.
% -
% -

% -
% -
\paragraph*{Abundance Separation.}
Because of the abundance separation, 
$\piInfy(n_y|\bL^\lambda)$ is given, to leading order, in Eq.~\eqref{eq:piInfy_poisson_mm} 
This allows us to express the 
average reaction rates $\{\avgInf{\hat \omega_{\rhoi} | \bL^x, \bL^\lambda}\}$ 
in Eq.~\eqref{eq:CME_post_tss_mm}
as
\begin{equation}
\begin{split}
    \avgInf{\hat \omega_{\rhoi} | \bL^x, \bL^\lambda}
    =
    \hat{\kappa}_{\rhoi} V 
    &\prod_x
    \frac{L^x ! \, \theta(L^x -\nu_{x,\rhoi})}
    {(L^x - \nu_{x,\rhoi})! \, V^{\nu_{x,\rhoi}}} \\
    \times&\prod_y
    \bigg(\frac{\tilde{\mathfrak c}_y( \bL^\lambda)}{V}\bigg)^{\nu_{y,\rhoi}}
    \,.
\end{split}
\label{eq:avg_rr_tss_mm}
\end{equation}
% -
% - 
Furthermore,
the terms $\{\tilde{\mathfrak c}_y( \bL^\lambda - \colS_{\lambda,\rhoi})\}$, 
defined in Eq.~\eqref{eq:cy_tilda}
and 
featuring the average reaction rates 
$\{\avgInf{\hat \omega_{\rhoi} | \bL^x - \colS_{x,\rhoi}, \bL^\lambda - \colS_{\lambda,\rhoi}}\}$,
satisfy 
$\tilde{\mathfrak c}_y( \bL^\lambda - \colS_{\lambda,\rhoi}) = 
\tilde{\mathfrak c}_y( \bL^\lambda) + \mathcal O(\Omega^{-1})$
since 
$a_Y( \bL^\lambda) = \mathcal{O}(\Omega)$
and
$\ln(a_Y( \bL^\lambda - \colS_{\lambda,\rhoi})) =  
\ln(a_Y( \bL^\lambda)) 
- \sum_{\lambda \in \Lambda(y)} \icle^{Y}_\lambda S_{\lambda,\rhoi} /  a_Y( \bL^\lambda)
+ \mathcal O(\Omega^{-2})
= \ln(a_Y( \bL^\lambda)) + \mathcal O(\Omega^{-1})$
(see App.~\ref{app:w_mm} for details).
% -
This implies that, by taking the $\mathcal{O}(1)$ contributions only,
as $\Omega \to \infty$,
the average reactions rates can still be approximated as in Eq.~\eqref{eq:rr_dcc0},
upon replacing $\bL^y$ with $\bL^\lambda$.
Namely,
\begin{subequations}\small
    \begin{align}
        \avgInf{\hat \omega_{\rhoi} | \bL^x, \bL^\lambda}
        & \approx \hat R_{\rhoi}(\bL^x, \bL^\lambda)
        \,,\\
        \avgInf{\hat \omega_{\rhoi} | \bL^x - \colSxi, \bL^\lambda - \colS_{\lambda, \rhoi}}
        & \approx \hat R_{\rhoi}(\bL^x- \colSxi, \bL^\lambda)
        \,,
    \end{align}
    \label{eq:rr_dcc0_mm}%
\end{subequations}
with $\hat R_{\rhoi}(\bL^x, \bL^\lambda)$ now reading
\begin{equation}\small
    \hat R_{\rhoi}(\bL^x, \bL^\lambda)
    \equiv 
    \hat{\kappa}_{\rhoi} V 
    \prod_x
    \frac{L^x ! \, \theta(L^x -\nu_{x,\rhoi})}
    {(L^x - \nu_{x,\rhoi})! \, V^{\nu_{x,\rhoi}}} 
    \prod_y
    \bigg(\frac{\tilde{\mathfrak c}_y( \bL^\lambda)}{V}\bigg)^{\nu_{y,\rhoi}}
    \,.
    \label{eq:rr_dcc1_mm}
\end{equation}
% -
% -
Note that, according to Eq.~\eqref{eq:avg_ny_m},
each factor $\tilde{\mathfrak c}_y( \bL^\lambda) / {V}$ in Eq.~\eqref{eq:rr_dcc1_mm}
represents the average concentration of species $y$ when the $\Rexc$ reactions are multimolecular,
just as
each factor $L^y \, \mathfrak p_y / {V}$ in Eq.~\eqref{eq:rr_dcc1}
represents the average concentration of species $y$ when the $\Rexc$ reactions are unimolecular.
% -
% -

% -
% -
Finally, the abundance separation ensures again that the probability $p_{T}^{(0)} (\bL^\lambda | \bL^x)$ 
satisfies a large deviation principle:
it is sharply peaked around time-independent and $\bL^x$-independent most probable values 
given by $\boldsymbol \eta^\lambda_{\ast} \Omega$
with $\boldsymbol \eta^\lambda_{\ast} = (\dots, \eta^\lambda_{\ast}, \dots) = \mathcal{O} (1)$ 
as $\Omega \to \infty$~\footnote{
Note that, for a given initial probability distribution of the form
$p_{0}(\bn) = \delta(\bn - \bn^{0})$,
the most probable values of $p_{T}^{(0)} (\bL^\lambda | \bL^x)$
are $\boldsymbol \eta^\lambda_{\ast} \Omega = \overline{\bL}^\lambda$,
with $\overline{L}^\lambda  
= \ell^\lambda_y n^{0}_y
+ 
\sum_{Y \in \SpeYY(y)} \ell^\lambda_Y n^{0}_{Y}$,
since they are time independent and are thus entirely determined
by the parameters of the initial probability
}.
% -
% -
Equation~\eqref{eq:CME_post_tss_mm} thus boils down to Eq.~\eqref{eq:emeCME}
where the concentrations $[\boldsymbol y] = (\dots, [y], \dots)$,
featuring the reaction rates $\hat r_{\rhoi}(\bL^x|[\boldsymbol y])$ in Eq.~\eqref{eq:emeMassActionOpen},
now read
\begin{equation}
    [y]\equiv 
    \frac{\tilde{\mathfrak c}_y(\boldsymbol \eta^\lambda_{\ast} \, \Omega)}{V}
    \,.
    \label{eq:y_conc_mm}
\end{equation}
% -
% -

% -
% -
\subsection{Emergent Open-CRN Thermodynamics}
We now recognize that the results of Sec.~\ref{sec:open_like_thermo}
are invariant under the stoichiometry change for the $\Rexc$ reactions.
% -
% -
Indeed, the emergence of the local detailed balance
in Eq.~\eqref{eq:emeLDB}
relies on the expression of the reaction rates
in Eq.~\eqref{eq:emeMassActionOpen}.
% -
The equivalence between the average entropy production rate of the emergent open CRN 
and (the leading-order contribution to) the average entropy production rate of the underlying closed CRN
in Eq.~\eqref{eq:emeEPR}
relies on the equilibration of the $\Rexc$ reactions and the large deviation principle. 
The only caveat is explained in App.~\ref{app:epr_mm}.
% -
Finally, the emergence of balance equation for the average Gibbs potential
in Eq.~\eqref{eq:eme2law}
relies on the the local detailed balance condition~\eqref{eq:emeLDB}.

%%%%%%%%%%%%%%%%%%%%%%%%%%%%%%%%%%%%%%%%%%%%%%%%%%%%%%%
%%%%%%%%%%%%%%%%%%%%%%%%%%%%%%%%%%%%%%%%%%%%%%%%%%%%%%%
%%%%%%%%%%%%%%%%%%%%%%%%%%%%%%%%%%%%%%%%%%%%%%%%%%%%%%%
%%%%%%%%%%%%%%%%%%%%%%%%%%%%%%%%%%%%%%%%%%%%%%%%%%%%%%%
%%%%%%%%%%%%%%%%%%%%%%%%%%%%%%%%%%%%%%%%%%%%%%%%%%%%%%%
%%%%%%%%%%%%%%%%%%%%%%%%%%%%%%%%%%%%%%%%%%%%%%%%%%%%%%%
%%%%%%%%%%%%%%%%%%%%%%%%%%%%%%%%%%%%%%%%%%%%%%%%%%%%%%%
\section{Conclusions \& perspectives}
\label{Sec:Conclusion}

We have identified precise conditions under which the dynamics and thermodynamics of open CRNs emerge from an underlying closed CRN:
a time-scale separation (between the $\Rexc$ and $\Rint$ reactions) and 
an abundance separation (between the $\Spex$ and $\Spey$ species, on the one hand, 
and the $\SpeY$ species, on the other hand).
% -
In the intermediate time window
after the rapid equilibration of the fast $\Rexc$ reactions 
and before the abundant $\SpeY$ species are significantly affected by the reactions,
closed CRNs enter a well-defined regime in which their
leading-order dynamics and thermodynamics reduce
to those of an open CRN.
% -
This time window becomes infinitely long in the asymptotic regime
obtained by first taking $\epsilon\to0$ and then $\Omega\to\infty$.
% -
For large but finite $\Omega$, 
deviations from the open-CRN behavior generally arise on a characteristic ultra-slow time scale
$T = \mathcal{O}(\Omega)$.
% -
% -
Crucially, our results do not rely on specific stoichiometric constraints,
and are instead systematically derived using
a perturbation ansatz (to account for the time-scale separation)
and 
a WKB ansatz (to account for the abundance separation).
% -
% -
We have thus shown that chemostats do not need to be postulated 
as external idealizations;
rather, they arise naturally as emergent thermodynamic structures
within closed CRNs under precise physical conditions.
% -
% -

% -
% -
Within the emergent open-CRN regime,
the dynamics and thermodynamics of CRNs can be analyzed using the same methods developed for open CRNs.
% -
In particular,
the analysis of the stoichiometry of the emergent open CRN
in terms of conservation laws and cycles,
as developed in Refs.~\cite{Polettini2014, rao16, rao18},
characterizes the role of the $\SpeY$ species
in generating the emergent nonconservative forces
driving the nonequilibrium dynamics.
% -
% -
For completeness,
we mention that conservation laws have also been used
in the analysis of experimental chemical data,
providing information on the underlying CRN topology and kinetics~\cite{Blokhuis2025, blokhuis2025datadimensionchemistry, blokhuis2026emergentconservedquantitiesirreversibility}.
% -
% -

% -
% -
The emergence of open-CRN behavior can be understood as a form of coarse-graining:
the fast $\Rexc$ reactions and the abundant $\SpeY$ species are effectively eliminated,
yielding an effective description in terms of 
the slow $\Rint$ reactions and the remaining species,
which is equivalent (to leading order) to that of an open CRN.
% -
% -
In the context of CRNs,
other coarse-graining approaches have been developed,
but they typically rely on a single explicit scale-separation limit.
% -
For instance, time-scale separation between reactions 
(partitioned in a similar way as in Sec.~\ref{sec:closedCRNs})
has been used to derive a closed description in terms of conserved quantities~\cite{Shimada2024}.
% -
This work may resemble ours at first sight.
There are, however, two crucial differences.
% -
First, 
both papers consider reactions of the form in Eq.~\eqref{eq:CRN_ExchangeLinear}.
% -
However, 
$\ch{Z}_{Y(y)}$ is an explicit chemical species in a closed CRN in our paper,
whereas it represents an idealized chemostat in Ref.~\cite{Shimada2024}.
% -
% -
Second, 
both papers assume a time-scale separation between two sets of reactions.
% -
However, 
the reactions of the form in Eq.~\eqref{eq:CRN_ExchangeLinear} are fast in our paper,
whereas the corresponding exchange reactions are slow in Ref.~\cite{Shimada2024}.
% -
% -
Time-scale
separation between chemical species 
has also been used to derive a closed (and thermodynamically consistent)
description in terms of the slowly evolving species interconverted by effective reactions~\cite{avan2020, avan23, avan2025}.
% -
% -
Abundance separation between chemical species was the only explicitly imposed asymptotic limit
in our previous contribution~\cite{reml25},
where we showed that closed CRNs can behave as open ones,
but only under restrictive stoichiometric constraints.
% -
These constraints automatically generated a time-scale separation:
the reactions involving only the species acting as chemostatted species
were faster by a factor of order $\Omega$
than those also involving the species acting as internal species.
% -
Thus,
both physical ingredients were present,
although the time-scale separation did not need to be introduced
through an additional independent limit.
% -
% -
Here, we have treated the time-scale and abundance separations explicitly
through two distinct asymptotic limits.
% -
Their combined action removes the need for the restrictive stoichiometric conditions
imposed in Ref.~\cite{reml25}.
% -
% -

% -
% -
From a theoretical perspective, 
our results provide a unified foundation 
for the nonequilibrium thermodynamics of open CRNs:
they 
clarify under which physical conditions 
its predictions are quantitatively valid.
% -
% -
From an experimental perspective,
the two dynamical conditions identified here
translate into physically measurable requirements on reaction time scales
as well as on species abundances,
thereby offering practical criteria for interpreting experiments
in fueled chemical systems.
For instance, such criteria predict the characteristic time scale
over which finite closed systems can effectively behave as open CRNs.
% -
% -
More broadly,
our results may guide the interpretation and design
of autonomous fueled systems operating under batch conditions,
including autonomous molecular machines~\cite{
Wilson:2016aa,Borsley2021,Amano:2021aa,Borsley:2022aa}
and chemical oscillators~\cite{Zhao2005,Frerichs2019,Frerichs2021}.
% -
% -
Beyond batch conditions,
chemical systems can also be maintained out of equilibrium,
or at least their equilibration can be delayed,
through alternative experimental strategies,
whose implementation can be arduous and, in some cases, impractical.
% -
These include 
semi-batch reactors~\cite{Nicholas2025},
membrane reactors~\cite{Sorrenti:2017aa},
continuous-flow stirred-tank reactors~\cite{Dev2023,Li:2024aa,Kriukov:2024aa},
microfluidic reactors~\cite{Juneja2025},
phase-separated liquids combined with
continuous injection of chemical species~\cite{Howlett:2022aa,Zhiheng2024},
and diffusion through gel matrices~\cite{Kar2024,Kar2025}.
% -
% -
% -

% -
% -
We focused here on CRNs at the microscopic scale, 
but our approach extends to CRNs at the macroscopic scale,
as the latter arise from a large-volume limit, $V \to \infty$, of the former~\cite{fala25}. 
Specifically, 
for our approach to hold, this limit must preserve the abundance separation,
which requires 
to take the $\Omega \to \infty$ limit before the $V \to \infty$ limit.
% -
% -

% -
% -
Our results open several directions for future investigation,
of which we mention only two.
% -
A first step would be to analyze the next-to-leading-order contributions 
to the dynamics and thermodynamics of closed CRNs,
thereby quantifying the corrections arising from the finite abundances
of the species that effectively act as chemostats.
% -
A second direction would be to relax the assumption of independent $\Rexc$ reactions,
which may lead to emergent chemostatted concentrations
that evolve on the slow time-scale of the $\Rint$ reactions.
% -
% -

% -
% -
By revealing how open-CRN behavior arises from closed CRNs,
our work shows that the nonequilibrium thermodynamics of open CRNs
is not merely an idealization,
but an emergent property of closed out-of-equilibrium systems.
It thereby provides a unified and physically grounded bridge between 
idealized open-CRN descriptions
and real chemical systems.
% -
% -

%%%%%%%%%%%%%%%%%%%%%%%%%%%%%%%%%%%%%%%%%%%%%%%%%%%%%%%%%%%%
%%%%%%%%%%%%%%%%%%%%%%%%%%%%%%%%%%%%%%%%%%%%%%%%%%%%%%%%%%%%
%%%%%%%%%%%%%%%%%%%%%%%%%%%%%%%%%%%%%%%%%%%%%%%%%%%%%%%%%%%%
%%%%%%%%%%%%%%%%%%%%%%%%%%%%%%%%%%%%%%%%%%%%%%%%%%%%%%%%%%%%

\section*{Acknowledgments}
BR and ME are supported by Project
INTER/ANR/25/19593353-NERD, funded by Fond National de la Recherche (FNR) Luxembourg and Agence Nationale de la Recherche (ANR) France. 
FA is supported by the project P-
DiSC\#BIRD2023-UNIPD funded by the Department of Chemical Sciences of the University of Padova (Italy).
This work was supported in part by the high-performance computing infrastructure developed under the project “CONVECS”, funded by the PR Veneto FESR 2021-2027 program, Priority 1 – Specific Objective 1.1 – Action 1.1.2.

%%%%%%%%%%%%%%%%%%%%%%%%%%%%%%%%%%%%%%%%%%%%%%%%%%%%%%%%%%%%
%%%%%%%%%%%%%%%%%%%%%%%%%%%%%%%%%%%%%%%%%%%%%%%%%%%%%%%%%%%%
%%%%%%%%%%%%%%%%%%%%%%%%%%%%%%%%%%%%%%%%%%%%%%%%%%%%%%%%%%%%
%%%%%%%%%%%%%%%%%%%%%%%%%%%%%%%%%%%%%%%%%%%%%%%%%%%%%%%%%%%%

\section*{Data Availability}

The data and code that support the findings of this study are openly available in Ref.~\cite{Remlein2026Data}.

%%%%%%%%%%%%%%%%%%%%%%%%%%%%%%%%%%%%%%%%%%%%%%%%%%%%%%%%%%%%
%%%%%%%%%%%%%%%%%%%%%%%%%%%%%%%%%%%%%%%%%%%%%%%%%%%%%%%%%%%%
%%%%%%%%%%%%%%%%%%%%%%%%%%%%%%%%%%%%%%%%%%%%%%%%%%%%%%%%%%%%
%%%%%%%%%%%%%%%%%%%%%%%%%%%%%%%%%%%%%%%%%%%%%%%%%%%%%%%%%%%%

\appendix

%%%%%%%%%%%%%%%%%%%%%%%%%%%%%%%%%%%%%%%%%%%%%%%%%%%%%%%%%%%%
%%%%%%%%%%%%%%%%%%%%%%%%%%%%%%%%%%%%%%%%%%%%%%%%%%%%%%%%%%%%
%%%%%%%%%%%%%%%%%%%%%%%%%%%%%%%%%%%%%%%%%%%%%%%%%%%%%%%%%%%%
%%%%%%%%%%%%%%%%%%%%%%%%%%%%%%%%%%%%%%%%%%%%%%%%%%%%%%%%%%%%

\section{Hierarchy of Equations Resulting from the Time-Scale Separation
\label{Appendix:CMETimeScale}}

We use here 
the time-scale separation assumption introduced in Subs.~\ref{subs:ASSUMPTION_tss} 
to rewrite the chemical master equation~\eqref{eq:CME3} 
describing the evolution of 
the probability~$p_t(\bn)$ of there being~$\bn = (\dots, n_x, \dots, n_y, \dots, n_Y, \dots) \in \mathcal{N}$
molecules at time~$t$
in closed CRNs (with species and reactions partitioned as in Sec.~\ref{sec:closedCRNs}).
% -
% -
% -
We start by
substituting  
$\mathrm d_t =
\mathrm d_\tau  
+ \epsilon \mathrm d_T$,
as well as Eqs.~\eqref{eq:Wepsilon} and~\eqref{eq:PerturbAnsatz},
into the chemical master equation~\eqref{eq:CME3},
which leads to
\begin{equation}
\begin{split}
     \big(\mathrm d_\tau&+ \epsilon \mathrm d_T\big)
     \Big(p_{\tau, T}^{(0)}(\bn) + \sum_{q=1}^\infty p_{\tau, T}^{(q)}(\bn) \epsilon^q \Big)
     = \\
     =
     &\sum_{\bnm} \Wexc(\bn,\bnm)
     \Big(p_{\tau, T}^{(0)}(\bnm) + \sum_{q=1}^\infty p_{\tau, T}^{(q)}(\bnm) \epsilon^q \Big)\\
     &+ \epsilon\, \sum_{\bnm} \Wint(\bn,\bnm)
     \Big(p_{\tau, T}^{(0)}(\bnm) + \sum_{q=1}^\infty p_{\tau, T}^{(q)}(\bnm) \epsilon^q \Big)
     \,.
\end{split}
\label{eq:CME_all_orders}
\end{equation}
By collecting terms of equal order in~$\epsilon$,
Eq.~\eqref{eq:CME_all_orders} splits into 
a hierarchy of equations of successive orders in~$\epsilon$.
% -
% -
The $\mathcal O(1)$ equation reads
\begin{equation}
    \mathrm d_\tau p_{\tau, T}^{(0)}(\bn) 
    = 
    \sum_{\bnm} \Wexc(\bn,\bnm) \, p_{\tau, T}^{(0)}(\bnm)
    \label{eq:App:CME_fast}
\end{equation}
and is given in Eq.~\eqref{eq:CME_fast} too,
while the $\mathcal O(\epsilon)$ equation reads
\begin{equation}
    \begin{split}
    \mathrm d_\tau p_{\tau, T}^{(1)}(\bn) 
    + \mathrm d_T p_{\tau, T}^{(0)}(\bn)  
    =&\sum_{\bnm}\Wexc(\bn,\bnm)p_{\tau, T}^{(1)}(\bnm)\\ 
    &+\sum_{\bnm}\Wint(\bn,\bnm)p_{\tau, T}^{(0)}(\bnm)
    \end{split}
    \label{eq:App:CMEOrder1}
\end{equation}
and is given in Eq.~\eqref{eq:CMEOrder1} too.
% -
% - 
We examine Eq.~\eqref{eq:App:CME_fast} in App.~\ref{Appendix:CMEFastTimeScale}
and Eq.~\eqref{eq:App:CMEOrder1} in App.~\ref{Appendix:CMESlowTimeScale}.

%%%%%%%%%%%%%%%%%%%%%%%%%%%%%%%%%%%%%%%%%%%%%%%%%%%%%%%%%%%%

\subsection{Fast Time Scale \label{Appendix:CMEFastTimeScale}}
As discussed in Subs.~\ref{subs:ASSUMPTION_tss}, 
Eq.~\eqref{eq:App:CME_fast} 
describes the fast dynamics arising from the $\Rexc$ reactions
which conserves the quantities $\bL^y(\bn)$ and $\bL^x(\bn)$ 
given in Eqs.~\eqref{eq:Ly} and~\eqref{eq:Lx}, respectively. 
Hence, the zero-order probability $p_{\tau, T}^{(0)}(\bn)$
can be written as in Eq.~\eqref{eq:p0_factorized} 
and Eq.~\eqref{eq:App:CME_fast} becomes
\begin{equation}
    \mathrm d_\tau  \pi_{\tau}(\bn | \bL^x, \bL^y) =
    \sum_{\bnm \in \mathcal N (\bL^x, \bL^y)} 
    \Wexc(\bn,\bnm)
    \pi_{\tau}(\bnm | \bL^x, \bL^y)
    \label{eq:App:CME_fast_2}
\end{equation}
for all $\bn \in \mathcal N (\bL^x, \bL^y)$.
Note that $\Wexc(\bn,\bnm) = 0$ if $\bn$ and $\bnm$ do not belong to the same 
stoichiometric compatibility class $\mathcal N (\bL^x, \bL^y)$.
% -
% - 
Equation~\eqref{eq:App:CME_fast_2} corresponds to the chemical master equation of 
a detailed-balanced (deficiency-zero) CRN 
where the species $\Spey$ and $\SpeY$ are interconverted via the $\Rexc$ reactions
(whose stoichiometry is specified in Eq.~\eqref{eq:CRN_ExchangeLinear}).
% -
Such chemical master equation has a well defined steady state solution 
which, according to Ref.~\cite{Anderson2010}, 
can be written as in Eq.~\eqref{eq:Binomial}
where
\begin{align}
    \piInfy(n_y|L^y) = 
    \frac{
    \frac{
    (\mathfrak c_{y})^{n_y} (\mathfrak c_{Y(y)})^{L^y - n_y}
    }{
    (n_{y})!(L^y - n_{y})! 
    }
    }{
    \sum_{n_y' = 0}^{L^y}
    \frac{
    (\mathfrak c_{y})^{n_y'}(\mathfrak c_{Y(y)})^{L^y - n_y'}
    }{
    (n_{y}')!(L^y - n_{y}')! 
    }
    }
    \,,
    \label{eq:App:SS_fast_y_1}
\end{align}
is a Poisson-like distribution with $n_y \leq L^y$ and
\begin{equation}
    \frac{\mathfrak c_{Y(y)}}{\mathfrak c_{y}} 
    = 
    \frac{k_{\rhoe(y)}}{k_{-\rhoe(y)}}
    \,.
\end{equation}
By now applying the binomial theorem to the denominator in Eq.~\eqref{eq:App:SS_fast_y_1},
namely, 
\begin{equation}
    (\mathfrak c_{y} + \mathfrak c_{Y(y)})^{L^y} = L^y !
    \sum_{n_y' = 0}^{L^y}
    \frac{
    (\mathfrak c_{y})^{n_y'}(\mathfrak c_{Y(y)})^{L^y - n_y'}
    }{
    (n_{y}')!(L^y - n_{y}')! 
    }\,
\end{equation}
the probability distribution $\piInfy(n_y|L^y)$ becomes
\begin{equation}
    \piInfy(n_y|L^y) =
    \frac{L^y !
    \bigg(\frac{
    \mathfrak c_{y}
    }{
    \mathfrak c_{y} + \mathfrak c_{Y(y)}
    }\bigg)^{n_y}
    \bigg(\frac{
    \mathfrak c_{Y(y)}
    }{
    \mathfrak c_{y} + \mathfrak c_{Y(y)}
    }\bigg)^{L^y - n_y}
    }{
    (n_{y})!(L^y - n_{y})!}
    \,,
    \label{eq:App:SS_fast_y_2}
\end{equation}
which matches the binomial distribution in Eq.~\eqref{eq:Binomialy} 
upon identifying 
\begin{equation}
    \mathfrak p_y =
    \frac{
    \mathfrak c_{y}
    }{
    \mathfrak c_{y} + \mathfrak c_{Y(y)}
    }
    \quad
    \text{ and }
    \quad
    \mathfrak p_{Y(y)} = \frac{
    \mathfrak c_{Y(y)}
    }{
    \mathfrak c_{y} + \mathfrak c_{Y(y)}
    }
    \,.
\end{equation}

%%%%%%%%%%%%%%%%%%%%%%%%%%%%%%%%%%%%%%%%%%%%%%%%%%%%%%%%%%%%
\subsection{Slow Time Scale \label{Appendix:CMESlowTimeScale}}
As discussed in Subs.~\ref{subs:IMPLICATIONS_tss},
Eq.~\eqref{eq:App:CMEOrder1} becomes Eq.~\eqref{eq:CMEOrder1_a}
by using the conservation of the quantities $\bL^y(\bn)$ and $\bL^x(\bn)$ 
given in Eqs.~\eqref{eq:Ly} and~\eqref{eq:Lx}, respectively, on the fast time scale.
Indeed, this implies that $ \sum_{\bn \in \mathcal N (\bL^x, \bL^y)} \Wexc(\bn,\bnm) = 0$.
% -
% -
Then, in the long-time limit of the fast time scale, i.e., $\tau \to \infty$,
implying that
$\lim_{\tau \to \infty} \mathrm d_\tau p_{\tau, T}^{(1)}(\bn) = 0$ 
and  
$\lim_{\tau \to \infty} p_{\tau, T}^{(0)}(\bn) = 
\piInf(\bn | \bL^x(\bn), \bL^y(\bn)) \, 
p_{T}^{(0)} (\bL^x(\bn), \bL^y(\bn))$,
Eq.~\eqref{eq:CMEOrder1_a} simplifies to
\begin{widetext}
\begin{equation}
    \mathrm d_T p_{T}^{(0)}(\bL^x, \bL^y) 
    =
    \sum_{\bn \in \mathcal N (\bL^x, \bL^y)}
    \sum_{\bnm}
    \Wint(\bn,\bnm)
    \piInf(\bnm | \bL^x(\bnm), \bL^y(\bnm)) \, 
    p_{T}^{(0)} (\bL^x(\bnm), \bL^y(\bnm))
    \,,
    \label{eq:CMEOrder1_b}
\end{equation}
\end{widetext}
by also using 
$ \sum_{\bn \in \mathcal N (\bL^x, \bL^y)} \piInf(\bn | \bL^x, \bL^y) = 1 $~\footnote{
Mathematically, the derivation of Eq.~\eqref{eq:CMEOrder1_b} is equivalent to applying the Fredholm alternative theorem to Eq.~\eqref{eq:App:CMEOrder1}
together with the long-time limit of the fast time scale, i.e., $\tau\to\infty$}.
% -
% -

% -
% -
We proceed now 
i) by using the explicit expression of the stochastic generator $\Wint(\bn,\bnm)$
given in Eq.~\eqref{eq:Wint}
and
ii) by recognizing that 
$\bL^x(\bn - \colSi) = \bL^x(\bn) - \colSxi$
and 
$\bL^y(\bn - \colSi) = \bL^y(\bn) - \colSyi$.
% -
Indeed,
according to the definitions in Eqs.~\eqref{eq:Lx} and~\eqref{eq:Ly}
together with the stoichiometry of the $\Rint$ reactions given in Subs.~\ref{sec:closedCRNs},
we have that 
$L^x(\bn - \colSi) = n_x - S_{x,\rhoi} = L^x(\bn) - S_{x,\rhoi}$
and 
$L^y(\bn - \colSi) =  n_y + n_{Y(y)} - S_{y ,\rhoi} = L^y(\bn) - S_{y ,\rhoi}$
(since $S_{Y ,\rhoi} = 0$ $\forall Y \in \SpeY$).
% -
We can thus rewrite Eq.~\eqref{eq:CMEOrder1_b} as
\begin{widetext}
\begin{equation}
\begin{split}
    \mathrm d_T p_{T}^{(0)}(\bL^x, \bL^y)
    =& \sum_{\bn \in \mathcal N (\bL^x, \bL^y)}
    \sum_{\rhoi} \bigg\{ 
    \hat \omega_{\rhoi}(\bn - \colSi) \, 
    \piInf(\bn - \colSi | \bL^x(\bn) - \colSxi, \bL^y(\bn) - \colSyi) \,
    p_{T}^{(0)} (\bL^x(\bn) - \colSxi, \bL^y(\bn) - \colSyi)
    \bigg\}\\
    &-\sum_{\bn \in \mathcal N (\bL^x, \bL^y)}
    \sum_{\rhoi} \bigg\{ \hat \omega_{\rhoi}(\bn) \, 
    \piInf(\bn | \bL^x(\bn), \bL^y(\bn)) \, 
    p_{T}^{(0)} (\bL^x(\bn), \bL^y(\bn))
    \bigg\}\,.
\end{split}
\label{eq:CMEOrder1_c}
\end{equation}
\end{widetext}
% -
% -
% -
% -
In the second line of Eq.~\eqref{eq:CMEOrder1_c}, for every $\rhoi$,
the sum $\sum_{\bn \in \mathcal N (\bL^x, \bL^y)}$ performs 
an average of the reaction rate $\hat \omega_{\rhoi}(\bn)$ over the fast dynamics,
weighted by the steady state distribution $\piInf(\bn | \bL^x, \bL^y)$.
Namely,
\begin{equation}\small
    \sum_{\bn \in \mathcal N (\bL^x, \bL^y)}
    \hat \omega_{\rhoi} (\bn)
    \,
    \piInf(\bn | \bL^x, \bL^y)
    = \avgInf{\hat \omega_{\rhoi} | \bL^x, \bL^y}
    \,.
    \label{eq:App:avg_rr_tss}
\end{equation}
Recall that $\bL^x(\bn)$ and $\bL^y(\bn)$ are constant and equal to $\bL^x$ and $\bL^y$, respectively,
for any $\bn \in \mathcal N (\bL^x, \bL^y)$.
% -
% -
% -
% -
In the first line of Eq.~\eqref{eq:CMEOrder1_c}, for every $\rhoi$,
the sum $\sum_{\bn \in \mathcal N (\bL^x, \bL^y)}$ performs an analogous average.
% - 
% -
This can be recognized by noting that,
if $\bn \in \mathcal N (\bL^x, \bL^y)$
then $(\bn -\colSi) \in \mathcal N (\bL^x - \colSxi, \bL^y - \colSxi)$.
Hence,
by applying the change of variable $\bn -\colSi \to \bn$ in the first line of Eq.~\eqref{eq:CMEOrder1_c},
the sum $\sum_{\bn \in \mathcal N (\bL^x, \bL^y)}$ can be rewritten as
the sum $\sum_{\bn \in \mathcal N (\bL^x - \colSxi, \bL^y -\colSyi)}$,
which performs 
the average of the reaction rate $\hat \omega_{\rhoi}(\bn)$ over the fast dynamics,
weighted by the steady state distribution $\piInf(\bn | \bL^x - \colSxi, \bL^y - \colSyi)$.
% -
We denote this average by $\avgInf{\hat \omega_{\rhoi} | \bL^x - \colSxi, \bL^y -\colSyi}$,
which is given in Eq.~\eqref{eq:App:avg_rr_tss} upon replacing 
$\bL^x$ with $\bL^x - \colSxi$
and
$\bL^y$ with $\bL^y -\colSyi$.
% -
% -

% -
% -
Let us now derive the explicit expression of $\avgInf{\hat \omega_{\rhoi} | \bL^x, \bL^y}$.
By using the steady state distribution of the fast dynamics 
(given in Eq.~\eqref{eq:Binomial})
in Eq.~\eqref{eq:App:avg_rr_tss},
we obtain 
\begin{equation}
\begin{split}
    &\avgInf{\hat \omega_{\rhoi} | \bL^x, \bL^y}
    =
    \hat{\kappa}_{\rhoi} V 
    \prod_x
    \frac{L^x ! \, \theta(L^x -\nu_{x,\rhoi})}
    {(L^x - \nu_{x,\rhoi})! \, V^{\nu_{x,\rhoi}}} \\
    &\times\prod_y 
    \sum_{n_y = 0}^{L^y}
    \frac{n_y ! \, \theta(n_y -\nu_{y,\rhoi})}
    {(n_y - \nu_{y,\rhoi})! \, V^{\nu_{y,\rhoi}}} 
    \piInfy(n_y|L^y)
    \,.
\end{split}
\label{eq:App:avg_rr_tss_a}
\end{equation}
Each factor of the product in the second line of Eq.~\eqref{eq:App:avg_rr_tss_a} 
can be rewritten as follows:
\begin{equation} \small
\begin{split}
    &\sum_{n_y = 0}^{L^y}
    \frac{n_y ! \, \theta(n^y -\nu_{y,\rhoi})}
    {(n_y - \nu_{y,\rhoi})! \, V^{\nu_{y,\rhoi}}} 
    \piInfy(n_y|L^y) = \\
    % - - - - - - - - - - - - - - - - - - - - - - - - - %
    % - - - - - - - - - - - - - - - - - - - - - - - - - %
    % - - - - - - - - - - - - - - - - - - - - - - - - - %
    &=\sum_{n_y = 0}^{L^y}
    \frac{n_y ! \, \theta(n^y -\nu_{y,\rhoi})}
    {(n_y - \nu_{y,\rhoi})! \, V^{\nu_{y,\rhoi}}} 
    % - - - - % 
    \frac{L^y!
    \big(\mathfrak p_y\big)^{n_y}
    \big(\mathfrak p_{Y(y)}\big)^{L^y - n_y}}
    {n_y! (L^y - n_y)!}\\
    % - - - - - - - - - - - - - - - - - - - - - - - - - %
    % - - - - - - - - - - - - - - - - - - - - - - - - - %
    % - - - - - - - - - - - - - - - - - - - - - - - - - %
    &=\sum_{n_y = \nu_{y,\rhoi}}^{L^y}
    \frac{L^y!\, 
    \theta(L^y -\nu_{y,\rhoi})}
    {(n_y - \nu_{y,\rhoi})! \, V^{\nu_{y,\rhoi}}} 
    % - - - - % 
    \frac{
    \big(\mathfrak p_y\big)^{n_y}
    \big(\mathfrak p_{Y(y)}\big)^{L^y - n_y}}
    {(L^y - n_y)!}\\
    % - - - - - - - - - - - - - - - - - - - - - - - - - %
    % - - - - - - - - - - - - - - - - - - - - - - - - - %
    % - - - - - - - - - - - - - - - - - - - - - - - - - %
    &=\sum_{m_y = 0 }^{L^y - \nu_{y,\rhoi}}
    \frac{L^y!\, \theta(L^y -\nu_{y,\rhoi})}
    {m_y! \, V^{\nu_{y,\rhoi}}} 
    % - - - - % 
    \frac{
    \big(\mathfrak p_y\big)^{m_y + \nu_{y,\rhoi}}
    \big(\mathfrak p_{Y(y)}\big)^{(L^y - \nu_{y,\rhoi}) - m_y}}
    {((L^y - \nu_{y,\rhoi}) - m_y)!}
    \,,
\end{split} 
\end{equation}
where we used 
i) the distribution $\piInfy(n_y|L^y)$ given  in Eq.~\eqref{eq:Binomialy} in the first equality,
ii) the properties of the Heaviside step function requiring that 
$n_y \geq \nu_{y,\rhoi}$ as well as $L^y \geq \nu_{y,\rhoi}$
in the second equality,
and 
iii)
the change of variables $ n_y - \nu_{y,\rhoi} \to m_y$ in the third equality.
% -
% -
% -
% -
Because of the binomial theorem
and Eq.~\eqref{eq:py_prop},
we have that 
\begin{equation}\small
\begin{split}
    &\sum_{m_y = 0 }^{L^y - \nu_{y,\rhoi}} 
    \frac{ (L^y - \nu_{y,\rhoi}) !
    \big(\mathfrak p_y\big)^{m_y}
    \big(\mathfrak p_{Y(y)}\big)^{(L^y - \nu_{y,\rhoi}) - m_y}}
    {m_y!((L^y - \nu_{y,\rhoi}) - m_y)!} =\\
    % - - - - - - - - - - - - - - - - - - - - - - - - - %
    % - - - - - - - - - - - - - - - - - - - - - - - - - %
    % - - - - - - - - - - - - - - - - - - - - - - - - - %
    & = \big(\mathfrak p_y + \mathfrak p_{Y(y)}\big)^{L^y - \nu_{y,\rhoi}}
    = 1
    \,,
\end{split}
\end{equation}
implying that
\begin{equation}
\begin{split}
    &\sum_{n_y = 0}^{L^y}
    \frac{n_y ! \, \theta(n^y -\nu_{y,\rhoi})}
    {(n_y - \nu_{y,\rhoi})! \, V^{\nu_{y,\rhoi}}} 
    \piInfy(n_y|L^y) = \\
    % - - - - - - - - - - - - - - - - - - - - - - - - - %
    % - - - - - - - - - - - - - - - - - - - - - - - - - %
    % - - - - - - - - - - - - - - - - - - - - - - - - - %
    &= \frac{L^y!\, \theta(L^y -\nu_{y,\rhoi})}
    {(L^y - \nu_{y,\rhoi}) ! \, V^{\nu_{y,\rhoi}}}
    (\mathfrak p_y)^{\nu_{y,\rhoi}}
    \,.
\end{split}
\label{eq:App:avg_rr_tss_d}
\end{equation}
By plugging Eq.~\eqref{eq:App:avg_rr_tss_d} into Eq.~\eqref{eq:App:avg_rr_tss_a},
we obtain the expression of average reaction rate $\avgInf{\hat \omega_{\rhoi} | \bL^x, \bL^y}$
given in Eq.~\eqref{eq:avg_rr_tss},
namely,
\begin{equation}\small
\begin{split}
    \avgInf{\hat \omega_{\rhoi} | \bL^x, \bL^y}
    =
    \hat{\kappa}_{\rhoi} V 
    &\prod_x
    \frac{L^x ! \, \theta(L^x -\nu_{x,\rhoi})}
    {(L^x - \nu_{x,\rhoi})! \, V^{\nu_{x,\rhoi}}}  \\
    \times&\prod_y
    \frac{L^y ! \, \theta(L^y -\nu_{y,\rhoi})}
    {(L^y - \nu_{y,\rhoi})! \, V^{\nu_{y,\rhoi}}}
    (\mathfrak p_y)^{\nu_{y,\rhoi}} 
    \,.
\end{split}
\end{equation}

%%%%%%%%%%%%%%%%%%%%%%%%%%%%%%%%%%%%%%%%%%%%%%%%%%%%%%%%%%%%
%%%%%%%%%%%%%%%%%%%%%%%%%%%%%%%%%%%%%%%%%%%%%%%%%%%%%%%%%%%%
%%%%%%%%%%%%%%%%%%%%%%%%%%%%%%%%%%%%%%%%%%%%%%%%%%%%%%%%%%%%

\section{Reservoirs \label{Appendix:IdealReservoirs}}
We identify the minimal conditions under which a system acts
as a thermal reservoir, i.e., a thermostat (App.~\ref{Appendix:IdealReservoirs_T}), 
or as a molecule reservoir, i.e., a chemostat (App.~\ref{Appendix:IdealReservoirs_C}).

%%%%%%%%%%%%%%%%%%%%%%%%%%%%%%%%%%%%%%%%%%%%%%%%%%%%%%%%%%%%
\subsection{Thermostats\label{Appendix:IdealReservoirs_T}}
Consider an isolated supersystem composed of two (weakly interacting) systems, 
labeled~${\sys}$ and~${\res}$, 
with constant volumes.
% -
The two systems exchange energy,
while the total energy of the supersystem $E \equiv E_{\sys} + E_{\res}$ 
remains conserved, 
with~$E_{\sys}$ and~$E_{\res}$ being the energies of system ${\sys}$ and ${\res}$, 
respectively. 
% -
The two systems are closed and do not exchange molecules:
both~$n_{\sys}$ and~$n_{\res}$ are conserved 
with~$n_{\sys}$ and~$n_{\res}$ being the numbers of molecules of system ${\sys}$ and ${\res}$, 
respectively. 
% -
% -
The system ${\res}$ acts as a thermostat if
it is always at equilibrium with a well defined temperature~$T_{\res}$
which does not change upon energy exchanges with system~${\sys}$.
% -
In thermodynamic terms, this requires that
the inverse heat capacity of system~${\res}$ vanishes,
\begin{equation}
    \frac{1}{C_{T, \res}} = \frac{\partial T_{\res}}{\partial E_{\res}} = 0 
    \,,
    \label{eq:diverging_C_T}
\end{equation}
or, equivalently, that the heat capacity of system~$\res$ diverges, i.e., $C_{T, \res} \to \infty$.
% -
% -

% -
% -
Under these conditions,
the probability $p(E_{\sys})$ that
system $\sys$ has energy $E_{\sys}$ is given by the canonical distribution:
$p(E_{\sys}) \propto \exp(-E_{\sys}/\kb\, T_{\res})$
(with $\kb$ being the Boltzmann constant).
% -
Indeed, 
according to the microcanonical distribution, 
$p(E_{\sys}) \propto \exp(S_{\res}(E - E_{\sys}) / \kb)$
where 
$S_{\res}(E - E_{\sys})$ is the entropy of system $\res$ when its energy is $E_{\res} = E - E_{\sys}$,
reading 
\begin{equation}
    S_{\res}(E - E_{\sys}) 
    = \int_{0}^{E - E_{\sys}}\frac{\mathrm dE_{\res}}{T_{\res}} 
    = \frac{E - E_{\sys}}{T_{\res}} 
    \,.
\end{equation}
Here, 
the first equality results from the fundamental thermodynamic relation 
for closed systems at constant volume,
while the second equality results from Eq.~\eqref{eq:diverging_C_T},
ensuring that $T_{\res}$ is independent of $E_{\res}$.
% -
% -

% -
% -
\remark
If system~$\res$ were an ideal gas,
its energy would be given by $E_{\res} = (f/2) \kb n_{\res} T_{\res}$
(with $f$ being the number of degrees of freedom of a single molecule)
and the condition in Eq.~\eqref{eq:diverging_C_T} would specialize to 
\begin{equation}
    \frac{1}{C_{T, \res}} = \frac{1}{(f/2)\kb n_{\res}} = 0
    \,.
\end{equation}
% -
This physically means that system~$\res$ must have   
a diverging number of molecules, i.e., $n_{\res} \to \infty$,
for its temperature to be unaffected by the energy exchanges
and thus acts as a thermostat.

%%%%%%%%%%%%%%%%%%%%%%%%%%%%%%%%%%%%%%%%%%%%%%%%%%%%%%%%%%%%
\subsection{Chemostats\label{Appendix:IdealReservoirs_C}}

Consider an isolated supersystem composed of a single molecular species,
partitioned into two (weakly interacting) systems, 
labeled~${\sys}$ and~${\res}$, 
with constant volumes.
% -
The two systems exchange energy,
while the total energy of the supersystem $E \equiv E_{\sys} + E_{\res}$ 
remains conserved, 
with~$E_{\sys}$ and~$E_{\res}$ being the energies of system ${\sys}$ and ${\res}$, 
respectively. 
% -
The two systems exchange molecules too,
while the total number of molecules of the supersystem $n \equiv n_{\sys} + n_{\res}$
remains conserved,
with~$n_{\sys}$ and~$n_{\res}$ being the numbers of molecules of system ${\sys}$ and ${\res}$, 
respectively. 
% -
% -
The system ${\res}$ acts as a chemostats 
if, in addition to acting as a thermostat (see App.~\ref{Appendix:IdealReservoirs_T}), 
it is always at equilibrium with a well defined chemical potential~$\mu_{\res}/T_{\res}$
which does not change upon exchanges of molecules with system~${\sys}$.
% -
In thermodynamic terms, this requires that
what we name here inverse chemical capacity of system~${\res}$ vanishes,
\begin{equation}
    \frac{1}{C_{\mu, \res}} = \frac{\partial (\mu_{\res}/T_{\res})}{\partial n_{\res}} = 0 
    \,,
    \label{eq:diverging_C_N}
\end{equation}
or, equivalently, that the chemical capacity of system~$\res$ diverges, i.e., $C_{\mu, \res} \to \infty$.
% -
% -

% -
% -
Under these conditions,
the probability $p(E_{\sys}, n_{\sys})$ that
system $\sys$ has energy $E_{\sys}$ and number of molecules $n_{\sys}$ 
is given by the grandcanonical distribution:
$p(E_{\sys}, n_{\sys}) \propto \exp(-(E_{\sys} -\mu_{\res} n_{\sys})/\kb\, T_{\res})$
(with $\kb$ being the Boltzmann constant).
% -
Indeed, 
according to the microcanonical distribution, 
$p(E_{\sys}, n_{\sys}) \propto \exp(S_{\res}(E - E_{\sys}, n - n_{\sys}) / \kb)$
where 
$S_{\res}(E - E_{\sys}, n - n_{\sys})$ is the entropy of system $\res$ when 
its energy is $E_{\res} = E - E_{\sys}$
and its number of molecules is $n_{\res} = n - n_{\sys}$,
reading 
\begin{align}
\begin{split}
    S_{\res}(E - E_{\sys},n - n_{\sys} ) 
    &= \int_{0}^{E - E_{\sys}}\frac{\mathrm dE_{\res}}{T_{\res}} 
    - \int_{0}^{n - n_{\sys}}\frac{\mu_{\res}}{T_{\res}} \mathrm dn_{\res} \\
    &= \frac{E - E_{\sys}}{T_{\res}} - \frac{\mu_{\res}}{T_{\res}}(n-n_{\sys})
    \,.
\end{split}
\end{align}
Here, 
the first equality results from the fundamental thermodynamic relation 
for systems at constant volume,
while the second equality results from Eq.~\eqref{eq:diverging_C_T} and Eq.~\eqref{eq:diverging_C_N},
ensuring that 
$T_{\res}$ is independent of $E_{\res}$
and ${\mu_{\res}}/{T_{\res}}$ is independent of $n_{\res}$, 
respectively.
% -
% -

% -
% -
\remark
If system~$\res$ were an ideal system
with a constant temperature,
its chemical potential would be given by Eq.~\eqref{eq:chem_pot}
and the condition in Eq.~\eqref{eq:diverging_C_N} specializes to 
\begin{equation}
    \frac{1}{C_{\mu, \res}} = \frac{\kb}{n_{\res}} = 0
    \,.
    \label{eq:cc_ideal}
\end{equation}
% -
This physically means that system~$\res$ must have   
a diverging number of molecules, i.e., $n_{\res} \to \infty$,
for its chemical potential to be unaffected by the molecule exchanges
and thus acts as a chemostat.
% -
% -
If, on the contrary, the number of molecules $n_{\res}$ is large but finite,
the chemical potential of system~$\res$ changes according to
\begin{equation}
    \mathrm{d}\bigg(\frac{\mu_{\res}}{T_{\res}}\bigg)
    = \frac{\mathrm{d}n_{\res}}{C_{\mu, \res}}
    \label{eq:cp_ideal_change_app}
\end{equation}
upon molecule exchanges with system~${\sys}$.
Assuming that the rate of these exchanges $\mathrm{d}_T n_{\res}$ is $\mathcal O(1)$,
Eq.~\eqref{eq:cp_ideal_change_app}, 
together with ${C_{\mu, \res}} = \mathcal O(n_{\res})$ from Eq.~\eqref{eq:cc_ideal}, 
implies that the characteristic time required for
$\mu_{\res}/T_{\res}$ to undergo an $\mathcal O(1)$ change is
\begin{equation}
    T = \mathcal O(n_{\res})
    \,.
\end{equation}
% -
Thus, on time scales of order $\mathcal O(n_{\res})$ or longer,
system~$\res$ can no longer act as a chemostat.
% -
% -

%%%%%%%%%%%%%%%%%%%%%%%%%%%%%%%%%%%%%%%%%%%%%%%%%%%%%%%%%%%%
%%%%%%%%%%%%%%%%%%%%%%%%%%%%%%%%%%%%%%%%%%%%%%%%%%%%%%%%%%%%
%%%%%%%%%%%%%%%%%%%%%%%%%%%%%%%%%%%%%%%%%%%%%%%%%%%%%%%%%%%%

\section{Large Deviation Principle \label{App:LargeDeviation}}

We show here that a large deviation principle emerges from the abundance separation 
and that, consequently, 
the master equation~\eqref{eq:CME_post_tss} boils down to the master equation~\eqref{eq:emeCME}.
% -
Specifically, we consider the limit $\Omega \to \infty$ 
which corresponds to diverging total numbers of molecules of the $\Spey$ and $\SpeY$ species 
(i.e., $\bL^y = \mathcal{O}(\Omega)$),
while the average numbers of molecules of the $\Spey$ species remain finite
(i.e., $\bL^y \mathfrak p_y = \mathcal{O}(1)$).
% -
% -

% -
% -
We start by rewriting the master equation~\eqref{eq:CME_post_tss}
with the rates in Eq.~\eqref{eq:rr_dcc0} 
plus contributions $\mathcal{O}(\Omega^{-1})$
thus obtaining 
\begin{widetext}
    \begin{equation}
    \mathrm d_T p_{T}^{(0)} (\bL^x, \bL^y) =  
    \sum_{\rhoi} 
    \big\{
    \big(\hat R_{\rhoi}(\bL^x- \colSxi, \bL^y) + \mathcal{O}(\Omega^{-1}) \big) \, 
    p_{T}^{(0)} (\bL^x - \colS_{x,\rhoi}, \bL^y - \colS_{y,\rhoi})  
    -  
    \big(\hat R_{\rhoi}(\bL^x, \bL^y) + \mathcal{O}(\Omega^{-1}) \big) \,  
    p_{T}^{(0)} (\bL^x, \bL^y)
    \big\}
    \,.
    \label{eq:rr_dcc00app}
    \end{equation}
\end{widetext}
% -
% -

% -
% -
We then write the probability $p_{T}^{(0)} (\bL^x, \bL^y)$ as
\begin{equation}
    p_{T}^{(0)} (\bL^x, \bL^y) =
    p_{T}^{(0)} (\bL^y | \bL^x)
    p_{T}^{(0)} (\bL^x)
    \label{eq:conditional_LxLy}
\end{equation}
and use the change of variables from $\bL^y$ to $\boldsymbol \eta^y $
(with $\boldsymbol \eta^y  = (\dots, \eta^y, \dots)$ 
and $\eta^y \equiv L^y/ \Omega = \mathcal{O} (1)$ 
as $\Omega \to \infty$)
to define
\begin{equation}\small
    P^{(\Omega)}_{T} (\boldsymbol \eta^y | \bL^x)
    P^{(\Omega)}_{T} (\bL^x)
    \equiv
    \Omega^{N_y}
    p_{T}^{(0)} (\Omega \boldsymbol \eta^y | \bL^x)
    p_{T}^{(0)} (\bL^x)
    \,,
    \label{eq:def_probOmega}
\end{equation}
where $N_y$ denotes the number of $\Spey$ species 
and the superscript $\Omega$ highlights that the probabilities 
on the left-hand side of Eq.~\eqref{eq:def_probOmega} depend on the large parameter $\Omega$.
% -
% -
In the $\Omega \to \infty$ limit,
we use the Wentzel–Kramers–Brillouin (WKB) ansatz~\cite{risken,touc09,bres14,touc18,qian21,fala25},
implementing a large deviation principle, 
for the conditional probability distribution
$P^{(\Omega)}_{T} (\boldsymbol \eta^y | \bL^x)$:
\begin{equation}
	P^{(\Omega)}_{T} (\boldsymbol \eta^y | \bL^x) 
    = 
    C_\Omega 
    e^{ -\Omega I_T(\boldsymbol \eta^y|\bL^x)}
    \sum_{q=0}^\infty  U^{(q)}_T(\boldsymbol \eta^y|\bL^x) \Omega^{-q}
    \,,
    \label{eq:WKB}
\end{equation}
where 
$C_\Omega$ is the normalization constant,
$I_T(\boldsymbol \eta^y|\bL^x)$ is the zero-order rate function,
and each $U^{(q)}_T(\boldsymbol \eta^y|\bL^x)$ is 
a subexponential contribution of order $\Omega^{-q}$.
% -
Furthermore, 
we use a multiscale expansion
for the marginalized distribution 
$P^{(\Omega)}_{T} (\bL^x)$
in powers of $\Omega^{-1}$:
\begin{equation}
	P^{(\Omega)}_{T} (\bL^x) 
    =
    P_{T} (\bL^x) 
    +
    \sum_{q=1}^\infty P^{(q)}_{T} (\bL^x)\Omega^{-q}
    \,.
    \label{eq:PerturbationAnsatz}
\end{equation}  
% -
% -
By using Eqs.~\eqref{eq:WKB} and~\eqref{eq:PerturbationAnsatz},
the master equation~\eqref{eq:rr_dcc00app}
can be re-formulated in terms of a hierarchy of equations of successive orders in~$\Omega$.
The leading-order equation reads
\begin{equation}
    \mathrm d_T I_T(\boldsymbol \eta^y|\bL^x) 
    = 0 
    \,,
    \label{eq:0ratefinction}
\end{equation}
meaning that the zero-order rate function $I_T(\boldsymbol \eta^y|\bL^x)$ is time independent. 
% -
% -

% -
% -
Finally, we assume that the zero-order rate function $I_T(\boldsymbol \eta^y|\bL^x)$
has a unique minimum denoted $\boldsymbol \eta^y_{\ast}$.
% -
According to Eq.~\eqref{eq:0ratefinction},
$\boldsymbol \eta^y_{\ast}$ is time independent and 
can be set to be $\bL^x$-independent too by the initial condition. 
% -
% -
This, together with the WKB ansatz in Eq.~\eqref{eq:WKB}, implies that
the probability $P^{(\Omega)}_{T} (\boldsymbol \eta^y | \bL^x)$
is sharply peaked around $\boldsymbol \eta^y_{\ast}$ in the limit $\Omega \to \infty$.
% -
Hence, any average of an arbitrary function $f(\bL^x,\boldsymbol \eta^y )$
over the variables $\boldsymbol \eta^y$ 
can be expressed according to 
the Laplace's saddle-point approximation, i.e.,
\begin{equation}\small
    \int\mathrm d\boldsymbol \eta^y\, 
    f(\bL^x,\boldsymbol \eta^y ) 
    P^{(\Omega)}_{T} (\boldsymbol \eta^y | \bL^x)
    =
    f(\bL^x,\boldsymbol \eta^y_{\ast} ) 
    + 
    \mathcal{O}(\Omega^{-1})
    \,,
    \label{eq:LaplaceApprox}
\end{equation}
% -
which, when applied to the master equation~\eqref{eq:rr_dcc00app}
together with the multiscale expansion of $P^{(\Omega)}_{T} (\bL^x)$ in Eq.~\eqref{eq:PerturbationAnsatz},
leads to
\begin{equation}
\begin{split}
    \mathrm d_T P_{T} (\bL^x)  
    =  
    \sum_{\rhoi} 
    \big\{&
    \hat R_{\rhoi}(\bL^x- \colSxi, \Omega\boldsymbol \eta^y_{\ast})   \, 
    P_{T}(\bL^x - \colS_{x,\rhoi})  \\
    &-  
    \hat R_{\rhoi}(\bL^x,  \Omega\boldsymbol \eta^y_{\ast}) \,  
    P_{T} (\bL^x) 
    \big\}
    \,.
\end{split}
    \label{eq:rr_dcc1app}
\end{equation}
% -
% -
Crucially, 
Eq.~\eqref{eq:rr_dcc1app} is equivalent to Eq.~\eqref{eq:emeCME}
upon identifying
\begin{subequations}
\begin{align}
    \hat r_{\rhoi}(\bL^x|[\boldsymbol y]) &= \hat R_{\rhoi}(\bL^x, \Omega\boldsymbol \eta^y_{\ast})
    \,,
    \\
    \hat r_{\rhoi}(\bL^x - \colS_{x,\rhoi}|[\boldsymbol y]) &= \hat R_{\rhoi}(\bL^x - \colS_{x,\rhoi}, \Omega\boldsymbol \eta^y_{\ast})
    \,,
\end{align}    
\end{subequations}
with $[\boldsymbol y]\equiv {\eta^y_{\ast} \Omega \, \mathfrak p_y}/{V}$ given in Eq.~\eqref{eq:y_conc}.
% -
% -

% -
% -
\mremark
Consider the following average of an arbitrary function $f(\bL^x,\boldsymbol \eta^y )$:
\begin{equation}\small
    \int\mathrm d\boldsymbol \eta^y\
    f(\bL^x, \boldsymbol \eta^y)\,
    P^{(\Omega)}_{T} \bigg(\boldsymbol \eta^y - \frac{\colS_{y,\rhoi}}{\Omega} 
    \bigg| 
    \bL^x - \colS_{x,\rhoi}\bigg)
    \,.
    \label{eq:Integral_f_app_remakr1}
\end{equation}
By using the change of variables from $\boldsymbol \eta^y$ to 
$\tilde{\boldsymbol \eta}^y = \boldsymbol \eta^y - {\colS_{y,\rhoi}}/{\Omega}$,
the integral in Eq.~\eqref{eq:Integral_f_app_remakr1} becomes
\begin{equation}\small
    \int\mathrm d\tilde{\boldsymbol \eta}^y\
    f\bigg(\bL^x,\tilde{\boldsymbol \eta}^y + \frac{\colS_{y,\rhoi}}{\Omega}\bigg)\,
    P^{(\Omega)}_{T} (\tilde{\boldsymbol \eta}^y | \bL^x - \colS_{x,\rhoi})
    \,,
    \label{eq:Integral_f_app_remakr2}
\end{equation}
which can be rewritten as 
\begin{equation}\small
    \int\mathrm d\tilde{\boldsymbol \eta}^y\
    f(\bL^x,\tilde{\boldsymbol \eta}^y )\,
    P^{(\Omega)}_{T} (\tilde{\boldsymbol \eta}^y | \bL^x - \colS_{x,\rhoi})
    + 
    \mathcal{O}(\Omega^{-1})
    \,.
    \label{eq:Integral_f_app_remakr3}
\end{equation}
The integral in Eq.~\eqref{eq:Integral_f_app_remakr3} can now be evaluated, to leading order, 
using the Laplace's saddle-point approximation~\eqref{eq:LaplaceApprox}
and the fact that $\boldsymbol \eta^y_{\ast}$ can be set to be $\bL^x$-independent.
% -
Thus,
\begin{equation}\small
    \int\mathrm d\boldsymbol \eta^y\
    f(\bL^x, \boldsymbol \eta^y)\,
    P^{(\Omega)}_{T} \bigg(\boldsymbol \eta^y - \frac{\colS_{y,\rhoi}}{\Omega} 
    \bigg| 
    \bL^x - \colS_{x,\rhoi}\bigg)
    \approx
    f(\bL^x, \boldsymbol \eta^y_{\ast})
    \,,
    \label{eq:Integral_f_app_remakr4}
\end{equation}
where all neglected terms are of order $\Omega^{-1}$.
% -
% -
% -
% -
Note that this procedure has been implicitly used when averaging the master equation~\eqref{eq:rr_dcc00app}
over the variables $\boldsymbol \eta^y$.
Indeed,
the term
$\hat R_{\rhoi}(\bL^x- \colSxi, \bL^y)\,
p_{T}^{(0)} (\bL^x - \colS_{x,\rhoi}, \bL^y - \colS_{y,\rhoi}) $
contributes as
\begin{equation}\small
    \int\mathrm d\boldsymbol \eta^y\
    \hat R_{\rhoi}(\bL^x- \colSxi, \boldsymbol \eta^y \Omega)\,
    P^{(\Omega)}_{T} \bigg(\boldsymbol \eta^y - \frac{\colS_{y,\rhoi}}{\Omega} 
    \bigg| 
    \bL^x - \colS_{x,\rhoi}\bigg)
    \,,
    \label{eq:Integral_rate_app1}
\end{equation}
which, under the same change of variables from $\boldsymbol \eta^y$ to 
$\tilde{\boldsymbol \eta}^y = \boldsymbol \eta^y - {\colS_{y,\rhoi}}/{\Omega}$,
becomes
\begin{equation}\small
    \int\mathrm d\tilde{\boldsymbol \eta}^y\
    \hat R_{\rhoi}(\bL^x- \colSxi,\tilde{\boldsymbol \eta}^y \Omega + \colS_{y,\rhoi})\,
    P^{(\Omega)}_{T} (\tilde{\boldsymbol \eta}^y | \bL^x - \colS_{x,\rhoi})
    \,.
    \label{eq:Integral_rate_app2}
\end{equation}
% -
% -
By using Eq.~\eqref{eq:rr_dcc1} and the fact that $\mathfrak p_y = \mathcal O(\Omega^{-1})$,
we recognize that the integral in Eq.~\eqref{eq:Integral_rate_app2}
can be rewritten as
\begin{equation}\small
    \int\mathrm d\tilde{\boldsymbol \eta}^y\
    \hat R_{\rhoi}(\bL^x- \colSxi,\tilde{\boldsymbol \eta}^y \Omega)\,
    P^{(\Omega)}_{T} (\tilde{\boldsymbol \eta}^y | \bL^x - \colS_{x,\rhoi})
    + 
    \mathcal{O}(\Omega^{-1})
    \,,
    \label{eq:Integral_rate_app3}
\end{equation}
in the same way as the integral in Eq.~\eqref{eq:Integral_f_app_remakr2} was rewritten as 
in Eq.~\eqref{eq:Integral_f_app_remakr3}.
% -
Finally, 
the integral in Eq.~\eqref{eq:Integral_rate_app3} can be evaluated, to leading order, 
using the Laplace's saddle-point approximation~\eqref{eq:LaplaceApprox}
and the fact that $\boldsymbol \eta^y_{\ast}$ can be set to be $\bL^x$-independent.
% -
% -

% -
% -

%%%%%%%%%%%%%%%%%%%%%%%%%%%%%%%%%%%%%%%%%%%%%%%%%%%%%%%%%%%%
%%%%%%%%%%%%%%%%%%%%%%%%%%%%%%%%%%%%%%%%%%%%%%%%%%%%%%%%%%%%
%%%%%%%%%%%%%%%%%%%%%%%%%%%%%%%%%%%%%%%%%%%%%%%%%%%%%%%%%%%%

\section{Leading Orders of the Average Entropy Production Rate\label{App:EPR}}
We now examine the average entropy production rate~\eqref{eq:EPR_0}
and determine the leading-order contributions to $\epre_{\tau, T}$ and $\epri_{\tau, T}$
that arise from the time-scale-separation assumption (Subs.~\ref{subs:ASSUMPTION_tss})
together with the abundance-separation assumption (Subs.~\ref{subs:ASSUMPTION_as}).

%%%%%%%%%%%%%%%%%%%%%%%%%%%%%%%%%%%%%%%%%%%%%%%%%%%%%%%%%%%%

\subsection{Contributions due to the $\Rexc$ Reactions \label{App:EPRe}}
We determine here the zero-order and first-order contributions to $\epre_{\tau, T}$
by using Eq.~\eqref{eq:PerturbAnsatz} in Eq.~\eqref{eq:EPRe_def}.
In particular, we show that both contributions vanish in the long time limit of the fast time scale 
independently of the large abundance-assumption. 
% -
% -

% -
% -
The zero-order contribution reads
\begin{equation}\small
    \eprez_{\tau,T} = 
    \sum_{\rhoe, \bn} 
    \hat\omega_{\rhoe}(\bn) p_{\tau, T}^{(0)}(\bn) 
    \ln \frac
    {\hat\omega_{\rhoe}(\bn) p_{\tau, T}^{(0)}(\bn) }
    {\hat\omega_{-\rhoe}(\bn + \colSe)p_{\tau, T}^{(0)}(\bn + \colSe) } 
    \,
\end{equation}
where $p_{\tau, T}^{(0)}(\bn)$ and $p_{\tau, T}^{(0)}(\bn + \colSe)$ can be written, 
according to Eq.~\eqref{eq:p0_factorized},
as
\begin{subequations}
\begin{align}
    p_{\tau, T}^{(0)}(\bn) & = 
    \pi_{\tau}^{(0)}(\bn | \bL^x, \bL^y) \, 
    p_{T}^{(0)} (\bL^x, \bL^y) 
    \,,
    \\
    p_{\tau, T}^{(0)}(\bn + \colSe) & = 
    \pi_{\tau}^{(0)}(\bn + \colSe | \bL^x, \bL^y) \, 
    p_{T}^{(0)} (\bL^x, \bL^y) 
    \,,
\end{align}
\label{eq:p0_factorized_app}%
\end{subequations}
for any $\bn \in \mathcal N (\bL^x, \bL^y)$
because of the definition of $\bL^x(\bn)$ and $\bL^y(\bn)$
in Eqs.~\eqref{eq:Lx} and~\eqref{eq:Ly}, respectively.
% -
In the long-time limit of the fast time scale, i.e.,  $\tau \to \infty$,
$\piInf(\bn|\bL^x,\bL^y)$ is the steady state distribution 
of the chemical master equation~\eqref{eq:App:CME_fast_2} of 
a detailed-balanced (deficiency-zero) CRN (as explained in App.~\ref{Appendix:CMEFastTimeScale}).
% -
This means that 
\begin{equation}\small
    {\hat\omega_{\rhoe}(\bn)}
    \piInf(\bn|\bL^x,\bL^y)
    =
    {\hat\omega_{-\rhoe}(\bn + \colSe)}
    \piInf(\bn + \colSe|\bL^x,\bL^y)
    \label{eq:DB_condition_app}
\end{equation}
and, consequently,
\begin{equation}
    \eprez_{\infty,T} = 0\,,
\end{equation}
with $\eprez_{\infty,T} = \lim_{\tau \to \infty} \eprez_{\tau,T}$.
% -
% -

% -
% -
The first-order contribution reads
\begin{equation}\small
\begin{split}
    \epref_{\tau,T} = 
    &\sum_{\rhoe, \bn} 
    \Bigg\{
    \hat\omega_{\rhoe}(\bn) p_{\tau, T}^{(1)}(\bn) 
    \ln \frac
    {\hat\omega_{\rhoe}(\bn) p_{\tau, T}^{(0)}(\bn) }
    {\hat\omega_{-\rhoe}(\bn + \colSe)p_{\tau, T}^{(0)}(\bn + \colSe) } \\
    &+ \hat\omega_{\rhoe}(\bn) p_{\tau, T}^{(1)}(\bn)\\
    &- \frac{\hat\omega_{\rhoe}(\bn) p_{\tau, T}^{(0)}(\bn) 
    \hat\omega_{-\rhoe}(\bn + \colSe)p_{\tau, T}^{(1)}(\bn + \colSe)}
    {\hat\omega_{-\rhoe}(\bn + \colSe)p_{\tau, T}^{(0)}(\bn + \colSe) }
    \Bigg\}
    \,,
\end{split}
\end{equation}
where $p_{\tau, T}^{(0)}(\bn)$ and $p_{\tau, T}^{(0)}(\bn + \colSe)$
can be written as in Eq.~\eqref{eq:p0_factorized_app}.
% -
In the long-time limit of the fast time scale, i.e.,  $\tau \to \infty$,
Eq.~\eqref{eq:DB_condition_app} holds,
implying that
\begin{equation}
\begin{split}
    \epref_{\infty,T} 
    = & \sum_{\rhoe, \bn} 
    \Big\{
    \hat\omega_{\rhoe}(\bn) p_{\infty, T}^{(1)}(\bn)\\
    &
    - \hat\omega_{-\rhoe}(\bn + \colSe)p_{\infty, T}^{(1)}(\bn + \colSe)
    \Big\}
    \,,
\end{split}
\label{eq:epref_intermidiet}
\end{equation}
with $\epref_{\infty,T} = \lim_{\tau \to \infty} \epref_{\tau,T}$.
% -
% -
Finally, 
by applying the changes of variable 
$\bn + \colSe \to \bn$
and 
$-\rhoe \to \rhoe$ 
to the second summation on the right-hand side of Eq.~\eqref{eq:epref_intermidiet},
we obtain that
\begin{equation}
    \epref_{\infty,T} = 0\,.
\end{equation}

%%%%%%%%%%%%%%%%%%%%%%%%%%%%%%%%%%%%%%%%%%%%%%%%%%%%%%%%%%%%
\subsection{Contributions due to the $\Rint$ Reactions \label{App:EPRi}}
We determine here the first-order contribution to $\epri_{\tau, T}$.
% -
% -
We start by using Eq.~\eqref{eq:PerturbAnsatz} in Eq.~\eqref{eq:EPRi_def},
which yields  
\begin{equation}\small
    \eprif_{\tau,T} = 
    \sum_{\rhoi, \bn} 
    \hat\omega_{\rhoi}(\bn) p_{\tau, T}^{(0)}(\bn) 
    \ln \frac
    {\hat\omega_{\rhoi}(\bn) p_{\tau, T}^{(0)}(\bn) }
    {\hat\omega_{-\rhoi}(\bn + \colSi)p_{\tau, T}^{(0)}(\bn + \colSi) } 
    \,
    \label{eq:eprif_app_0}
\end{equation}
where $p_{\tau, T}^{(0)}(\bn)$ and $p_{\tau, T}^{(0)}(\bn + \colSi)$ 
can be written, 
according to Eq.~\eqref{eq:p0_factorized},
as
\begin{subequations}\small
\begin{align}
    p_{\tau, T}^{(0)}(\bn) & = 
    \pi_{\tau}^{(0)}(\bn | \bL^x, \bL^y) \, 
    p_{T}^{(0)} (\bL^x, \bL^y) 
    \,,
    \\
    p_{\tau, T}^{(0)}(\bn + \colSi) & =     
    \begin{aligned}[t]
        &\pi_{\tau}^{(0)}(\bn + \colSi | \bL^x + \colSxi, \bL^y + \colSyi) \\
        &\times p_{T}^{(0)} (\bL^x + \colSxi, \bL^y + \colSyi) \,,
    \end{aligned}
\end{align}
\end{subequations}
for any $\bn \in \mathcal N (\bL^x, \bL^y)$
and any $(\bn + \colSi) \in \mathcal N (\bL^x + \colSxi, \bL^y + \colSyi)$
as discussed in App.~\ref{Appendix:CMESlowTimeScale}.
% -
% -
In the long-time limit of the fast time scale, i.e.,  $\tau \to \infty$,
the term
\begin{equation}
    \frac{
    \hat\omega_{\rhoi}(\bn)
    \piInf (\bn | \bL^x, \bL^y)
    }{
    \hat\omega_{-\rhoi}(\bn + \colSi)
    \piInf (\bn + \colSi | \bL^x + \colSxi, \bL^y + \colSyi)
    }
    \label{eq:ratio_rr_avg}
\end{equation}
entering the logarithm in Eq.~\eqref{eq:eprif_app_0} can be rewritten as
\begin{equation}
    \frac{\avgInf{\hat \omega_{\rhoi} | \bL^x, \bL^y}}   
    {\avgInf{\hat \omega_{-\rhoi} | \bL^x + \colS_{x,\rhoi}, \bL^y + \colS_{y,\rhoi}}} 
    \,
    \label{eq:ratio_rr_avg_avg}
\end{equation}
by
i) 
using Eq.~\eqref{eq:MassActionOpen} for closed CRNs
together with Eqs.~\eqref{eq:Binomial} and~\eqref{eq:Binomialy};
ii)
recalling that 
$\bn \in \mathcal N (\bL^x, \bL^y)$ (resp. $(\bn + \colSi) \in \mathcal N (\bL^x + \colSxi, \bL^y + \colSyi)$)
implies that $n_x = L^{x}$ (resp. $n_x + S_{x, \rhoi} = L^{x} + S_{x, \rhoi}$)
$\forall x \in \Spex$;
iii)
noting that $(n_y - \nu_{y,\rhoi}) = (n_y + S_{y, \rhoi} - \nu_{y,-\rhoi})$
as well as $(L^y - n_y) = ((L^y + S_{y, \rhoi}) - (n_y + S_{y, \rhoi} ))$;
iv)
multiplying and dividing by 
$\theta(L^y -\nu_{y,\rhoi}) = \theta(L^y + S_{y, \rhoi} - \nu_{y,-\rhoi})$
as well as
$(L^y -\nu_{y,\rhoi})! = (L^y + S_{y, \rhoi} - \nu_{y,-\rhoi})!$;
v)
recalling the expression of the average reaction rates in Eq.~\eqref{eq:avg_rr_tss}.
% -
% - 
Hence, according to Eq.~\eqref{eq:eprif_app_0},  
the average entropy production rate
$\eprif_{\infty,T} = \lim_{\tau \to \infty} \eprif_{\tau,T}$ reads
\begin{widetext}
\begin{equation}\small
    \eprif_{\infty,T} = 
    \sum_{\rhoi}
    \sum_{\bL^x, \bL^y}
    \underbrace{
    \sum_{\bn \in \mathcal N (\bL^x, \bL^y)}
    \hat\omega_{\rhoi}(\bn) 
    \piInf(\bn | \bL^x, \bL^y)}_{= \avgInf{\hat \omega_{\rhoi} | \bL^x, \bL^y}} 
    p_{T}^{(0)} (\bL^x, \bL^y) 
    \ln \frac
    {\avgInf{\hat \omega_{\rhoi} | \bL^x, \bL^y}
    p_{T}^{(0)} (\bL^x, \bL^y)}
    {\avgInf{\hat \omega_{-\rhoi} | \bL^x + \colS_{x,\rhoi}, \bL^y + \colS_{y,\rhoi}}
    p_{T}^{(0)} (\bL^x + \colSxi, \bL^y + \colSyi)} 
    \,,
    \label{eq:eprif_app_1}
\end{equation}
\end{widetext}
where we recognized that
the sum $\sum_{\bn \in \mathcal N (\bL^x, \bL^y)}$ performs 
the average of the reaction rate $\hat \omega_{\rhoi}(\bn)$ over the fast dynamics,
weighted by the steady state distribution $\piInf(\bn | \bL^x, \bL^y)$.
% -
% -
We thus obtain the expression of the first-order contribution to $\epri_{\tau, T}$ 
in the long time limit of the fast time scale
given in Eq.~\eqref{eq:EPRi_epsilon}.
% -
% -

% -
% - 
We now apply the abundance separation assumption to $\eprif_{\infty,T}$ in Eq.~\eqref{eq:eprif_app_1}.
% -
% -
First,
the average reaction rates are given, to leading order, in Eq.~\eqref{eq:rr_dcc0}.
This implies that
\begin{widetext}
\begin{equation}\small
    \eprif_{\infty,T} = 
    \sum_{\rhoi}
    \sum_{\bL^x, \bL^y}
    \hat R_{\rhoi}(\bL^x, \bL^y)
    p_{T}^{(0)} (\bL^x, \bL^y) 
    \ln \frac
    {\hat R_{\rhoi}(\bL^x, \bL^y)
    p_{T}^{(0)} (\bL^x, \bL^y)}
    {\hat R_{-\rhoi}(\bL^x+ \colSxi, \bL^y)
    p_{T}^{(0)} (\bL^x + \colSxi, \bL^y + \colSyi)} 
    + \mathcal{O}(\Omega^{-1})
    \,.
    \label{eq:epr_withR}
\end{equation}
\end{widetext}
% -
% -
Second,
we express $p_{T}^{(0)} (\bL^x, \bL^y)$ according to Eq.~\eqref{eq:conditional_LxLy}
and we use the change of variables from $\bL^y$ to $\boldsymbol \eta^y $
(with $\boldsymbol \eta^y  = (\dots, \eta^y, \dots)$ 
and $\eta^y \equiv L^y/ \Omega = \mathcal{O} (1)$
as $\Omega \to \infty$)
together with Eq.~\eqref{eq:def_probOmega}
thus obtaining 
\begin{widetext}
\begin{equation}\small
    \eprif_{\infty,T} \approx 
    \sum_{\rhoi}
    \sum_{\bL^x} \int \mathrm d \boldsymbol \eta^y \,
    \hat R_{\rhoi}(\bL^x, \Omega \boldsymbol \eta^y )
    P^{(\Omega)}_{T} (\boldsymbol \eta^y | \bL^x)
    P^{(\Omega)}_{T} (\bL^x)
    \ln \frac
    {\hat R_{\rhoi}(\bL^x, \Omega \boldsymbol \eta^y )
    P^{(\Omega)}_{T} (\boldsymbol \eta^y | \bL^x)
    P^{(\Omega)}_{T} (\bL^x)}
    {\hat R_{-\rhoi}(\bL^x + \colSxi, \Omega \boldsymbol \eta^y )
    P^{(\Omega)}_{T} \Big(\boldsymbol \eta^y + \frac{\colS_{y,\rhoi}}{\Omega} \Big| \bL^x + \colS_{x,\rhoi}\Big)
    P^{(\Omega)}_{T} (\bL^x + \colSxi)} 
    \,,
    \label{eq:eprif_inf_almost_pre_Laplace}
\end{equation}
\end{widetext}
where we neglected the $\mathcal{O}(\Omega^{-1})$ terms.
% -
Finally,
we apply the Laplace's saddle-point approximation~\eqref{eq:LaplaceApprox}
and use  
\begin{equation}\small
     P^{(\Omega)}_{T} \Big(\boldsymbol \eta^y + \frac{\colS_{y,\rhoi}}{\Omega} \Big| \bL^x + \colS_{x,\rhoi}\Big) = 
    P^{(\Omega)}_{T} (\boldsymbol \eta^y  | \bL^x + \colS_{x,\rhoi}) 
    + \mathcal{O}(\Omega^{-1})
    \,,
\end{equation}
together with the leading-order equivalence 
\begin{equation}
    P^{(\Omega)}_{T} (\boldsymbol \eta^y_{\ast} | \bL^x + \colS_{x,\rhoi})
    \approx 
    P^{(\Omega)}_{T} (\boldsymbol \eta^y_{\ast}  | \bL^x)
\end{equation}
which follows from the $\bL^x$-independence of $\boldsymbol \eta^y_{\ast}$ (see Ref.~\cite{menz12}).
% -
This, together with 
the multiscale expansion of $P^{(\Omega)}_{T} (\bL^x)$ in Eq.~\eqref{eq:PerturbationAnsatz},
transforms
Eq.~\eqref{eq:eprif_inf_almost_pre_Laplace} into
\begin{equation}
\begin{split}
    \eprif_{\infty,T} \approx 
    &\sum_{\rhoi}
    \sum_{\bL^x} 
    \hat R_{\rhoi}(\bL^x, \Omega \boldsymbol \eta^y_{\ast} )
    P_{T} (\bL^x)\\
    \times& \ln \frac
    {\hat R_{\rhoi}(\bL^x, \Omega \boldsymbol \eta^y_{\ast} )
    P_{T} (\bL^x)}
    {\hat R_{-\rhoi}(\bL^x + \colSxi, \Omega \boldsymbol \eta^y_{\ast} )
    P_{T} (\bL^x + \colSxi)} 
    \,,
    \label{eq:eprif_inf_almost_post_Laplace}
\end{split}
\end{equation}
where we neglected the $\mathcal{O}(\Omega^{-1})$ terms.
% -
% -
Crucially, Eq.~\eqref{eq:eprif_inf_almost_post_Laplace} is equivalent to Eq.~\eqref{eq:emeEPR} 
upon identifying 
\begin{subequations}
\begin{align}
    \hat r_{\rhoi}(\bL^x|[\boldsymbol y]) &= \hat R_{\rhoi}(\bL^x, \Omega\boldsymbol \eta^y_{\ast})
    \,,
    \\
    \hat r_{-\rhoi}(\bL^x + \colSxi|[\boldsymbol y]) &= \hat R_{-\rhoi}(\bL^x + \colSxi, \Omega\boldsymbol \eta^y_{\ast})
    \,,
\end{align}
\end{subequations}
with $[\boldsymbol y]\equiv {\eta^y_{\ast} \Omega \, \mathfrak p_y}/{V}$ given in Eq.~\eqref{eq:y_conc}.

%%%%%%%%%%%%%%%%%%%%%%%%%%%%%%%%%%%%%%%%%%%%%%%%%%%%%%%%%%%%
%%%%%%%%%%%%%%%%%%%%%%%%%%%%%%%%%%%%%%%%%%%%%%%%%%%%%%%%%%%%
%%%%%%%%%%%%%%%%%%%%%%%%%%%%%%%%%%%%%%%%%%%%%%%%%%%%%%%%%%%%

\section{Semi-Analytical Analysis of the Illustrative Example of 
the Emergent Open-CRN Thermodynamics }
\label{App:LinearReactions}

We analyze here the dynamics and thermodynamics 
of the closed unimolecular CRN introduced in Sec.~\ref{Sec:LinearExample}
as well as 
of the corresponding emergent open CRN
by using a semi-analytical approach.

%%%%%%%%%%%%%%%%%%%%%%%%%%%%%%%%%%%%%%%%%%%%%%%%%%%%%%%%%%%%
\subsection{Closed CRN}

%%%%%%%%%%%%%%%%%%%%%%%%%%%%%%%%%%%%%%%%%%%%%%%%%%%%%%%%%%%%
\paragraph*{Dynamics.} 
For the closed unimolecular CRN in Sec.~\ref{Sec:LinearExample},
the chemical master equation~\eqref{eq:CME} specializes to
\begin{widetext}    
\begin{equation}
    \begin{split}
        \mathrm d_t p_t(n_x,n_{y_1},n_{y_2},n_{Y_1},n_{Y_2}) 
        =&\epsilon \big\{
        \hat\kappa_{1_\mathrm{i}} (n_x + 1)p_t(n_x+1,n_{y_1}-1,n_{y_2},n_{Y_1},n_{Y_2}) 
        + \hat\kappa_{-1_\mathrm{i}}(n_{y_1} + 1)p_t(n_x-1,n_{y_1}+1,n_{y_2},n_{Y_1},n_{Y_2})\\
        &+ \hat\kappa_{2_\mathrm{i}} (n_x + 1)p_t(n_x+1,n_{y_1},n_{y_2}-1,n_{Y_1},n_{Y_2}) 
        + \hat\kappa_{-2_\mathrm{i}}(n_{y_2}+1)p_t(n_x-1,n_{y_1},n_{y_2}+1,n_{Y_1},n_{Y_2})\\
        &- (\hat\kappa_{1_\mathrm{i}} n_x 
        + \hat\kappa_{-1_\mathrm{i}}n_{y_1} 
        + \hat\kappa_{2_\mathrm{i}} n_x 
        + \hat\kappa_{-2_\mathrm{i}}n_{y_2})
        p_t(n_x,n_{y_1},n_{y_2},n_{Y_1},n_{Y_2})
        \big\}\\
        &+ \hat\kappa_{1_\mathrm{e}}(n_{y_1}+1)p_t(n_x,n_{y_1}+1,n_{y_2},n_{Y_1}-1,n_{Y_2}) 
        + \hat\kappa_{-1_\mathrm{e}}(n_{Y_1}+1)p_t(n_x,n_{y_1}-1,n_{y_2},n_{Y_1}+1,n_{Y_2}) \\
        &+ \hat \kappa_{2_\mathrm{e}}(n_{y_2}+1)p_t(n_x,n_{y_1},n_{y_2}+1,n_{Y_1},n_{Y_2}-1)  
        + \hat\kappa_{-2_\mathrm{e}}(n_{Y_2}+1)p_t(n_x,n_{y_1},n_{y_2}-1,n_{Y_1},n_{Y_2}+1)\\
        & -(\hat\kappa_{1_\mathrm{e}}n_{y_1}
        +\hat\kappa_{-1_\mathrm{e}}n_{Y_1}
        +\hat\kappa_{2_\mathrm{e}}n_{y_2}
        +\hat\kappa_{-2_\mathrm{e}}n_{Y_2})
        p_t(n_x,n_{y_1},n_{y_2},n_{Y_1},n_{Y_2})\,.
    \end{split}
    \label{eq:App:CMELinearTwo}
\end{equation}
\end{widetext}
Since the quantity ${L^M \equiv n_{x} + n_{y_1}+ n_{y_2}+ n_{Y_1}+ n_{Y_2}}$
is conserved by both the $\Rint$ and $\Rexc$ reactions 
in Eqs.~\eqref{eq:LinearInternal} and~\eqref{eq:LinearExchange}, respectivly,
the solution of Eq.~\eqref{eq:App:CMELinearTwo} can be written 
using a multinomial ansatz~\cite{gardiner},
\begin{widetext}
\begin{equation}
\begin{split}
    p_t(n_x,n_{y_1},n_{y_2},n_{Y_1},n_{Y_2}) = 
    \frac{L^M!}{n_x!n_{y_1}!n_{y_2}!n_{Y_1}!n_{Y_2}}
    \mathfrak p_x(t)^{n_x}
    \mathfrak p_{y_1}(t)^{n_{y_1}}
    \mathfrak p_{y_2}(t)^{n_{y_2}}
    \mathfrak p_{Y_1}(t)^{n_{Y_1}}
    \mathfrak p_{Y_2}(t)^{n_{Y_2}}\,,
    \end{split}
    \label{eq:App:AnsatzLinearTwo}
\end{equation}
where the success probabilities 
$\mathfrak p_x(t)$, $\mathfrak p_{y_1}(t)$,
$\mathfrak p_{y_2}(t)$,
$\mathfrak p_{Y_1}(t)$,
and
$\mathfrak p_{Y_2}(t)$ 
satisfy the following ordinary differential equation
\begin{equation}
    {\mathrm d_t }{\boldsymbol{\mathfrak p}}(t) = 
    \begin{bmatrix}
            -\epsilon\,(\hat\kappa_{1_\mathrm{i}}+\hat\kappa_{2_\mathrm{i}}) 
            &\epsilon\,\hat\kappa_{-{1_\mathrm{i}}}
            &\epsilon\,\hat\kappa_{-{2_\mathrm{i}}}
            &0 
            &0
            \\
            \epsilon\,\hat\kappa_{1_\mathrm{i}}
            &-(\epsilon\,\hat\kappa_{-{1_\mathrm{i}}} + \hat\kappa_{{1_\mathrm{e}}})
            &0
            &\hat\kappa_{-{1_\mathrm{e}}} 
            &0
            \\
            \epsilon\, \hat\kappa_{2_\mathrm{i}}
            &0
            &-(\epsilon\,\hat\kappa_{-{2_\mathrm{i}}} + \hat\kappa_{2_\mathrm{e}})
            &0
            &\hat\kappa_{-2_\mathrm{e}}
            \\
            0
            &\hat\kappa_{1_\mathrm{e}}
            &0
            &-\hat\kappa_{-1_\mathrm{e}}
            &0
            \\
            0
            &0
            &\hat\kappa_{2_\mathrm{e}}
            &0
            &-\hat\kappa_{-2_\mathrm{e}}        
    \end{bmatrix} {\boldsymbol{\mathfrak p}}(t)\,,\label{eq:App:LinearODE}
\end{equation}
\end{widetext}
with 
${\boldsymbol{\mathfrak p}(t) \equiv  
(\mathfrak p_x(t),
\mathfrak p_{y_1}(t),
\mathfrak p_{y_2}(t),
\mathfrak p_{Y_1}(t),
\mathfrak p_{Y_2}(t))}$.
% -
% -
Correspondingly, 
the vector collecting all average numbers of molecules,
i.e., $\braket{\bn}_t \equiv (
\braket{n_{x}}_t,
\braket{n_{y_1}}_t,
\braket{n_{y_2}}_t,
\braket{n_{Y_1}}_t,
\braket{n_{Y_2}}_t
)$
can be written as
\begin{equation}
    \braket{\bn}_t \equiv L^M {\boldsymbol{\mathfrak p}}(t)
    \,.
    \label{eq:App:LinearMean}
\end{equation}
% -
% -

% -
% -
The results illustrated in Sec.~\ref{Sec:LinearExample} are obtained
by numerically solving Eq.~\eqref{eq:App:LinearODE} 
with initial condition 
${\boldsymbol{\mathfrak p}}(0) =
(0, 0, 0, V[Y_1], V[Y_2]) / L^M$ and $L^M = V[Y_1]+V[Y_2]$.

%%%%%%%%%%%%%%%%%%%%%%%%%%%%%%%%%%%%%%%%%%%%%%%%%%%%%%%%%%%%
\paragraph*{Thermodynamics.} 
By using Eq.~\eqref{eq:App:AnsatzLinearTwo} in Eq.~\eqref{eq:EPR},
the average entropy production rate of the closed unimolecular CRN in Sec.~\ref{Sec:LinearExample}
can be written as a function of the success probabilities 
according to
\begin{widetext}
\begin{equation}
\begin{split}
    \braket{\dot \Sigma}_t 
    =&\underbrace{ \epsilon\bigg\{ 
    \big(
    \hat\kappa_{1_\mathrm{i}} \mathfrak p_x(t) 
    - \hat\kappa_{-1_\mathrm{i}}\mathfrak p_{y_1}(t)
    \big)
    \ln\frac{
    \hat\kappa_{1_\mathrm{i}} \mathfrak p_x(t)
    }{
    \hat\kappa_{-1_\mathrm{i}}\mathfrak p_{y_1}(t)
    }
    +\big(
    \hat\kappa_{2_\mathrm{i}} \mathfrak p_x(t) 
    - \hat\kappa_{-2_\mathrm{i}}\mathfrak p_{y_2}(t)
    \big)
    \ln\frac{
    \hat\kappa_{2_\mathrm{i}} \mathfrak p_x(t)
    }{
    \hat\kappa_{-2_\mathrm{i}}\mathfrak p_{y_2}(t)}\bigg\}
    }_{=\braket{\dot \Sigma_\mathrm{i}}_t}\\
    & + 
    \underbrace{
    \big(
    \hat\kappa_{1_\mathrm{e}} \mathfrak p_{y_1}(t) 
    - \hat\kappa_{-1_\mathrm{e}} \mathfrak p_{Y_1}(t)
    \big)
    \ln\frac{
    \hat\kappa_{1_\mathrm{e}}  \mathfrak p_{y_1}(t)
    }{
    \hat\kappa_{-1_\mathrm{e}} \mathfrak p_{Y_1}(t)
    }
    +\big(
    \hat\kappa_{2_\mathrm{e}}  \mathfrak p_{y_2}(t) 
    - \hat\kappa_{-2_\mathrm{e}}\mathfrak p_{Y_2}(t)
    \big)
    \ln\frac{
    \hat\kappa_{2_\mathrm{e}} \mathfrak p_{y_2}(t)
    }{
    \hat\kappa_{-2_\mathrm{e}}\mathfrak p_{Y_2}(t)}
    }_{= \braket{\dot \Sigma_\mathrm{e}}_t} \,.
\end{split}
\label{eq:App:LinearTwoEPRClosed}
\end{equation}
\end{widetext}
% -

% -
The results illustrated in Sec.~\ref{Sec:LinearExample} are obtained
by numerically solving Eq.~\eqref{eq:App:LinearODE}
and substituting the solution into Eq.~\eqref{eq:App:LinearTwoEPRClosed},
with time expressed in terms of the variable $T$,
i.e., $t = T / \epsilon$.

%%%%%%%%%%%%%%%%%%%%%%%%%%%%%%%%%%%%%%%%%%%%%%%%%%%%%%%%%%%%
\subsection{Emergent Open CRN}

%%%%%%%%%%%%%%%%%%%%%%%%%%%%%%%%%%%%%%%%%%%%%%%%%%%%%%%%%%%%
\paragraph*{Dynamics.}
For the emergent open CRN in Sec.~\ref{Sec:LinearExample},
the chemical master equation~\eqref{eq:emeCME} specializes to
\begin{widetext}
\begin{equation}
    \begin{split}
        \mathrm d_T P_T(n_x) 
        = &\hat\kappa_{1_\mathrm{i}}(n_x+1) P_T(n_x+1) 
        + V \hat\kappa_{-1_\mathrm{i}}[y_1]P_T(n_x-1) 
        - (\hat\kappa_{1_\mathrm{i}}n_x + V \hat\kappa_{-1_\mathrm{i}}[y_1]) P_T(n_x) \\
        & + \hat\kappa_{2_\mathrm{i}}(n_x+1) P_T(n_x+1) 
        + V \hat\kappa_{-2_\mathrm{i}}[y_2]P_T(n_x-1) 
        - (\hat\kappa_{2_\mathrm{i}}n_x + V \hat\kappa_{-2_\mathrm{i}}[y_2]) P_T(n_x)
        \,,
    \end{split}
    \label{eq:App:LinearCMEOpen}
\end{equation}
\end{widetext}
whose solution can be written using a Poisson ansatz~\cite{gardiner},
\begin{equation}
    P_T(n_x) = 
    \frac{\mathfrak c_x(T)^{n_x}}{n_x!}\exp\{-\mathfrak c_x(T)\}
    \,,
    \label{eq:App:AnsatzLinearTwoChemostat}
\end{equation}
where $c_x(T)$ satisfies the following ordinary differential equation
\begin{equation}
    \mathrm d_T \mathfrak c_x(T)
    =
    -(\hat\kappa_{1_\mathrm{i}}+\hat\kappa_{2_\mathrm{i}})\mathfrak c_x(T)
    +V (\hat\kappa_{-1_\mathrm{i}}[y_1] + \hat\kappa_{-2_\mathrm{i}}[y_2])
    \,.
    \label{eq:App:dot_cx}
\end{equation}
% -
% -

% -
% -
The results illustrated in Sec.~\ref{Sec:LinearExample} are obtained
by analytically solving Eq.~\eqref{eq:App:dot_cx} with initial condition $\mathfrak c_x(0) = 0$
yielding 
\begin{equation}
    \mathfrak c_x(T) = 
    V
    \frac{
    \hat\kappa_{-1_\mathrm{i}}[y_1] + \hat\kappa_{-2_\mathrm{i}}[y_2]
    }{
    \hat\kappa_{1_\mathrm{i}}+\hat\kappa_{2_\mathrm{i}}}
    \big\{ 
    1 -e^{-(\hat\kappa_{1_\mathrm{i}}+\hat\kappa_{2_\mathrm{i}})T}
    \big\}
    \,.
    \label{eq:App:LinearTwoMeanChemo}
\end{equation}

%%%%%%%%%%%%%%%%%%%%%%%%%%%%%%%%%%%%%%%%%%%%%%%%%%%%%%%%%%%%
\paragraph*{Thermodynamics.} 
By using Eq.~\eqref{eq:App:LinearCMEOpen} in Eq.~\eqref{eq:emeEPR},
the average entropy production rate of the emergent open CRN in Sec.~\ref{Sec:LinearExample}
becomes
\begin{equation}
    \begin{split}
        \braket{\hat{\dot\Sigma}}_T
         = &
         \{ \hat\kappa_{1_\mathrm{i}} \mathfrak c_x(T) - V\hat\kappa_{-1_\mathrm{i}} [y_1]\} 
         \ln\frac{ \hat\kappa_{1_\mathrm{i}}  \mathfrak c_x(T)}{V\hat\kappa_{-1_\mathrm{i}} [y_1]} \\
         & +
         \{ \hat\kappa_{2_\mathrm{i}} \mathfrak c_x(T) - V\hat\kappa_{-2_\mathrm{i}} [y_2]\} 
         \ln\frac{ \hat\kappa_{2_\mathrm{i}}  \mathfrak c_x(T)}{V\hat\kappa_{-2_\mathrm{i}} [y_2]}
         \,.
    \end{split}
    \label{eq:App:LinearTwoEPRChemo}
\end{equation}
% -

% -
The results illustrated in Sec.~\ref{Sec:LinearExample} are obtained
by 
using the analytical solution of Eq.~\eqref{eq:App:LinearTwoMeanChemo} in Eq.~\eqref{eq:App:LinearTwoEPRChemo}.

%%%%%%%%%%%%%%%%%%%%%%%%%%%%%%%%%%%%%%%%%%%%%%%%%%%%%%%%%%%%
%%%%%%%%%%%%%%%%%%%%%%%%%%%%%%%%%%%%%%%%%%%%%%%%%%%%%%%%%%%%
%%%%%%%%%%%%%%%%%%%%%%%%%%%%%%%%%%%%%%%%%%%%%%%%%%%%%%%%%%%%

\section{Generalization to Multimolecular $\Rexc$ Reactions\label{app:GeneralChemostatting}}
We examine here the differences in the derivations of
the emergent open-CRN dynamics and thermodynamics
when the $\Rexc$ reactions have the multimolecular stoichiometry given in Eq.~\eqref{eq:ExchangeReaction}
as compared with
the corresponding derivations based on the unimolecular $\Rexc$ reactions 
in Eq.~\eqref{eq:CRN_ExchangeLinear}.
% -
% -

% -
% - 
\subsection{Implications of the Abundance Separation on the Steady State Probability of the Fast Dynamics
\label{app:pss_mm}}
Each $\piInfy(n_y|\bL^\lambda)$ given in Eq.~\eqref{eq:piInfy_mm} 
and featuring 
the steady state probability of the fast dynamics in Eq.~\eqref{eq:pss_mm}
boils down to the Poisson distribution in Eq.~\eqref{eq:piInfy_poisson_mm} 
due to the abundance separation.
% -
% -
Indeed, $a_Y(\bL^\lambda) = \mathcal{O}(\Omega)$ and, therefore, 
each factor ${(\mathfrak c_Y)^{(a_Y(\bL^\lambda) - b_{Y,y} n_y)}}/{(a_Y(\bL^\lambda) - b_{Y,y} n_y)!}$ 
in Eq.~\eqref{eq:piInfy_mm}
can be simplified using Stirling's approximation to
\begin{equation}
    e^{\Big\{\big(a_Y(\bL^\lambda) - b_{Y,y} n_y\big)
    \big(\ln(\mathfrak c_Y) - \ln(a_Y(\bL^\lambda) - b_{Y,y} n_y) + 1\big)
    \Big\}} 
    \,.
\end{equation}
% -
Factorizing $a_Y(\bL^\lambda)$ yields
\begin{equation}
    e^{\Big\{a_Y(\bL^\lambda) 
    \Big(1 - \frac{b_{Y,y} \, n_y}{a_Y(\bL^\lambda)}\Big)
    \Big(\ln(\mathfrak c_Y) - \ln(a_Y(\bL^\lambda)) - \ln\Big(1 - \frac{b_{Y,y} \, n_y}{a_Y(\bL^\lambda)}\Big) + 1\Big)
    \Big\}} 
    \,,
\end{equation}
where ${b_{Y,y} n_y}/{a_Y(\bL^\lambda)} = \mathcal{O}(\Omega^{-1})$,
since both $n_y$ and $b_{Y,y}$ are $\mathcal{O}(1)$
as $\Omega \to \infty$:
$n_y$ by assumption,
while $b_{Y,y}$ because it arises from products
of matrices defined by the conservation laws.
% -
A Taylor expansion in ${b_{Y,y} n_y}/{a_Y(\bL^\lambda)}$ then yields
\begin{equation}\small
\begin{split}
    \exp\Big\{a_Y(\bL^\lambda) 
    \Big(&\big(\ln(\mathfrak c_Y) - \ln(a_Y(\bL^\lambda)) + 1\big) \\
    -&\big(\ln(\mathfrak c_Y) - \ln(a_Y(\bL^\lambda)) \big)\frac{b_{Y,y} \, n_y}{a_Y(\bL^\lambda)}
    + \mathcal{O}(\Omega^{-2})
    \Big)
    \Big\}
    \,.
\end{split}
\end{equation}
Therefore, to leading order, 
\begin{equation}\small
\begin{split}
    \frac{(\mathfrak c_Y)^{(a_Y(\bL^\lambda) - b_{Y,y} n_y)}}
    {(a_Y(\bL^\lambda) - b_{Y,y} n_y)!}
    &\approx
    \exp\Big\{
    - \big(\ln (\mathfrak c_Y) - \ln (a_Y(\bL^\lambda))\big)  b_{Y,y} n_y \\
    & + (a_Y(\bL^\lambda))\big(\ln (\mathfrak c_Y) - \ln (a_Y(\bL^\lambda)) + 1\big)
    \Big\}\,.
\end{split}
\label{eq:simplified_factor_piInfy_mm}
\end{equation}
By now plugging Eq.~\eqref{eq:simplified_factor_piInfy_mm} into Eq.~\eqref{eq:piInfy_mm}
and 
recognizing that the factor in the second line of Eq.~\eqref{eq:simplified_factor_piInfy_mm} is 
$n_y$-independent,
we obtain
\begin{equation}
    \piInfy(n_y|\bL^\lambda)
    \propto 
    \frac{(\tilde{\mathfrak c}_y (\bL^\lambda))^{n_y}}{n_y !}
\end{equation}
with $\tilde{\mathfrak c}_y (\bL^\lambda)$ defined in Eq.~\eqref{eq:cy_tilda}.
% -
% -
Finally, the normalization constant of $\piInfy(n_y|\bL^\lambda)$
can be approximated as $\sum_{n_y = 0}^{\infty} {(\tilde{\mathfrak c}_y (\bL^\lambda))^{n_y}}/ {n_y !} $
because, as a result of the abundance separation,
the support of $\piInfy(n_y|\bL^\lambda)$ is effectively confined within its domain 
$N_y (\bL^\lambda) =
\big\{ n_y \in \mathbb N : 
(a_Y(\bL^\lambda) - b_{Y,y} n_y) \geq 0\,\, \forall Y \in \SpeYY(y)
\big\}$ 
so that, for normalization purposes, 
$N_y (\bL^\lambda)$ can be identified with the set of non-negative integers 
$\mathbb N$.
% -
% -

% -
% - 
\subsection{Implications of the Abundance Separation on the Average Reaction Rates \label{app:w_mm}}
The terms $\{\tilde{\mathfrak c}_y( \bL^\lambda - \colS_{\lambda,\rhoi})\}$ 
featuring the average reaction rates 
$\{\avgInf{\hat \omega_{\rhoi} | \bL^x - \colS_{x,\rhoi}, \bL^\lambda - \colS_{\lambda,\rhoi}}\}$
in Eq.~\eqref{eq:avg_rr_tss_mm}
can be approximated, to leading order, by $\{\tilde{\mathfrak c}_y( \bL^\lambda)\}$.
% -
% -
Indeed, according to Eq.~\eqref{eq:cy_tilda},
\begin{equation}
    \tilde{\mathfrak c}_y( \bL^\lambda - \colS_{\lambda,\rhoi}) =
    e^{
    \{\ln (\mathfrak c_y) - \sum_{Y \in \SpeYY(y)}
    (\ln (\mathfrak c_Y) - \ln (a_Y(\bL^\lambda  - \colS_{\lambda,\rhoi})))  b_{Y,y}\}
    }
    \,,
    \label{eq:cy_tilda_S}
\end{equation}
where, according to Eq.~\eqref{eq:aY},
\begin{equation}\small
    a_Y(\bL^\lambda  - \colS_{\lambda,\rhoi}) 
    = \sum_{\lambda \in \Lambda(y)} \icle_\lambda^Y (L^\lambda  - S_{\lambda,\rhoi})
    = {a_Y(\bL^\lambda)}
    - \sum_{\lambda \in \Lambda(y)} {\icle_\lambda^Y S_{\lambda,\rhoi}}\,,
\end{equation}
with the first and second terms on the right-most hand side being 
$\mathcal{O}(\Omega)$ and $\mathcal{O}(1)$, respectively, 
as $\Omega \to \infty$
.
% -
This implies that
\begin{equation}\small
\begin{split}
    \ln (a_Y(\bL^\lambda  - \colS_{\lambda,\rhoi}))
    & = \ln \bigg\{a_Y(\bL^\lambda)  
    - \sum_{\lambda \in \Lambda(y)} {\icle_\lambda^Y S_{\lambda,\rhoi}}\bigg\}\\
    = \ln & (a_Y(\bL^\lambda))
    + \ln \bigg\{1   
    - \sum_{\lambda \in \Lambda(y)} \frac{\icle_\lambda^Y S_{\lambda,\rhoi}}{a_Y(\bL^\lambda)}\bigg\}\\
    = \ln & (a_Y(\bL^\lambda)) 
    - \sum_{\lambda \in \Lambda(y)} \frac{\icle_\lambda^Y S_{\lambda,\rhoi}}{a_Y(\bL^\lambda)}
    + \mathcal{O}(\Omega^{-2})
    \,,
\end{split}
\label{eq:aY_mm_approx}
\end{equation}
since ${\icle_\lambda^Y S_{\lambda,\rhoi}}/{a_Y(\bL^\lambda)} = \mathcal{O}(\Omega^{-1})$.
% -
Finally, 
by plugging Eq.~\eqref{eq:aY_mm_approx} into Eq.~\eqref{eq:cy_tilda_S},
we obtain
\begin{equation}
    \tilde{\mathfrak c}_y( \bL^\lambda - \colS_{\lambda,\rhoi}) = 
    \tilde{\mathfrak c}_y( \bL^\lambda) + \mathcal{O}(\Omega^{-1}) 
    \,.
    \label{eq:cy_tilda_S_leading}
\end{equation}
% -
% -

% -
% - 
\subsection{Implications of the Abundance Separation on the Derivation of the Average Entropy Production Rate \label{app:epr_mm}}
As mentioned in the main text,
the derivation of the leading-order contribution to the the average entropy production rate
follows same steps as in App.~\ref{App:EPR}. 
% -
The only minor difference is that the term in Eq.~\eqref{eq:ratio_rr_avg},
which reads
\begin{equation}
    \frac{
    \hat\omega_{\rhoi}(\bn)
    \piInf (\bn | \bL^x, \bL^\lambda)
    }{
    \hat\omega_{-\rhoi}(\bn + \colSi)
    \piInf (\bn + \colSi | \bL^x + \colSxi, \bL^\lambda + \colS_{\lambda, \rhoi})
    }
\end{equation}
when the $\Rexc$ reactions are multimolecular reactions, 
becomes, to leading order,
\begin{equation}
    \frac
    {\hat R_{\rhoi}(\bL^x, \bL^\lambda)}
    {\hat R_{-\rhoi}(\bL^x+ \colSxi, \bL^\lambda)} 
\end{equation}
instead of Eq.~\eqref{eq:ratio_rr_avg_avg} 
by using Eqs.~\eqref{eq:pss_mm} and~\eqref{eq:piInfy_poisson_mm} 
together with 
$\tilde{\mathfrak c}_y( \bL^\lambda - \colS_{\lambda,\rhoi}) \approx \tilde{\mathfrak c}_y( \bL^\lambda) $
according to Eq.~\eqref{eq:cy_tilda_S_leading}.
% -
As a consequence, 
the first-order contribution $\eprif_{\tau,T}$ given in Eq.~\eqref{eq:eprif_app_0} 
to the dissipation due to the $\Rint$ reactions
can be directly written as in Eq.~\eqref{eq:epr_withR} 
by simultaneously using the time-scale separation and the abundance separation.

%%%%%%%%%%%%%%%%%%%%%%%%%%%%%%%%%%%%%%%%%%%%%%%%%%%%%%%%%%%%
%%%%%%%%%%%%%%%%%%%%%%%%%%%%%%%%%%%%%%%%%%%%%%%%%%%%%%%%%%%%
%%%%%%%%%%%%%%%%%%%%%%%%%%%%%%%%%%%%%%%%%%%%%%%%%%%%%%%%%%%%

\bibliography{refs.bib}

\end{document}